\documentclass[twocolumn,aps,showpacs,floatfix]{revtex4-1}
\usepackage{graphicx}
\usepackage{epsf}
\usepackage{amsmath}
\usepackage{amsfonts}
\usepackage{amssymb}
\usepackage{dcolumn}
\usepackage{bm}
\usepackage{colordvi}           
\usepackage{bigstrut}           
\usepackage{multirow}           
\usepackage{epsfig} 
\usepackage[figuresright]{rotating}
\DeclareMathOperator{\sgn}{sgn} 

\def\gsim{\lower -0.3ex \hbox{$>$} \kern -0.75em \lower 0.7ex
\hbox{$\sim$}}
\def\lsim{\lower -0.3ex \hbox{$<$} \kern -0.75em \lower 0.7ex \hbox{$\sim$}}
\def\Journal #1,#2,#3,#4#5#6#7{#1 {\bf #2}, #3 (#4#5#6#7)}
\begin{document}
\title{Topological Phases in Two-Dimensional Materials: A Brief Review}
\author{Yafei Ren}
\affiliation{ICQD, Hefei National Laboratory for Physical Sciences at the Microscale, University of Science and Technology of China, Hefei, Anhui 230026, China}
\affiliation{Department of Physics, and Synergetic Innovation Center of Quantum Information and Quantum Physics, University of Science and Technology of China, Hefei, Anhui 230026, China}
\author{Zhenhua Qiao}
\thanks{qiao@ustc.edu.cn}
\affiliation{ICQD, Hefei National Laboratory for Physical Sciences at the Microscale, University of Science and Technology of China, Hefei, Anhui 230026, China}
\affiliation{Department of Physics, and Synergetic Innovation Center of Quantum Information and Quantum Physics, University of Science and Technology of China, Hefei, Anhui 230026, China}
\author{Qian Niu}
\thanks{niu@physics.utexas.edu}
\affiliation{Department of Physics, The University of Texas at Austin, Austin, Texas 78712, USA}

\begin{abstract}
  Topological phases with insulating bulk and gapless surface or edge modes have attracted much attention because of their fundamental physics implications and potential applications in dissipationless electronics and spintronics. In this review, we mainly focus on the recent progress in the engineering of topologically nontrivial phases (such as $\mathbb{Z}_2$ topological insulators, quantum anomalous Hall effects, quantum valley Hall effects \textit{etc.}) in two-dimensional material systems, including quantum wells, atomic crystal layers of elements from group III to group VII, and the transition metal compounds.

  \textit{This article was invited by Prof. M.-Y. Chou.}
\end{abstract}

\maketitle
\tableofcontents

\section{Introduction}\label{introduction}
Topological phases of condensed matter have attracted immense attention ever since the first proposal in the 1980s~\cite{TKNN_82} to use this concept to explain the intriguing properties of the quantum Hall effect in two-dimensional (2D) electronic systems under a strong external magnetic field~\cite{QHE_Klitzing_80}. The quantum Hall effect is manifested by a vanishing longitudinal conductance but nonzero quantized Hall conductance in a six-terminal Hall-bar measurement~\cite{QHE_Klitzing_80,QHE_Klitzing_04}. The vanishing longitudinal conductance originates from the insulating bulk while the quantized Hall conductance arises intrinsically from the Berry curvatures of the filled magnetic Bloch bands, as reported by Thouless \textit{et al}~\cite{TKNN_82}. The Berry-curvature integration over the filled bands in the magnetic Brillouin-zone gives rise to an integer named as Thouless-Kohmoto-Nightingale-Nijs (TKNN) number. Later, this expression was recognized as the first Chern class of a \textit{U}(1) principal fiber bundle on a torus, where the fibers and torus correspond respectively to the magnetic Bloch waves and the magnetic Brillouin zone~\cite{Chern_Simon_83,Chern_Simon1_83}. Therefore, the TKNN number is also known as the Chern number, which is a topological invariant in the sense that the integer will not change as long as the bulk band gap, wherein the Fermi level lies, is not closed by the external perturbations~\cite{TKNN_Niu_85}. The Chern number is closely related to the amplitude of the quantized Hall conductance in units of $e^2/h$, reflecting the topological nature of quantum Hall effect, i.e. the Hall conductance is quantized as an integer as long as the bulk band gap remains open.

For an experimental sample with a finite size, the topology of the filled bands is reflected by the one-dimensional (1D) gapless chiral edge states according to the principle of the \textit{bulk-edge} correspondence, where the Chern number counts the number of edge states localized at each boundary between the quantum Hall effect and the vacuum~\cite{Bulk_edge_correspondence_Hatsugai_93}. Due to the topological protection of the one-way propagating characteristic at each boundary, the edge modes are robust against weak disorders in any form. The high-precision of the quantized Hall resistance plateau has let the quantum Hall effect become a new method to determine the fine structure constant, a fundamental physical constant~\cite{QHE_Klitzing_04}. Moreover, the robustness feature together with the vanishing longitudinal resistance has attractive practical potential in design of dissipationless or low-power electronic devices.

However, to realize the quantum Hall effect, the most crucial requirement is to apply a strong magnetic field. Unfortunately, it is beyond the current state of technology to generate such a huge magnetic field outside the laboratory. Therefore, the question naturally arises: is it possible to achieve the quantum Hall effect in the absence of a strong magnetic field? It is known that the necessary condition to induce a Hall effect is to break the time-reversal invariance, which can be achieved by either an external magnetic field or the intrinsic ferromagnetism~\cite{rev_AHE_Ong_10}. The former results in the ordinary Hall effect, while the latter produces the anomalous Hall effect. Both effects actually exhibits the same transport characteristics.
In the 2D limit, the strong magnetic field can lead to Landau-level quantization, hence the quantum Hall effect~\cite{QHE_Klitzing_80,QHE_Klitzing_04}. Therefore, the anomalous Hall effect is also expected to become quantized in 2D systems with ferromagnetism by some means. Hereinbelow, we refer to the quantum Hall effect in the absence of the external magnetic field as the quantum \textit{anomalous} Hall effect (QAHE). In 1988, Haldane~\cite{QAHE_Haldane_88} theoretically achieved this expectation by an exquisite toy model in a 2D honeycomb-lattice system by considering alternating magnetic fields with zero net flux. However, since the host material of the 2D honeycomb-lattice was believed to be unrealistic up until 2004 and the experimental realization of alternative magnetic fluxes is also extremely difficult, the further progress towards the realization of the QAHE was seriously hampered. Only two alternative proposals in Kagom\'e lattice and disordered 2D ferromagnetic metals were reported~\cite{QAHE_Nagaosa_00,QAHE_Nagaosa_03}. Nevertheless, such a theoretical proposal not only raised the hope of achieving QAHE for dissipationless applications, but also immediately inspired great interest when the era of 2D materials finally comes.

On the other hand, during that period investigations of the dynamics of the magnetic Bloch electrons within the partially occupied magnetic Bloch bands revealed the underlying importance of the Berry curvature in the dynamical processes~\cite{Niu_95,Niu_96}. The further generalization to ordinary Bloch electrons in crystals without a strong magnetic field opened up new grounds in the study of the topological properties of 2D electron gases~\cite{Niu_99}. Especially, this inspired a revisit to the old issues of the anomalous Hall effect including the spin-Hall effect~\cite{rev_AHE_Ong_10}. Apart from the fundamental interest of the anomalous/spin-Hall effect, potential applications in spintronics to generate dissipationless spin current further stimulated research, which led to the discovery of the topological contribution to the spin-Hall effect~\cite{SHE_ZhangSC_03,SHE_MacDonald_04,SHE_ZhangSC_05,SHE_expAwschalom_04, SHE_expJungwirth_05}. However, it is shown that the longitudinal charge current in the spin-Hall effect is nonzero hence dissipative. To overcome this difficulty, a spin-Hall insulator was proposed by Murakami \textit{et al} in 2004, wherein a finite spin-Hall conductance is expected in the bulk insulating/zerogap materials~\cite{QSHE_ZhangSC_04}. These pioneering works established a solid foundation for future research on topological phases.

The era of 2D materials started from the year of 2004, when graphene, a single layer of carbon atoms arranged in a honeycomb-lattice structure, was first successfully exfoliated~\cite{graphene_Novoselov_04}. Its unique mechanical, electrical, and optical properties and its special linear-Dirac dispersion soon attracted great interest from various research fields, making it a \textit{star} material~\cite{rev_Graphene_Neto_10}. However, its half-filled conduction bands with gapless Dirac dispersion limits its applications of graphene in semiconductor-based electronics. Different binary degrees of freedom (i.e. real spin, AB sublattices, KK$^\prime$ valleys, and top/bottom layers) have been adopted to engineer bulk band gaps that are able to harbour various topological phases. In particular, graphene provides a real 2D honeycomb-lattice platform to revive Haldane's proposal of the QAHE.

Soon after the discovery of graphene, based on Haldane's model, Kane and Mele made a great stride forward towards the realization of QAHE~\cite{QAHE_Haldane_88}. They showed that through the introduction of the next-nearest neighbor hopping, which induces an intrinsic spin-orbit coupling and opens a bulk band gap at the Dirac points. This insulating phase hosts two copies of QAHE with opposite spin polarization and chiralities. That is, opposite Chern numbers are present for spin-up and spin-down electrons, respectively. The resulting spin-helical gapless edge states counter-propagate along the same boundary with opposite spins, which results in a vanishing charge Hall conductance yet quantized spin-Hall conductance. This phase was therefore called the quantum spin Hall effect (QSHE)~\cite{QSHE_Mele_05}, and it is preserved even when the spin is no longer a good quantum number, indicating its topological nature that is characterized by the $\mathbb{Z}_2$ topological order~\cite{QSHE_Mele_05-1}.
The Karmers-degenerate spin-helical edge states are robust against weak disorders due to the topological protection from time-reversal symmetry. This insulating phase is therefore also called a topological insulator (TI), to embrace a broader connotation in addition to the QSHE~\cite{QSHE_Mele_05-1}. Although the intrinsic spin-orbit coupling in graphene was later shown to be extremely weak, making the Kane-Mele model unrealistic~\cite{SOC_G_MacDonald_06,SOC_G_Brataas_06,SOC_G_Fang_07,SOC_G_Fabian_09,SOC_G_Trickey_07}, much effort has been put in producing a strong intrinsic spin-orbit coupling in graphene through some external means (e.g. by doping heavy atoms)~\cite{QSHE_G_Wu_11} and in searching for new graphene-like materials that possess strong intrinsic spin-orbit coupling (e.g. low-buckled honeycomb-lattice materials)~\cite{QSHE_Sn_Zhang_13}.

Almost parallel to the proposal of Kane and Mele, Bernevig \textit{et al} suggested another route, i.e. band inversion by spin-orbit coupling, to realize QSHE in strained zinc-blende semiconductors~\cite{QSHE_ZhangSC_06} and HgTe quantum wells~\cite{QSHE_QW_HgTe_ZhangSC_06}. In 2007, one year after these theoretical proposals, QSHE was indeed experimentally observed in inverted HgTe quantum wells~\cite{QSHE_QW_HgTe_exp_ZhangSC_07}. These pioneering studies laid the foundations for the field of the $\mathbb{Z}_2$ TI, which was subsequently extended to three-dimensional (3D) materials~\cite{rev_TI_Bernevig,rev_TI_Kane_10,rev_TI_ShenSQ,rev_TI_Zhang_11}.

Both the theoretical and experimental advances of 2D and 3D TIs have also inspired the exploration of the QAHE by breaking time-reversal symmetry.
In 2010, it was theoretically predicted that the QAHE can be achieved after a band inversion that originates from the ferromagnetism in the 3D TI thin films~\cite{QAHE_MagTI_FangZh_10}. Following this theoretical recipe, in 2013 after three years of continuous effort, the QAHE was finally experimentally observed in Cr-doped (Bi, Sb)$_2$Te$_3$ thin films~\cite{QAHE_MagTI_exp_XueQK_13}. Since then, there have been three other follow-up observations of the QAHE in the same host materials of (Bi, Sb)$_2$Te$_3$ doped with Cr~\cite{QAHE_MagTI_exp_Checkelsky_14,QAHE_MagTIexp_WangKL_14} or V~\cite{QAHE_MagTI_exp_Moodera_15} atoms. However, all the experimental observations were achieved at extremely low temperatures (i.e. $30$-$100~$mK). Therefore, how to raise the temperature of the QAHE is a critical challenge for the communities of both condensed matter physics and materials science, from both the theoretical and experimental sides.

Engineering band gaps at the Dirac points of graphene is another rewarding route to realize the QAHE, which, in general, requires both spin-orbit coupling and intrinsic ferromagnetism~\cite{QAHE_MagTI_FangZh_10}. Although the intrinsic spin-orbit coupling in graphene is extremely weak, there exists another extrinsically tunable spin-orbit coupling, i.e. Rashba spin-orbit coupling~\cite{2D_electron_hole_SOC}, which arises from the mirror-symmetry breaking about the graphene plane, e.g. by applying a perpendicular electric field~\cite{QAHE_G_Qiao_10}. In 2010, it was proposed that the QAHE can occur in graphene by simultaneously considering the Rashba spin-orbit coupling and out-of-plane exchange field (or Zeeman field), which can be introduced via the proximity effect through doping magnetic atoms or utilizing ferromagnetic insulating substrates~\cite{QAHE_G_adatom_Qiao_11,QAHE_G_5d_Mokrousov_12,QAHE_G_AFM_Qiao_14}. Although the QAHE has not been experimentally realized in graphene yet, there has already been striking progress towards the ultimate realization of the QAHE. Recently, a sizable anomalous Hall conductance of $\sigma_{xy}\approx 0.2 e^2/h$ has been observed in graphene proximity-coupled with a magnetic thin film YIG~\cite{QAHE_G_AFM_exp_Shi_15}. To expedite the definitive observation of QAHE in graphene, a crucial issue is to enlarge the Rashba spin-orbit coupling that is strongly dependent on the strength of the van der Waals interaction between the graphene sheet and the magnetic substrate.

In graphene, the binary KK$^\prime$ valley degree of freedom can also be adopted to design topological \textit{valleytronics} similar to spintronics by leveraging the valley-pseudospin in the manner of electron spin~\cite{valleytronics_G_Beenakker_07}. The analogy between valley-pseudospin and electron spin also inspired the prediction of the quantum valley-Hall effect (QVHE), for which the K and K$^\prime$ valleys carry nonzero Chern numbers but with opposite signs~\cite{QVHE_G_NiuQ_07}. Therefore, this phase is well-defined only when KK$^\prime$ valleys are decoupled in the absence of short-range scattering. This new phase can be realized by breaking the inversion symmetry, e.g. introducing the staggered AB sublattice potentials in monolayer graphene or applying a perpendicular electric field in the Bernal-stacked multilayer graphene~\cite{QVHE_G_NiuQ_07,QSHE_BLG_Qiao_11,QSHE_TLG_Qiao_12}. It is noteworthy that, different from the QSHE, the bulk-edge correspondence is absent in monolayer graphene but present in multilayer graphene.

Interestingly, when the QVHE coexists with the QAHE or TI, the resulting edge modes can even be robust against weak short-range scattering since the QVHE originates from the inversion symmetry breaking that is compatible with the preservation or breaking of the time-reversal invariance. In addition to the edge states, gapless interface modes can also be generated along the interface between two QVHEs with opposite valley topologies (except for the exact armchair case that exhibits an unavoidable band gap). These interface states are also known as topological confinement states, kink states, zero modes, or zero-line modes (ZLMs) in the literature. In order to manifest the formation along the line of zero electric field, for clarity, we shall refer to this as ZLMs. This kind of ZLMs have been shown to exist naturally at the interfaces between different topological phases~\cite{ZLM_Yang_15}.

Apart from these topologically nontrivial phases induced by spin-orbit coupling via the band inversion or band gap opening in Dirac semi-metals, the strong Coulomb interaction can also induce QSHE or QAHE in honeycomb-lattice systems~\cite{TMI_Raghu_08}, Kagom\'e-lattice systems~\cite{QAHE_Nagaosa_00}, and bilayer graphene~\cite{QAHE_BLG_Interaction_Levitov_10,QSHE_BLG_Fertig_11, QSHE_BLG_MacDonald_11,QSHE_BLG_MacDonald1_11}. Moreover, in 2011, the topological family was further extended to include the fractional QAHE in fractionally-filled flat bands~\cite{FQAHE_Neupert_11,FQAHE_Sun_11,FQAHE_Tang_11}, the Floquet topological insulator driven by the time-dependent periodic potential with gapless edge states in the quasi-energy spectrum~\cite{FTI_Lindner_11,FTI_Cayssol_13}, and the topological crystalline insulators that are topologically protected by mirror-reflection symmetry~\cite{TCI_Fu_11,rev_TCI_FuL_15}.

There are several excellent reviews related to the topics mentioned above, such as the anomalous Hall effect~\cite{rev_AHE_Ong_10, rev_AHE_Nagaosa_06}, the Berry-phase effect~\cite{rev_Berryphase_Niu_10}, 2D/3D TIs~\cite{rev_TI_Kane_10, rev_QSHE_Zhang_11, rev_QSHE_Zhang_10, rev_TI_ShenSQ, rev_TI_Bernevig}, topological crystalline insulator~\cite{rev_TCI_FuL_15}, and the QAHE in magnetic TIs~\cite{rev_QAHE_Kou_15,rev_QAHE_Wang_14}. In this review, we shall focus on the recent theoretical progress in studies of the 2D topological phases (including $\mathbb{Z}_2$ TI, QAHE and QVHE) of the noninteracting particles in atomic crystal layers and quasi-2D quantum wells in the absence of strong magnetic field. In Sec.~\ref{TI}, we review the recent work on TIs built from atomic layers of the group-IV and -V elements, and quasi-2D quantum wells. In Sec.~\ref{QAHE}, the QAHE is reviewed, beginning with its theoretical prediction and experimental realization in magnetic TIs, followed by the honeycomb-lattice based QAHE. In section~\ref{QVHE}, we review the QVHE in graphene and related systems. The electronic structure and transport properties of ZLMs at the interfaces between two QVHE with different Chern numbers at each valley are also described. In Sec.~\ref{summary} we give a short discussion and summary.

\section{2D Topological Insulators (TIs)} \label{TI}
To realize the 2D $\mathbb{Z}_2$ TI, two typical proposals are raised independently. One, Kane-Mele model in graphene, was to open a bulk band gap at the two inequivalent Dirac points~\cite{QSHE_Mele_05} as a generalization of Haldane's model to spinful system with time reversal symmetry~\cite{QAHE_Haldane_88}. The other was the Bernevig-Hughes-Zhang (BHZ) proposal to induce a band inversion in a 2D semiconductor~\cite{QSHE_QW_HgTe_ZhangSC_06}. These two models not only pioneer the investigation of TIs, but also represent two general routes for realizing 2D TIs, i.e. by opening up a band gap in the 2D Dirac semi-metals or by inducing a band inversion in narrow-gap semiconductors.

In the following, we briefly describe the basic physical origins of these two models and survey recent theoretical suggestions of possible materials that would have large topologically nontrivial band gaps and would be experimentally feasible with potential practical applications. In Subsec.~\ref{TI_Haldane} and Subsec.~\ref{TI_IV}, the basic physics of the Kane-Mele model is reviewed, and then extended to the atomic-crystal layers of group-IV elements and organic honeycomb-lattice structures. The band inversion in the BHZ model is introduced in Subsec.~\ref{TI_BHZ_QW}. Based on these two basic formations, the theoretical proposals to create TIs within various 2D materials, such as graphene, atomic crystal layers of group-V elements, group III-V and IV-VI compounds, and transition metal dichalcogenides are reviewed in Subsecs.~\ref{TI_IV_Gexp}-\ref{TI_Fun}. The topological Anderson insulator as well as the time-reversal symmetry breaking QSHE are discussed in Subsec.~\ref{TI_TAI} and Subsec.~\ref{TI_TRB-QSHE}.

\subsection{The Honeycomb lattice and the Haldane model} \label{TI_Haldane}
\begin{figure*}
  \centering
  \includegraphics[width=12cm]{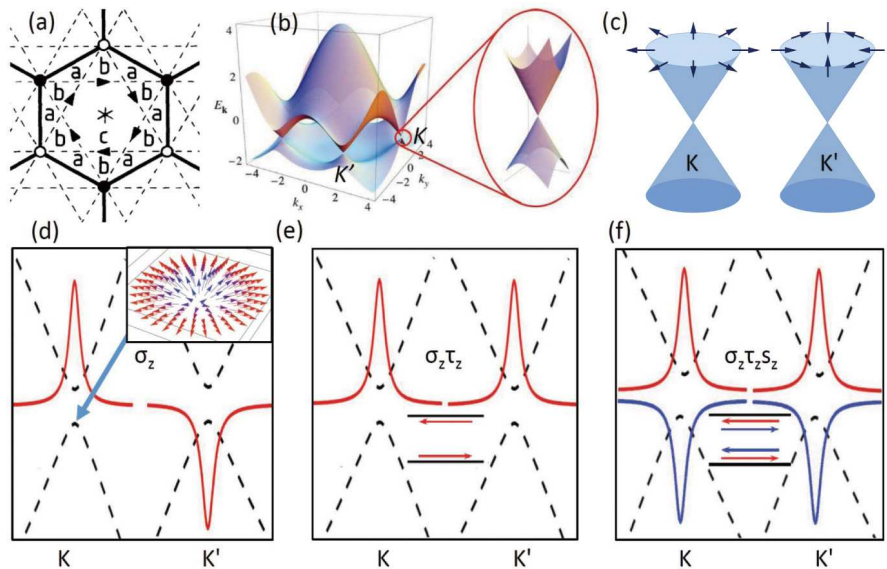}
  \caption{(color online). (a), (b), and (c) show the real-space lattice structure, electronic structure and pseudospin texture of pristine graphene, respectively.
  (d) Black dashed lines show the band structure of spinless particles in a honeycomb lattice with the staggered sublattice potential denoted by $\sigma_z$. The mass term opens a band gap at the K/K$^\prime$ points. Inset: pseudospin texture of valence band at K valley, corresponding to a meron. Red solid lines denote the Berry curvature profile, which has opposite signs at the K/K$^\prime$ valleys. (e) Band structure of Haldane's model in the absence of staggered sublattice potential where the Berry curvatures at the K/K$^\prime$ valleys have the same sign. The band gap is topologically nontrivial and shows the QAHE with a Chern number $\mathcal{C}=1$; the corresponding edge state is plotted in the inset. In this model, an alternating magnetic flux is applied to the honeycomb lattice. As shown in (a), the whole 2D plane is divided into three regions labeled by a, b, and c where the magnetic flux through a and b are of the same magnitude but opposite sign while region c has no magnetic flux. (f) Kane-Mele model: The intrinsic spin-orbit coupling of graphene $\sigma_z\tau_z s_z$ gives rise to two copies of Haldane's model with spin-up and -down bands of opposite Berry curvature as shown by red and blue solid lines, respectively. The corresponding Chern numbers of 1 and $-1$ lead to counter-propagating helical edge states shown in the inset. Fig.~(a) reprinted with permission from~[\onlinecite{QAHE_Haldane_88}], copyright 1988 by the American Physical Society. Fig.~(b) reprinted with permission from~[\onlinecite{rev_Graphene_Neto_10}], copyright 2010 by the American Physical Society. Inset of Fig.~(d) reprinted with permission from~[\onlinecite{meron_Ezawa_12}].}\label{Haldane_KaneMele}
\end{figure*}

As mentioned above, the single layer honeycomb-lattice structure plays an important role in both 2D TI and QAHE systems~\cite{QSHE_Mele_05,QSHE_Mele_05-1,QAHE_Haldane_88, QAHE_G_Qiao_10}. Let us first briefly describe the $\pi$-band electronic structure of a spinless particle in the planar honeycomb lattice as displayed in Fig.~\ref{Haldane_KaneMele}(a), where two sets of inequivalent triangular lattices are present, namely the AB sublattices labelled by the empty and solid circles, respectively. Sublattice symmetry, also known as chiral symmetry, occurs when the nearest neighbor hopping $t_1$ between these two sublattices is present~\cite{ChiralSymm_Koshino_14}. Therefore, only the off-diagonal terms that couple the AB sublattices are nonzero in the tight-binding momentum-space Hamiltonian~\cite{rev_Graphene_Neto_10}. When the coupling between the sublattices A and B vanishes, accidental degeneracy occurs with doubly degenerate zero-energy eigenstates that appear at the K and K$^{\prime}$ points, as shown in Fig.~\ref{Haldane_KaneMele}(b), where linear Dirac dispersions appear~\cite{rev_G_spinger}. In the long wavelength limit, the low-energy continuum model Hamiltonian of the Dirac dispersions can be expressed as
\begin{eqnarray}
  H(\bm{k}) &= & v(\tau_z k_x\sigma_x + k_y\sigma_y),
\end{eqnarray}
where $v=3t_1/2$ is the Fermi-velocity, and $\bm{\sigma}$ and $\bm{\tau}$ are the Pauli matrices for the sublattice and valley pseudospins, respectively. For a single valley, the effective Hamiltonian can be written as $H(\bm{k}) = \sum_i d_i \sigma_i $ ($i=x,y,z$) with $\bm{d}=(d_x,d_y,d_z)$ being the pseudospin texture. The chiral symmetry guarantees the vanishing of $d_z$ hence the pseudospin texture is in-plane as displayed in Fig.~\ref{Haldane_KaneMele}(c). Such a gapless linear dispersion can only become gapped by introducing a diagonal mass term, which breaks the sublattice symmetry.

The simplest way to break the sublattice symmetry is to consider a staggered sublattice potential $M\sigma_z$ that is momentum independent and opens bulk band gaps at K and K$^{\prime}$ points, as can be seen in Fig.~\ref{Haldane_KaneMele}(d). The out-of-plane pseudospin textures near the gapped points, i.e. nonvanishing $d_z=M$, leads to two merons [See the inset of Fig.~\ref{Haldane_KaneMele}(d)], which carry half Chern numbers~\cite{meron_Ezawa_11,meron_Moon_95,meron_Petkovic_07} that can be calculated by
\begin{equation}
\mathcal{C}=-\frac{1}{8\pi^2} \int d^2 k \bm{d} \cdot \partial_{k_x} \bm{d} \times \partial_{k_y} \bm{d}.
\label{CMeron}
\end{equation}
These two merons carry opposite topological charges, i.e. $\mathcal{C}_K=-\mathcal{C}_{K'}=0.5 \sgn(M)$, since these two valleys are related to each other by the time-reversal symmetry, which guarantees the vanishing total Chern number. Alternatively, the Chern number carried by different valleys can also be obtained by integrating the Berry curvatures in the momentum space around the two Dirac points with a more generalized definition:
\begin{equation}
\mathcal{C}=\frac{1}{2\pi}\sum_n \int d^2 k \Omega_n(\bm{k}),
\label{ChernNumber}
\end{equation}
where $\Omega_n$ is the $z$-component Berry curvature of the $n$-th occupied band. As a counterpart of the magnetic field in real space, the Berry curvature in momentum space is defined as $\Omega_n=(\nabla \times \bm{A}_n)_z$, where the Berry connection $\bm{A}_n(\bm{k}) = i \langle u_n(\bm{k})\vert\nabla_k | u_n(\bm{k})\rangle $ corresponds to the vector potential with $\vert u_n(\bm{k})\rangle$ denoting the periodic part of the Bloch function of the $n$-th band . More explicitly, the Berry curvature can be further expressed as
\begin{eqnarray}
\Omega_n(\bm{k})=-{\sum_{n^{\prime} \neq n}} {\frac{2 {\rm {Im}} \langle u_n |v_x|u_{n^\prime}\rangle \langle u_{n^\prime}|v_y|u_n \rangle } {(\omega_{n^\prime}-\omega_{n})^2}},
\label{BerryCurvature}
\end{eqnarray}
where the summation is over all the occupied valence bands below the gap, $\omega_n\equiv E_n/ \hbar$, and $v_{x(y)}$ is the velocity operator. In Fig.~\ref{Haldane_KaneMele}(d), the profile of the Berry curvature for the gapped Dirac cones is plotted in red solid lines . The time-reversal symmetry that relates valley K to K$^{\prime}$ requires that the two valleys have opposite Berry curvatures, leading to a vanishing Chern number. Although the total Chern number is zero, the difference between the Chern numbers that stem from the Berry curvature localized around K/K$^\prime$ valleys is quantized, which gives rise to the QVHE as will be discussed in Sec.~\ref{QVHE}.

On the other hand, a nonzero total Chern number can only be obtained if the mass terms at valleys K and K$^{\prime}$ have opposite signs, which breaks the time-reversal symmetry. Such a mechanism has been proposed by Haldane in an elegant model by applying alternating out-of-plane magnetic fields through a honeycomb-lattice structure~\cite{QAHE_Haldane_88}. As displayed in Fig.~\ref{Haldane_KaneMele}(a), the magnetic fields of opposite direction are applied in the ``a" and ``b" regions, i.e. $\phi_a=-\phi_b$. The zero net magnetic flux in a closed path surrounding the unit cell does not affect the nearest-neighbor hopping amplitude $t_1$. However, for the next-nearest-neighbor hopping amplitude $t_2$, the net flux is nonzero, and is given by $\phi=2\pi(2\phi_a+\phi_b)/\phi_0$ with $\phi_0=h/e$. Thus, $t_2$ acquires a phase factor to become $t_2e^{i\phi}$~\cite{QAHE_Haldane_88}. Because this factor is position-independent, the system also possesses translational symmetry. The corresponding low-energy continuum spinless Hamiltonian can be expressed as~\cite{rev_TI_ShenSQ}:
\begin{eqnarray}
    H(\bm{k}) & =& v(\tau_zk_x\sigma_x+k_y\sigma_y) + M \sigma_z + m \tau_z \sigma_z \\ \nonumber
                    &= & v \left[
                             \begin{array}{cccc}
                                M+m     &   \pi^{\dag} &                     &        \\
                               \pi  &   -(M+m)                 &                      &        \\
                                    &                     &    M-m                  & -\pi  \\
                                    &                     &  -\pi^{\dag}  &   -(M-m)      \\
                             \end{array}
                           \right],
\end{eqnarray}
where the second and third terms originate from the staggered sublattice potentials and next-nearest-neighbor hopping terms, respectively, with $m=-3\sqrt{3}t_2\sin\phi$. The mass terms at K and K$^{\prime}$ points are modified to be $(M+m)$ and $(M-m)$. As a consequence, the Chern numbers for valleys KK$^\prime$ are changed to  $\mathcal{C}_K=0.5 \sgn(M+m)$ and  $\mathcal{C}_{K'}=0.5\sgn(m-M)$. When $|m|>|M|$, the signs of the Merons at valleys K and K$^{\prime}$ are identical and thus $\mathcal{C}_K=\mathcal{C}_{K^{\prime}}$, giving rise to the QAHE characterized by a nonzero Chern number $\mathcal{C}=\mathcal{C}_{K}+\mathcal{C}_{K'}=\sgn(m)$. This corresponds to a pair of chirally propagating edge states along the boundaries as displayed in Fig.~\ref{Haldane_KaneMele}(e), exhibiting exactly the same transport properties as those of the magnetic field induced quantum Hall effect, for example, robustness against any kind of weak disorder~\cite{QAHE_Haldane_88,QSHE_Mele_05}.

\subsection{Honeycomb lattices of group-IV elements} \label{TI_IV}
\begin{figure*}[h]
  \includegraphics[width=12cm]{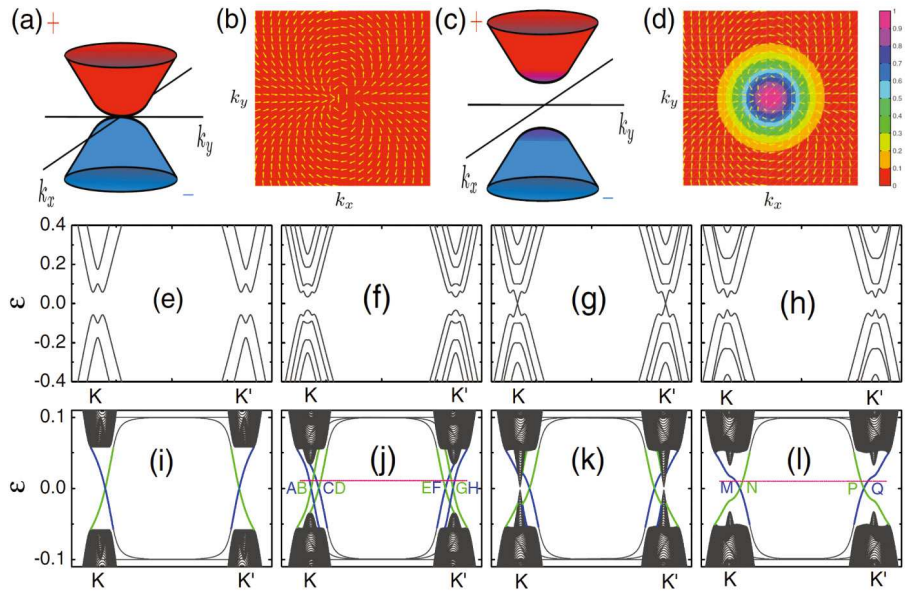}
  \caption[] {(color online). (a) and (c): Electronic structure of Bernal stacked bilayer graphene without and with different layer potentials. (b) and (d): Pseudospin texture of bilayer graphene corresponding to (a) and (c). (e)-(h): Bulk band structures along high-symmetry lines for different strengthes of Rashba spin-orbit coupling. (i)-(l): Band structure of zigzag bilayer graphene nanoribbon corresponding to Figs.~(e)-(h).
  Figs.~(a)-(d) reprinted with permission from~[\onlinecite{pseudospin_texture_MacDonald_12}]. Figs.~(e)-(l) reprinted figure with permission from~[\onlinecite{QSHE_BLG_Qiao_11}], copyright 2011 by the American Physical Society.}
 \label{QSHE_BLG}
\end{figure*}
\subsubsection{Graphene: Kane-Mele model} \label{TI_IV_G}
The Haldane model has proved to be a great success in advancing the design of dissipationless electronics in the absence of a magnetic field. However, in the 1980s it was unimaginable that such an ideal toy model could be realized in practice. The primary reasons were that (i) the 2D materials were believed to be unstable in nature~\cite{graphene_flat_unstable_Krishnan_97}, and (ii) the alternating magnetic fluxes were extremely difficult to impose in experiment. Therefore, for a rather long time, very little progress was made following Haldane's proposal until the graphene was successful exfoliated. Graphene is a 2D atomic crystal layer composed of carbon atoms arranged on a honeycomb lattice, making it an excellent test platform. Based on the earlier work of DiVincenzo and Mele on the spin-orbit coupling of graphite~\cite{QSHE_G_SOC_DiVincenzo_84}, in 2005 Kane and Mele in a seminal paper proposed that, in the long wavelength limit, the intrinsic spin-orbit coupling of graphene due to the next-nearest-neighbor hopping can be written as two copies of the Haldane model with the mass terms of the spin-up and spin-down electrons having opposite signs:
\begin{eqnarray}
  H(\bm{k})= v(\tau_zk_x\sigma_x+k_y\sigma_y) + m \tau_z \sigma_z s_z,
\end{eqnarray}
where $\bm{s}=(s_x,s_y,s_z)$ are the spin-Pauli matrices. For each single spin, the time-reversal symmetry is effectively broken and the mass term gives rise to a nonzero Chern number according to the Haldane model. Specifically, for the spin-up and -down bands, the mass terms are respectively $m$ and $-m$, leading to the Chern numbers $\mathcal{C}_{\uparrow}=\sgn(m)$ and $\mathcal{C}_{\downarrow}=-\sgn(m)$ corresponding to the counter-propagating edge states of opposite spins [see inset of Fig.~\ref{Haldane_KaneMele}(f)].
This leads to a vanishing charge Hall conductance but quantized spin-Hall conductance when $s_z$ is a good quantum number. Thus, this is called the QSHE. The two copies of edge modes with opposite spins are related to each other by the Kramers degeneracy theorem since time-reversal invariance is retrieved by the combination of $\tau_z \sigma_z$ with a Zeeman term $s_z$.

Compared with the quantum Hall effect, where the spatial separation of counter-propagating edge states protects them from any weak disorder induced backscattering, the counter-propagating edge modes of opposite spins are spatially overlapped and thus the backscattering is possible. However, by studying the four-terminal transport properties of this system, Sheng \textit{et al} found that the spin-related transport properties are rather robust/insensitive to the sample boundary conditions (such as zigzag or armchair-type boundaries) and exhibit a quantized spin-Hall conductance in the presence of relatively large disorder strengths~\cite{QSHE_Haldane_05}. Moreover, the robust time-reversal symmetry protected edge states are also present even when $s_z$ is no longer a good quantum number, as with Rashba spin-orbit coupling~\cite{QSHE_Mele_05-1}.

This robustness reveals a most important feature of the QSHE, i.e. the elastic backscattering between the two states within a Kramers degenerate pair is forbidden due to the protection from time-reversal symmetry~\cite{rev_TI_Zhang_11}.
However, backscattering between states in different Kramers degenerate pairs is allowed, making them annihilate together. Therefore, the system will become a trivial insulator if there is an even number of Kramers degenerate pairs at each boundary and the topologically nontrivial phase occurs only when the number of Kramers degenerate pair is odd, hence the name ``topological insulator" coined by Kane and Mele~\cite{QSHE_Mele_05-1}. Such a property intrinsically classifies the time-reversal symmetric insulators into two classes, characterized by the $\mathbb{Z}_2$ topological invariant~\cite{QSHE_Mele_05-1}:
\begin{equation}
\mathbb{Z}_2 = \frac{1}{2\pi}\left[\oint_{\partial\,
    \mathrm{HBZ}}\mathrm{d}\bm{k}\cdot\bm{A}(\bm{k})-\int_{\mathrm{HBZ}}\mathrm{d}^2k\, \Omega_z(\bm{k})\right]\,\mathrm{mod}(2).
\label{Z2}
\end{equation}
The presence of the $\mathrm{mod}(2)$ term makes the topological invariant can only take two values ``0" and ``1" indicating topologically trivial and nontrivial respectively, which reflects the binary classification of insulator with even or odd numbers of Kramers pairs. In addition, there are other definitions of the $\mathbb{Z}_2$ topological indices, which have been well described in Refs.~[\onlinecite{QSHE_G_Haldane_06}] and [\onlinecite{Z2_Inversion_Kane_07}]. For clarity and correctness of the definitions, hereinbelow, we refer to the QSHE as a ``2D topological insulator (2D TI)" no matter whether the spin is a good quantum number or not.

\subsubsection{Low-buckled honeycomb lattice} \label{TI_IV_LowBuckle}
The Kane-Mele model can also be applied to the low-buckled honeycomb-lattice structures of other group-IV element based 2D atomic crystal layers, e.g. silicene~\cite{QSHE_Si.Ge_Cahangirov_09,QSHE_Si_Lay_10,QSHE_Si_Lay1_10,QSHE_Si_Aufray_10}, germanene~\cite{QSHE_Si.Ge_Cahangirov_09,QSHE_Ge_Huang_12}, and stanene~\cite{QSHE_Sn_Zhang_13}, which are the respective counterparts of silicone, germanium, and tin. Similar structures exist in the 2D alloys of these elements~\cite{QSHE_Si.Ge_Cahangirov_09,QSHE_Si_Fazzio_13,QSHE_SiC3_Zhang_14}. The low-buckled honeycomb structure originates from the larger interatomic distances in these systems~\cite{QSHE_Si.Ge_Yao_11,QSHE_Si.Ge.Sn_Yao_11} and makes the atomic orbitals mix the $sp^3$ hybridization with the $sp^2$ one, which results in a first-order contribution of the atomic spin-orbit coupling to the intrinsic spin-orbit coupling of the Bloch electrons [see Fig.~\ref{QSHE_Bi_functionalized_Yao_14_Fig4}(c)]. Because of the higher atomic numbers of Si, Ge, and Sn, their larger intrinsic spin-orbit coupling induced bulk gaps can reach the orders of 1, 10, and 100 meV, respectively, making the $\mathbb{Z}_2$ TIs measurable under experimentally achievable temperatures~\cite{QSHE_Si.Ge.Sn_Yao_11,QSHE_Ge_Huang_12,QSHE_Sn_Zhang_13}. In addition, although the low-buckled structure naturally breaks the mirror symmetry about the plane leading to an intrinsic Rashba-type spin-orbit coupling, this is not detrimental to the 2D $\mathbb{Z}_2$ TIs since the intrinsic Rashba spin-orbit coupling is momentum-dependent and vanishes at the Dirac K/K$^\prime$ points~\cite{QSHE_Si.Ge.Sn_Yao_11}. Another striking property of the low-buckled structure is the external tunability when an electric field~\cite{QSHE_Si_Li_13,QSHE_Si_Ezawa_12,QSHE_Si_Falko_12,QSHE_Si_Schwingerschlogl_13} or  strain~\cite{QSHE_IV-V_Bansil_14,QSHE_GeCH3_Whangbo_14} is applied.

\subsubsection{Multilayer graphene} \label{TI_MLG}
Although the Bernal-stacked multilayer graphene is also a zero-gap semi-metal, the interlayer coupling modifies the linear Dirac dispersion at valleys KK$'$ of the monolayer graphene to become non-linear in the multilayer case~\cite{rev_BLG_McCann_13}. We begin with bilayer graphene as an example that has quadratic bands touching at K and K$^{\prime}$ points as displayed in Fig.~\ref{QSHE_BLG}(a), which gives rise to a different pseudospin texture from that of the monolayer graphene of Fig.~\ref{QSHE_BLG}(b)~\cite{pseudospin_texture_MacDonald_12}. In the presence of intrinsic spin-orbit coupling, even though either top or bottom layer can form a separate TI, the interlayer coupling induced combination of these two TIs gives rise to a trivial insulator~\cite{QSHE_G_Chen_14,QSHE_BLG_Fertig_11}. However, it is found that the extrinsic Rashba spin-orbit coupling due to the breaking of the mirror reflection symmetry $z \rightarrow -z$ by, e.g. applying a perpendicular electric field~\cite{2D_electron_hole_SOC}, adsorbing atoms~\cite{SOC_G_Rader_12}, or placing on top of a metallic substrate~\cite{SOC_G_Dedkov_08,SOC_G_Varykhalov_08,SOC_G_Rader_12}, can induce a $\mathbb{Z}_2$ TI assisted by the interlayer potential difference~\cite{QSHE_BLG_Qiao_11,QSHE_BLG_Qiao_13}. Such a TI phase can be understood in two limits as described in the following.

When the Rashba spin-orbit coupling $\lambda_{\rm{R}}$ is much larger than the interlayer potential difference $U$, the low-energy continuum model Hamiltonian can be expressed as
\begin{eqnarray}
H^{{\mathrm{eff}}}_{K}
=\frac{1}{\lambda_R}\left[
\begin{array}{cccc}
U{\lambda_R} & {i v^2 k^2_-} &  0 & {2i  t_\perp v k_-} \\
{-i v^2 k^2_+} & U{\lambda_R} & 0 & 0 \\
0 & 0 & -U{\lambda_R} & {i v^2 k^2_-} \\
{-2i  t_\perp v k_+} & 0 & {-i v^2 k^2_+} & -U{\lambda_R}
\end{array}
\right], \nonumber \\ \label{effctiveFormK1}
\end{eqnarray}
where the basis functions are mainly decided by \{$A_{1\uparrow}$, $B_{1\downarrow}$, $A_{2\uparrow}$, $B_{2\downarrow}$\}~\cite{QSHE_BLG_Qiao_13}. In the absence of a perpendicular electric field, the strong Rashba spin-orbit coupling lifts the spin degeneracy of the bands in the K and K$^{\prime}$ valleys by mixing the upward and downward spins as well as the layer pseudospin. However, the gapless character is preserved leading to both linear and quadratic band touching, which can be lifted by either a Zeeman term or an inequivalent layer potential resulting in a QAHE [see Sec.~\ref{QAHE}] or TI~\cite{QSHE_BLG_Qiao_13}. The weak inequivalent layer potential lifts both linear and quadratic band touching and gives rise to a Chern number at valley K $\mathcal{C}_K=\sgn (U)$, mainly attributed to the gapped linear dispersion. The symmetric time-reversal counterpart lies in K$^{\prime}$ valley with $\mathcal{C}_{K^\prime}=-\sgn (U)$. As a result, the two copies of QAHE possess opposite spin-orientations and valley indices, i.e. opposite momenta, corresponding to a TI with helical edge states. Different from the TI in monolayer graphene where the inversion symmetry and time-reversal symmetry guarantee the spin degeneracy of edge states, which have dispersion curves crossing the Brillouin zone and hence do not have a well-defined valley index, spin degeneracy is absent in bilayer graphene due to the breaking of inversion symmetry and edge states have a well-defined valley index, as can be seen in Fig.~\ref{QSHE_BLG}(l).
As a consequence, these edge states are protected not only by the time-reversal symmetry but also by their large momentum separation, displaying both TI and QVHE characteristics [see also Sec.~\ref{QVHE}].

On the other hand, in bilayer graphene, chiral symmetry also appears with the sublattices separately coming from the top and bottom layers. The separate contributions of sublattices from different layers makes it possible to induce staggered sublattice potentials by applying an external perpendicular electric field that can open a band gap, as shown in Fig.~\ref{QSHE_BLG}(c). Different from monolayer graphene where the pseudospin texture at each valley gives a half Chern number per spin by forming a meron, the quadratic coupling between these two sublattices leads to different pseudospin textures for gapped bilayer graphene as displayed in Fig.~\ref{QSHE_BLG}(d), that leads to a unit Chern number per spin and hence QVHE with gapless edge modes for zigzag nanoribbon [see also Sec.~\ref{QVHE}]. Nevertheless, even numbers of Kramers pairs at each boundary from both the spin and valley degeneracies makes this phase topologically trivial with $\mathbb{Z}_2=0$~\cite{QSHE_BLG_Qiao_11}. When the Rashba spin-orbit coupling is included, the spin-up and -down states are mixed to lift the spin degeneracy of the energy bands. The increase of $\lambda_{\rm{R}}$ induces an inversion between the lowest conduction and highest valence bands by closing and reopening the band gap at the Dirac points, as shown in Figs.~\ref{QSHE_BLG}(e)-\ref{QSHE_BLG}(h). During the phase transition, the two-fold spin degenerate edge states at each boundary become coupled and split by the Rashba spin-orbit coupling. This eliminates two edges states at each boundary and keeps only one Kramers degenerate pair that is topologically protected by the time-reversal symmetry as displayed in Figs.~\ref{QSHE_BLG}(i)-\ref{QSHE_BLG}(l). Such a topological phase transition from a trivial insulator to a $\mathbb{Z}_2$ TI can be well interpreted by the two-band BHZ model, as will be reviewed in the following Subsection~\cite{QSHE_BLG_Qiao_13,QSHE_BLG_LiJ_12}.

\begin{figure*}
  \centering
  \includegraphics[width=14 cm]{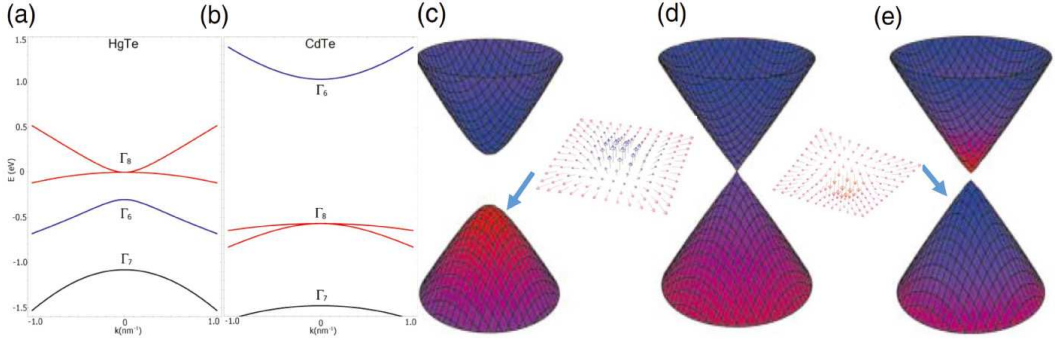}
  \caption{(color online). (a) and (b): Bulk energy bands of HgTe and CdTe near the $\Gamma$ point. (c)-(e) Low energy band structure of CdTe/HgTe/CdTe quantum well in the normal regime, critical point, and inverted regime, respectively. Each band is double degenerate due to inversion symmetry and time reversal symmetry. The inset figures between (c)-(e) show the pseudospin texture of the valence band corresponding to the upper panel. Left and right inset figures show two merons of opposite sign indicating the change of Chern number by 1.
  Figs.~(a)-(e) reprinted figure with permission from~[\onlinecite{QSHE_QW_HgTe_ZhangSC_06}]. }
  \label{QSHE_QW_HgTe_ZhangSC_06_Fig1}
\end{figure*}

Similar results can be reached for the ABC-stacked trilayer graphene~\cite{QSHE_TLG_Qiao_12}. However, different from the bilayer case with one pair of states at each valley, there is an unbalanced number of edge states at the two boundaries of a zigzag-terminated nanoribbon. Specifically, there are five pairs of spin-helical edge states located at one boundary with three pairs at the other boundary. When inter-valley scattering is introduced in the armchair ribbon, scattering destroys extra Kramers degenerate pairs and leaves only one pair of spin-helical edge states in both boundaries. Note that the number of spin-helical pairs at any boundary is strictly consistent with the requirement of \textit{odd} pairs of spin-helical edge modes in $\mathbb{Z}_2$ TIs~\cite{QSHE_TLG_Qiao_12}.

\subsection{Band-inversion in quantum wells} \label{TI_BHZ_QW}
Parallel to the idea of TI based on the Dirac semi-metal in graphene-like materials with half-filled $\pi$-bands, another route to realize 2D $\mathbb{Z}_2$ TIs is to utilize a semiconductor with fully-filled valence bands~\cite{QSHE_QW_HgTe_ZhangSC_06,QSHE_QW_HgTe_exp_ZhangSC_07,QSHE_InAs_ZhangSC_08,QSHE_InAs_exp_DuRR_11, QSHE_QW_InAs_DuRR_12,QSHE_QW_InAs_exp_DuRR_15,QSHE_QAHE_QW_Junction_ZhangSC_14}. 
Following the realization of the 4D generalization of the quantum Hall effect and spurred by the requirement of dissipationless spin current~\cite{SHE_ZhangSC_03,SHE_MacDonald_04, SHE_ZhangSC_05,SHE_expAwschalom_04,SHE_expJungwirth_05}, it was suggested to look for the QSHE in zinc-blende semiconductors~\cite{QSHE_ZhangSC_06}. Soon after, a quantum-well based $\mathbb{Z}_2$ TI was theoretically proposed in a CdTe/HgTe/CdTe heterostructure with proper tuning of the quantum-well thickness~\cite{QSHE_QW_HgTe_ZhangSC_06} that has been successfully observed~\cite{QSHE_QW_HgTe_exp_ZhangSC_07, QSHE_HgTe_exp_Moler_13}.

For conventional semiconductors composed of light elements, the filled valence bands are gapped from the conduction bands as shown in Fig.~\ref{QSHE_QW_HgTe_ZhangSC_06_Fig1}(b) for CdTe. By contrast, for HgTe, an inversion between $\Gamma_6$ and $\Gamma_8$ bands makes this material a zero-gap semiconductor as displayed in Fig.~\ref{QSHE_QW_HgTe_ZhangSC_06_Fig1}(a). In the symmetric quantum well of CdTe/HgTe/CdTe, a topologically trivial insulating phase occurs for the thin HgTe layer because of the dominating contribution from CdTe. When the thickness of HgTe increases, a band inversion occurs with increasing contribution from HgTe. At the critical point, the low-energy continuum model Hamiltonian can be written as
\begin{eqnarray}
    H_{\rm{eff}}(\bm{k}) = \epsilon(k) + \left[
                             \begin{array}{cccc}
                                m(k)    &   A\pi^{\dag} &                     &        \\
                               A^{*}\pi  &   -m(k)                 &                      &        \\
                                    &                     &    m(-k)             & -A^*\pi  \\
                                    &                     &  -A\pi^{\dag}  &   -m(-k)      \\ \nonumber
                             \end{array}
                           \right], \\
\end{eqnarray}
where the symmetric mass term is $\epsilon(k)=C-D(k_x^2+k_y^2)$ while the asymmetric mass term is $m(k)=M-B(k_x^2+k_y^2)$. The two decoupled blocks are related by the time-reversal operation and hence we only consider the upper-block in our following discussion. The $2\times 2$ effective Hamiltonian be regarded as the kinetic energy of a massive Dirac fermion expressed in the pseudospin Hilbert space. However, different from the previously introduced mass term in graphene, the mass term $m(k)$ has a momentum dependence that gives rise to different pseudospin textures, depending on the sign of $M$ and $B$. For a positive $B$, when $M$ is also positive, the mass term $m(k)$ is positive around $\bm{k}=0$ but negative for a large momentum. The corresponding pseudospin texture is schematically plotted in the inset of Fig.~\ref{QSHE_QW_HgTe_ZhangSC_06_Fig1}(c), representing two merons with opposite signs corresponding to a trivial insulator. When $M$ decrease to zero, the gapless Dirac dispersion appears as displayed in Fig.~\ref{QSHE_QW_HgTe_ZhangSC_06_Fig1}(d). When $M$ is negative, the band gap reopens and the mass term is negative for both the zero and large momenta, corresponding to a skyrmion that is composed of two merons with the same sign, as displayed in Fig.~\ref{QSHE_QW_HgTe_ZhangSC_06_Fig1}(e), and carries a unit Chern number $\mathcal{C}=1$. The time-reversal counterpart of this block possesses a Chern number of $\mathcal{C}=-1$. The two decoupled blocks therefore result in a vanishing total Chern number but counter-propagating edge states, forming a Kramers degenerate pair corresponding to a $\mathbb{Z}_2$ TI~\cite{QSHE_QW_HgTe_ZhangSC_06,rev_QSHE_Zhang_08}. 

In the symmetric quantum well of CdTe/HgTe/CdTe, by neglecting the bulk inversion asymmetry whose effect is small for HgTe based structure~\cite{QSHE_HgTe_Novik_05}, the inversion symmetry is preserved, allowing the eigenstates to possess well-defined parity. In the trivial insulating regime, the conduction band edge has odd parity while the valence edge has even parity~\cite{QSHE_QW_HgTe_ZhangSC_06}. The inversion symmetry guarantees the occurrence of band inversion for the time-reversal symmetric momentum, which induces a parity exchange, leading to a TI~\cite{QSHE_QW_HgTe_ZhangSC_06, Z2_Inversion_Kane_07}. For the structural inversion asymmetric quantum wells of AlSb/InAs/GaSb/AlSb, the band inversion can also occur and induce a $\mathbb{Z}_2$ TI phase, as theoretically predicted in Ref.~[\onlinecite{QSHE_InAs_ZhangSC_08}], and later experimentally realized~\cite{QSHE_InAs_exp_DuRR_11,QSHE_QW_InAs_DuRR_12, QSHE_QW_InAs_exp_DuRR_15}. However, at the critical point of the band inversion, two non-degenerate Dirac cones occur at the momenta of $\bm{k}_0$ and $-\bm{k}_0$~\cite{topo_PhaseTrans_Murakami_07}, which is different from that in the inversion symmetric system, where the doubly degenerate Dirac cones exist around the time-reversal symmetric momentum. The case of bilayer graphene is similar to the asymmetric quantum well with $\bm{k}_0$ and $-\bm{k}_0$ corresponding to the K and K$^\prime$ points, respectively~\cite{QSHE_BLG_Qiao_11,QSHE_BLG_Qiao_13}.

In addition to materials with inverted band structure, e.g. HgTe and InAs/GaSb heterostructure, band inversion induced TIs can also appear in conventional semiconductor composed quantum wells, like GaN/InN/GaN and GaAs/Ge/GaAs quantum wells~\cite{QSHE_QW_InN_Miao_12,QSHE_QW_Ge_Zhang_13}. Due to the sizable band gap of the conventional semiconductors, large enough electric field is required to reduce the band gap and induce band inversion. This large electric field is difficult to achieve via gate technology but may be realized by fabricating high-quality semiconductor heterostructure. For example, in GaN/InN/GaN quantum wells, the large strain due to the lattice mismatch between InN and GaN can induce sizable electric field via the strong piezoelectric effect~\cite{QSHE_QW_InN_Miao_12}. Alternatively, in GaAs/Ge/GaAs quantum wells, the strong electric field appears since the charge transfer at As-Ge interface is different from that at Ge-Ga interface~\cite{QSHE_QW_Ge_Zhang_13}. Simultaneously, the large electric field also enhance the spin-orbit coupling and hence the nontrivial band gap of TIs originating from the band inversion.

\subsection{Graphene-based experimental prototypes} \label{TI_IV_Gexp}
Although the 2D $\mathbb{Z}_2$ TI was first predicted in monolayer graphene, it is regarded as extremely unrealistic since the intrinsic spin-orbit coupling in graphene is very minute due to the small atomic number of carbon and the distinctive planar honeycomb lattice~\cite{SOC_G_MacDonald_06,SOC_G_Brataas_06,SOC_G_Fang_07,SOC_G_Trickey_07, SOC_G_Fabian_09}. Specifically, in ideal planar graphene, the spin-orbit coupling of the $\pi$-band is only about $1~\mu$eV due to the second-order contribution via the virtual transitions to the $\sigma$-orbit~\cite{SOC_G_Brataas_06,SOC_G_Fabian_09}. Even though this value can be further enhanced to about $24~\mu$eV through virtual transitions to $d$-orbital, the intrinsic spin-orbit coupling is still too weak to open an experimentally observable band gap~\cite{SOC_G_Fabian_09}. Therefore, one has to turn to external means to tune the intrinsic-type spin-orbit coupling. For example, Weeks \textit{et al} reported that the adsorption of indium (In) or thallium (Tl) atoms is able to significantly increase the intrinsic spin-orbit coupling~\cite{QSHE_G_Wu_11}. To achieve this, several necessary conditions must be satisfied: First, the impurity bands should be farther away from the Fermi level; Second, the magnetization must not be allowed to preserve the time-reversal symmetry; Third, the Rashba spin-orbit coupling due to the breaking of the mirror reflection about the graphene plane from adsorption should be much smaller than the enhanced intrinsic-type spin-orbit coupling. Although the Rashba term plays a detrimental role, the intrinsic spin-orbit coupling induced TI phase can be stabilized as long as the Rashba term is not dominant.

As concrete examples for the In and Tl adsorption, the energy of the outer $p$-shell electrons is far away from the Dirac point, and their coupling with the graphene's $\pi$-band electrons can transmit their strong spin-orbit coupling to graphene via the second-order perturbation. The resulting sizable intrinsic spin-orbit coupling opens a bulk band gap of about 7 (21) meV in the In (Tl) adsorption case~\cite{QSHE_G_Wu_11}. Although this numerical calculation is based on a periodic adsorption that is beyond current experimental capabilities, it is shown that even when the adatoms are randomly distributed, the enhanced spin-orbit coupling from the adsorption is rather stable while the inter-valley scattering that is unfavorable for TIs is effectively suppressed~\cite{QAHE_G_Qiao_randomAdat_12}. Therefore, these findings indicate a high possibility of realizing $\mathbb{Z}_2$ TIs in graphene. Yet, experimentally, it has been found that the adatoms are inclined to form clusters rather than to distribute individually~\cite{G_cluster_Sutter_11}. But surprisingly, Cresti \textit{et al} reported that even when the adsorbed heavy atoms (e.g. Tl) form islands or clusters on top of graphene, a nearly quantized plateau of spin-Hall conductance can still be formed~\cite{QSHE_G_Roche_14}.

In contrast to the In and Tl with outer $p$ shell, the outer $d$-orbitals of the 5$d$-transition metal adatoms are not far away from the Dirac point of graphene, such as Os, Ir and Re. These outer $d$-orbitals can strongly couple and hybridize with graphene $\pi$ bands, and thus greatly alter the linear Dirac dispersion in the high adsorption case~\cite{QSHE_G_Franz_12,QSHE_G_Duan_13}. The dominate contribution from $d$-orbitals at the Fermi energy leads to large spin-orbit coupling and hence large topologically nontrivial band gap~\cite{QSHE_G_Franz_12,QSHE_G_Duan_13}. Similar effect can also be found in graphene with Ru adatom in $2\times 2$ supercell, which is a $4d$-transition metal~\cite{QSHE_G_Fazzio_14}. For these adatoms with outer $d$-orbital, the spontaneous magnetism may be formed due to the strong electron-electron correlation, which yet can be suppressed by external electric field~\cite{QSHE_G_Franz_12}.

\begin{figure}
  \centering
  \includegraphics[width=8 cm]{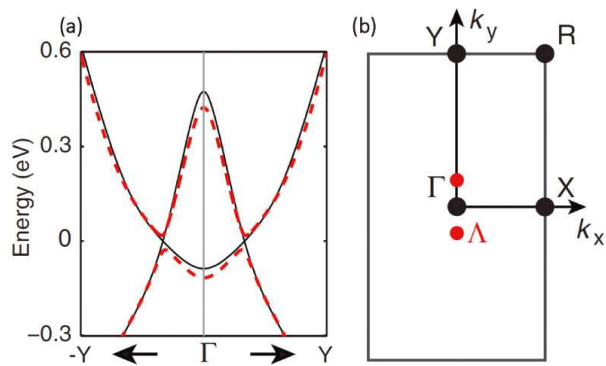}
  \caption{(color online). (a) Band structure and (b) Brillouin zone of 1T$'$-MoS$_2$. (a) Band structures with (red dashed line) and without (black solid line) spin-orbit coupling. (B) Four time-reversal invariant momenta are marked by black dots and labeled as $\Gamma$, X, Y, and R. Red dots: the two Dirac points labelled as $\Lambda$.
  Reprinted figure with permission from~[\onlinecite{QSHE_TMD_Li_14}].}
  \label{QSHE_TMD_Li_14_Fig2}
\end{figure}

Apart from the adsorption of heavy atoms, the proximity effect of thin-film insulators with strong spin-orbit couplings is another effective method to enhance the spin-orbit coupling in graphene~\cite{G_TIprox_Jhi_13,G_TIprox_Zhong_13,QSHE_G_Rossi_14,QSHE_G_Frauenheim_13,QSHE_G_Chen_14,QSHE_G-BiTeX_Yan_14,QSHE_G_Chen_15}. The \textit{ab-initio} calculation shows that the spin-orbit coupling of graphene can be increased up to the order of meV through sandwiching with Sb$_2$Te$_3$ or MoTe$_2$ slabs, where  the two Dirac cones of graphene are folded into $\Gamma$ point due to the $\sqrt{3}\times \sqrt{3}$ superlattice and the inter-valley scattering induces a topologically trivial band gap. However, the folded Dirac cones are located inside the bulk band gap of the neighbor insulators, which can introduce strong spin-orbit coupling in graphene to overcomes the trivial band gap and results in a topologically nontrivial phase~\cite{QSHE_G_Chen_15}. When the top of the valence bands or the bottom of the conduction bands of the sandwiching materials is close to the Dirac points of graphene, for example Bi$_2$Se$_3$, the inversion between the bands of sandwiching materials and graphene's $\pi$ bands can also result in a $\mathbb{Z}_2$ TI~\cite{QSHE_G_Frauenheim_13}. Similar effect can be found in bilayer graphene sandwiched by Bi$_2$Se$_3$~\cite{QSHE_G_Chen_14}. In addition, the van der Waals heterostructure between graphene and BiTeX (X = Cl, Br, and I) chalcogenides can form another family of $\mathbb{Z}_2$ TIs with a bulk band gap of around 70-80~meV~\cite{QSHE_G-BiTeX_Yan_14}.

\subsection{Other Dirac materials}
The low-energy physics dominated by linear Dirac dispersion is not limited to graphene or other atomic crystal layers of group-IV elements~\cite{rev_DiracMaterial_2D_15}. In analogy to silicene, the 2D organometallic lattices composed of  triphenyl and lead (Pb) or bismuth (Bi) naturally form a low-buckled honeycomb lattice~\cite{QSHE_Orgainc_Liu_13}. For the triphenyl-lead honeycomb lattice, its electronic structure proximity to the Fermi energy shows linear Dirac dispersion similar to that in silicene. When the spin-orbit coupling is considered, the intrinsic Kane-Mele type spin-orbit coupling emerges, opening a bulk band gap and harbouring a $\mathbb{Z}_2$ TI phase. For the triphenyl-bismuth honeycomb lattice, however, although a topologically nontrivial band gap is also opened at the Dirac points, the Fermi-level is above the band gap since the bismuth atom has one more valence electron than the lead atom. Although these materials are theoretically shown to be TIs, they still await experimental synthesis. On the other hand, recent experimental progress has seen the synthesis of Kagom\'e organometallic lattices, such as lattices of Ni$_3$C$_{12}$S$_{12}$~\cite{QSHE_Organic_Kambe_13} and Ni$_3$(C$_{18}$H$_{12}$N$_6$)$_2$~\cite{QSHE_Organic_Sheberla_14}. These two lattices also exhibit linear Dirac dispersion at K/K$^\prime$ points, which can give rise to a TI phase in the presence of spin-orbit coupling. However, the topologically nontrivial band gap is much higher than the Fermi energy level, making the experimental observation of TI difficult~\cite{QSHE_Organic_kagome_WangZF_13,QSHE_Organic_Yang_14}. In addition to these organometallic lattices, $\mathbb{Z}_2$ TI has also been reported in metal-atom free systems, e.g. honeycomb lattices of s-triazines~\cite{QSHE_G_Zhao_14} and $\delta$-graphyne~\cite{QSHE_G_Wang_13}, where the linear Dirac dispersions appear at the K/K$^\prime$ valleys close to the Fermi level.

In addition to lattices with hexagonal first Brillouin zone, Dirac dispersions have also been reported in systems with a rectangular first Brillouin zone where the Dirac points lie on the high symmetric line along $\Gamma$-$X$ or $\Gamma$-$Y$~\cite{QSHE_TMD_Li_14, QSHE_Zr.HfTe5_Fang_14, QSHE_Bi_Wang_15, QSHE_P_Zunger_15}. In 2D transition metal dichalcogenides with a 1T$^\prime$ structure~\cite{QSHE_TMD_Li_14}, two Dirac cones are present at $\Lambda$ points near the $\Gamma$ point in the absence of spin-orbit coupling, as denoted by the black solid lines in Fig.~\ref{QSHE_TMD_Li_14_Fig2}(a) and the $\Lambda$ points labelled by solid red circles in Fig.~\ref{QSHE_TMD_Li_14_Fig2}(b). The introduction of the spin-orbit coupling lifts the degeneracy at Dirac points and gives rise to a $\mathbb{Z}_2$ TI. For the other cases, like in Bi(110) bilayer~\cite{QSHE_Bi_Wang_15}, transition-metal pentatelluride~\cite{QSHE_Zr.HfTe5_Fang_14}, and 4-layer black phosphorene under an electric field~\cite{QSHE_P_Zunger_15}, although the physical origins of the Dirac cones in the absence of spin-orbit coupling are different, the band gaps opened by the spin-orbit coupling can host a $\mathbb{Z}_2$ TI. The resulting edge states of a Bi (110) bilayer has already been observed experimentally at a temperature of up to $77~$K~\cite{QSHE_Bi_Wang_15}.

\subsection{Buckled honeycomb lattice of group-V elements} \label{TI_V}
\subsubsection{Bismuth bilayer}\label{TI_Bi}
\begin{figure}
  \centering
  \includegraphics[width=8 cm]{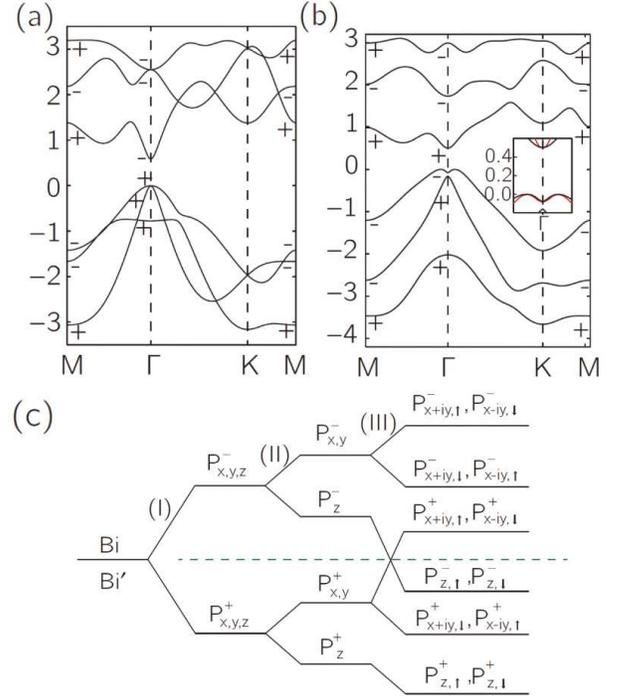} \\
  \caption{(color online). Band structure and corresponding wavefunction parity eigenvalues of a single Bi(111) bilayer (a) without and (b) with spin-orbit coupling. Inset of (b): zoom in on two inverted bands at $\Gamma$. The highest occupied energy level is set to zero. (c) Schematic of the evolution from the atomic $p_{x,y,z}$ orbitals of atoms into the conduction and valence bands of single Bi(111) bilayer at the $\Gamma$ point. Three stages (I), (II) and (III) take into account the effects from chemical bonding, crystal-field splitting and spin-orbit coupling, respectively. The green dashed line represents the chemical potential. Note that the $p_z$ orbitals slightly hybridize with the $p_{x,y}$ orbitals at stage (III).
  Reprinted with permission from~[\onlinecite{QSHE_Edge_Bi_FengJ_14}], copyright 2014 by the American Physical Society.}\label{QSHE_Edge_Bi_FengJ_14_Fig2}
\end{figure}

In addition to the semiconductor quantum well, a real 2D semiconductor with fully filled valence bands can also be realized in the atomic crystal layer structure of group-V elements such as black and blue phosphorene~\cite{QSHE_P_Tomanek_14}, arsenene~\cite{QSHE_As_Ezawa_15}, and antimony (Sb) and bismuth (Bi) bilayers in the (111) orientation~\cite{QSHE_Sb_ChiangTC_12,QSHE_Bi_Murakami_06, QSHE_Bi_Str_Hofmann_06, QSHE_Bi_structure_Hofmann_05}. Among these 2D layer structures, black phosphorene has a puckered structure while the others have a stable buckled honeycomb-lattice structure~\cite{QSHE_P_Tomanek_14,QSHE_As_Ezawa_15,QSHE_Sb_ChiangTC_12,QSHE_Bi_Murakami_06,QSHE_Bi_Str_Hofmann_06, QSHE_Bi_structure_Hofmann_05}.
However, in contrast to the honeycomb lattices composed of group-IV elements where the $\pi$-band from the $p_z$ orbital is half-filled to form a semi-metallic phase with the Fermi-level lying at the Dirac points, the group-V elements have five valence electrons that fill all the valence bands from the molecular orbitals, which leads to the insulating band structure. For these group-V element based structures, the increase in atomic number also decreases the electronegativity that weakens the valence bonds, which shrinks the band gap. Moreover, as the atomic number increases, the indirect band gaps for blue phosphorene~\cite{QSHE_P_Tomanek_14} and arsenene~\cite{QSHE_As_Ezawa_15} become the direct gap at the $\Gamma$ point for the Sb and Bi bilayer~\cite{QSHE_Sb_ChiangTC_12, QSHE_Bi_Murakami_06,QSHE_Bi_BandStr_Blugel_08}. For example, as shown in Fig.~\ref{QSHE_Edge_Bi_FengJ_14_Fig2}(a), the band gap for Bi bilayer is opened up at $\Gamma$ where the valence bands are dominated by $p_{x,y}$ orbitals with even parity, whereas the conductance bands are dominated by a $p_z$ orbital with odd parity~\cite{QSHE_Edge_Bi_FengJ_14}.

Another result of increasing the atomic number is to enhance the spin-orbit coupling of the Bloch electrons~\cite{2D_electron_hole_SOC}. As a result, in the Bi bilayer composed of the heaviest group-V element, the spin-orbit coupling is so strong to induce a band inversion between the conduction and valence bands and the formation of a TI phase [see Figs.~\ref{QSHE_Edge_Bi_FengJ_14_Fig2}(b) and (c)]~\cite{QSHE_Edge_Bi_FengJ_14, QSHE_Bi_Murakami_06, QSHE_Bi_Bihlmayer_11, QSHE_Bi-confinement_Schmidt_15, QSHE_Bi_Wang_13,QSHE_Bi_LiuF_14,QSHE_Bi_WangD_14}. Recently, the time-reversal symmetry-protected edge states have been experimentally observed in an exfoliated Bi (111) bilayer~\cite{QSHE_Bi_Palacios_13}, and in a Bi bilayer on a substrate, e.g. Bi$_2$Te$_3$~\cite{QSHE_Bi_Hasegawa_11,QSHE_Biexp_Jia_12} and Bi$_2$Te$_2$Se~\cite{QSHE_Bi_exp_Yeom_14}. However, for the Bi bilayer on a Bi substrate, Takayama \textit{et al} recently reported their ARPES observations seemed to indicate that their one-dimensional edge states of a Bi bilayer on a Bi substrate were of a non-topological nature~\cite{QSHE_Bi_Takahashi_15}, which is inconsistent with the conclusion from the scanning tunneling microscopy measurements by Drozdov \textit{et al}~\cite{QSHE_Biexp_Yazdani_14}.

In the above studies, all the large-gap 2D $\mathbb{Z}_2$ TIs were predicted for free-standing systems. However, in reality, substrates are a necessity in most situations, which will influence the intrinsic topological properties of certain materials due to interfacial and proximity effects such as the unexpected charge transfer and the resultant Rashba spin-orbit coupling~\cite{QSHE_Bi_Takahashi_15}. Therefore, suitable substrates to support room-temperature TIs are important for potential device applications~\cite{QSHE_Bi-SiSurface_Liu_14, QSHE_Bi-SiSurface_Liu1_14, QSHE_BiSb_Bansil_15, QSHE_Bi_Bansil_13, QSHE_Bi_on_Si_Bansil_14}. The influence of several traditional substrates, e.g. hexagonal boron-nitride (h-BN) and silicon-carbide~\cite{QSHE_Bi_Bansil_13}, has been intensively studied. It has been shown that h-BN is an ideal substrate for stabilizing the topologically nontrivial phase of a freestanding Bi bilayer.

\subsubsection{Other group-V elements} \label{TI_V_Other}
Similar to the Bi(111) bilayer, a Sb(111) bilayer also possesses a buckled honeycomb lattice but it is a robust trivial band insulator in the monolayer case due to its weaker spin-orbit coupling. However, stacking of the Sb(111) bilayers reduces the band gap and a TI phase occurs when the stacking reaches four layers~\cite{QSHE_Sb_Wang_13}. In addition to the buckled structure, as the atomic number decreases, a stable puckered structure can also be formed due to enhancement of the buckling via reduced atom size, for example in black phosphorus~\cite{QSHE_P_Zunger_15} and arsenene~\cite{QSHE_As_Ezawa_15}. However, both the buckled and puckered structures of monolayer arsenide and phosphorus are found to be topologically trivial insulators. Nevertheless, in the stacked black phosphorus with three to four layers, the external perpendicular electric field can drive a topological phase transition from a trivial insulator into a TI via a band inversion near the $\Gamma$ point~\cite{QSHE_P_Zunger_15}. Similar results may be expected for arsenide thin films, which to our knowledge has not been reported yet.

\subsubsection{III-V, VII-V, and IV-VI compounds} \label{TI_III-V_compound}
Another direct analogy to the low-buckled honeycomb structure of group-IV elements is the binary III-V compounds based on the B-In and N-Sb elements~\cite{QSHE_III-V_Sahin_09, QSHE_III-V_Hennig_13}. However, different from the honeycomb lattice of group-IV elements, the inversion-symmetry breaking in these compounds opens a large local band gap at the K and K$^\prime$ valleys. On the other hand, the conductance and valence band edges, which determine the low-energy physics, are mainly influenced by the $s$ and $p_{x,y}$ orbitals respectively for some materials and are located at the $\Gamma$ point. Although the spin-orbit coupling is not strong enough to induce band inversion for some materials composed of light elements, the buckled honeycomb lattice provides the possibility to tuning the topological phase via external means, like strain and electric field. As an example of As-based III-V compounds, the GaAs monolayer has been shown to be a $\mathbb{Z}_2$ TI with a band gap as wide as 257 meV under certain strain conditions. Different from Kane-Mele's model, the topologically nontrivial phase is induced by the $s$-$p$ band inversion via an external strain at the $\Gamma$ point~\cite{QSHE_GaAs_Zhang_15}.

Later, the binary III-V compound systems were extended to the Bi-based family~\cite{QSHE_Bi-III_Bansil_14}. The Bi based compounds with low-buckled structures, i.e. XBi (X= Ga, In, Tl), are natural $\mathbb{Z}_2$ TIs due to the strong spin-orbit coupling. In particular, the bulk band gap of TlBi can be as large as 560~meV, which is possible for room-temperature measurement of the topological phase. For X=B and Al, although the intrinsic systems are topologically trivial insulators, an externally applied strain can drive the systems to be $\mathbb{Z}_2$ TIs. Functionalization of the binary group III-V systems can further increase the bulk band gaps, for instance, chloridization of GaBi can enlarge the topological band gap to 650~meV~\cite{QSHE_BiGaCl_Zhao_15}. Very recently, the III-V family has been theoretically extended to include the Tl-based compounds $g$-TlX with X=N-Sb, which are found to be thermally stable under room-temperature~\cite{QSHE_III-V_HuangB_15}. Topologically nontrivial phases due to the band inversion at the $\Gamma$ point are found in TlAs and TlSb~\cite{QSHE_III-V_HuangB_15}. This kind of $s$-$p$ band inversion can also be applied to the group-V based compounds alloyed with the group-VII elements, e.g. Bi$_4$Br$_4$~\cite{QSHE_BiBr_Yao_14,QSHE_BiBr_Yao_15} and Bi$_4$F$_4$~\cite{QSHE_BiF_Xiang_15}, which, however, possess a quasi-cubic lattice structure rather than the honeycomb-lattice structure.

\subsection{Functionalized honeycomb lattices of group-IV and -V elements} \label{TI_Fun}
\subsubsection{Functionalized honeycomb lattices of group-IV elements} \label{TI_Fun_IV}

\begin{figure}
  \centering
  \includegraphics[width=8 cm]{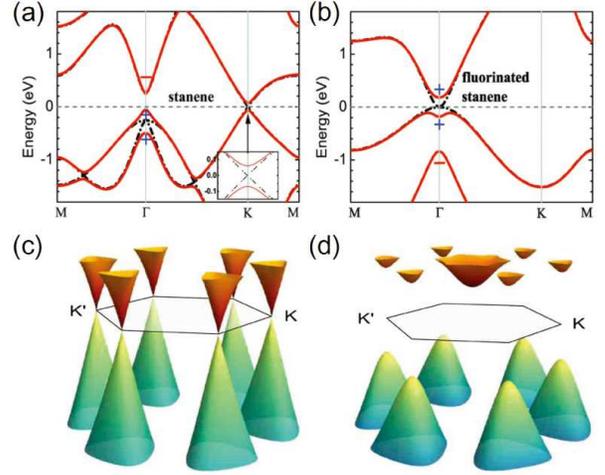} \\
  \caption{(color online). (a) and (b): Electronic structures of stanene and functionalized stanene. (c) and (d): Band structure of functionalized Bi bilayer without and with spin-orbit coupling, respectively.
  Figs.~(a)-(b) reprinted with permission from~[\onlinecite{QSHE_Sn_Zhang_13}], copyright 2013 by the American Physical Society. Figs.~(c)-(d) reprinted with permission from~[\onlinecite{QSHE_Bi_QAHE_Jhi_15}].}\label{Function_IV_V}
\end{figure}

In order to open up a larger bulk band gap and increase the tunability of the $\mathbb{Z}_2$ topological insulating phase in group-IV element based planar or low-buckled honeycomb systems, various schemes have been suggested. In addition to the heavy-atom adsorption scheme~\cite{QSHE_G_Wu_11}, functionalization is one of the most efficient approaches to enlarge the bulk band gap~\cite{QSHE_Ge_Huang_12,QSHE_Sn_Zhang_13}. This method was first reported by Ma \textit{et al} by considering halogenated germanene~\cite{QSHE_Ge_Huang_12} and then was generalized to stanene (i.e. the tin monolayer) and other honeycomb lattices~\cite{QSHE_Sn_Zhang_13,QSHE_BiGaCl_Zhao_15}. Here, we take the stanene (monolayer tin) as an example, which is intrinsically a Kane-Mele type TI with a topologically nontrivial band gap at the K/K$'$ points, as shown in Fig.~\ref{Function_IV_V}(a). The chemical function groups, such as -F, -Cl, -Br, -I, and -OH, are strongly coupled with the $p_z$ orbitals of the stannum atoms and lift the Dirac cones at the K/K$'$ points by a huge gap, which is the so-called ``orbital filtering effect".

In contrast, a band inversion can be induced between a small gapped $s$ and $p_{x,y}$ bands leading to a parity exchange between the occupied and unoccupied bands at the time-reversal point $\Gamma$, where a large nontrivial bulk gap of about 300 meV is opened due to the strong spin-orbit coupling~\cite{QSHE_Sn_Zhang_13}, as displayed in the lower panel of Fig.~\ref{Function_IV_V}(b). Moreover, although stanane~\cite{QSHE_GeH_Goldberger_13}, i.e. the hydrogenized monolayer stanene, is a topologically trivial insulator, the hydrogenized bilayer and trilayer stanenes are found to be $\mathbb{Z}_2$ TIs~\cite{QSHE_Sn_Bansil_14}. This kind of functionalization can not only enlarge the bulk band gap but also provide extra means to control the edge modes between two topologically distinct phases of stanene, e.g. the states between fluorinated stanene and stanane. Similar results have also been reported for germanene and the 2D counterpart of lead~\cite{QSHE_Ge_Duan_14,QSHE_GeCH3_Whangbo_14,ZLM_Ge_Fazzio_14} as well as the dumbbell stanene~\cite{QSHE_Sn_DB_Rubio_14}.

\subsubsection{Functionalized honeycomb lattices of group-V elements} \label{TI_Fun_V}
\begin{figure}
  \centering
  \includegraphics[width=8 cm]{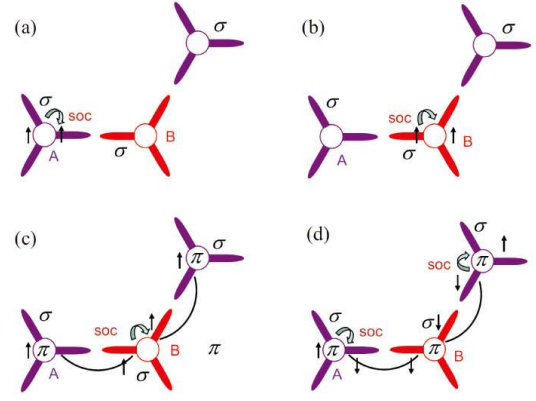} \\
  \caption{(color online). Origin of the Kane-Mele type spin orbit coupling. For functionalized Bi bilayer, the on-site spin orbit coupling $|A\uparrow \rangle \rightarrow |A\uparrow \rangle$ is nonzero since the orbitals are due to $p_x\pm i p_y$, which has a unit $z$-direction angular momentum with a sign opposite to that of the B sublattice, as shown in (a) and (b). For graphene with a $p_z$ dominated orbital, the term is zero since the  $z$-direction angular momentum $m=0$, as shown in (d), where the on-site spin mixing mediated intrinsic spin-orbit coupling intensity is on the second order of atomic spin-orbit coupling. However, for a low buckled structure, the $sp^3$-like hybridization induces coupling between the $p_z$ orbital of the A sublattice with the $p_{x,y}$ orbital of B sublattice, which mediates the intrinsic spin-orbit coupling on the first order intensity of atomic spin-orbit coupling.
  Reprinted with permission from~[\onlinecite{QSHE_Bi_functionalized_Yao_14}], copyright 2014 by the American Physical Society.} \label{QSHE_Bi_functionalized_Yao_14_Fig4}
\end{figure}

Similar to the case of group-IV elements, functionalization of honeycomb lattice composed of group-V elements is also an effective method to induce TIs with large band gaps where the orbital-filtering effect also plays a key role. For example, the functionalized Bi-bilayers have been intensively studied with reports of large nontrivial band gaps for various functional groups (such as -H, -F, -Cl, -Br, -I~\cite{QSHE_Bi_QAHE_QVHE_Mokrousov_15, QSHE_Bi_functionalized_Yao_14, QSHE_Bi.Sb_QVHE_Yao_14, QSHE_Bi_QAHE_Jhi_15, QSHE_Bi_Heine_15}, and -CH$_3$~\cite{QSHE_Bi_Heine_15}). However, filtering of the $p_z$-orbital has given different results. For the group-IV elements the functional groups lift the $p_z$-orbital around the KK$^\prime$ valleys and induce a band inversion at the $\Gamma$ point, whereas for the group-V elements the functional groups couple strongly with the $p_z$-orbital to induce a large local gap at the $\Gamma$ point, and then further flatten the buckled honeycomb-lattice structures~\cite{QSHE_Bi_QAHE_Jhi_15} to give rise to Dirac cones at the KK$^\prime$ valleys, which are dominated by the $p_{x,y}$-orbitals in the absence of spin-orbit coupling as displayed in Fig.~\ref{Function_IV_V}(c)~\cite{QSHE_Bi_functionalized_Yao_14,QAHE_Bi_WuCJ_14}. Due to the non-vanishing orbital angular momentum of the states at the KK$^\prime$ points, the intrinsic on-site spin-orbit coupling dominates and produces an effective Kane-Mele-type spin-orbit coupling, with a strength of the same order as that of the atoms as displayed in Fig.~\ref{Function_IV_V}(d). As a consequence, the topological band gap of the functionalized group-V elements is very large, for instance, the gap for the functionalized Bi bilayer can be as large as 1 eV. Figure \ref{QSHE_Bi_functionalized_Yao_14_Fig4} provides a clear comparison of the different origins of the Kane-Mele-type intrinsic spin-orbit coupling for the planar honeycomb-lattice structure with the $p_z$-orbital (e.g. graphene), the low-buckled honeycomb-lattice structure dominated by the $p_z$-orbital (e.g. silicene), and the planar honeycomb lattice with $p_{x,y}$-orbitals (e.g. functionalized Bi bilayer). Furthermore, the effect of the functional groups -H and -F has been extended from a single Bi bilayer film to 1-5 bilayers of Bi and Sb thin films, where various topological phase transitions have been demonstrated~\onlinecite{QSHE_Bi.Sb_JiS_15}.

Such an orbital-filtering effect can also find application when the Bi atoms are fabricated on some suitable substrate such as a Si(111) surface covered with halogen or hydrogen atoms, where the Bi atoms can self-assemble to form a honeycomb-lattice structure with high-kinetic and high-thermodynamic stability. Similar to the effect of the functionalization on Bi bilayers, the coupling between Bi $p_z$-orbital and the substrate can also lift the $p_z$-orbital away from the Fermi level and generate the linear Dirac cones at the KK$'$ points, where a topologically nontrivial gap can open when the spin-orbit coupling is further considered. This is known as the ``substrate-orbital-filtering-effect"~\cite{QSHE_Bi-SiSurface_Liu_14,QSHE_Bi-SiSurface_Liu1_14,QSHE_BiSb_Bansil_15,QSHE_Bi_Bansil_13,QSHE_Bi_on_Si_Bansil_14}.

\subsection{Topological Anderson insulator} \label{TI_TAI}
\begin{figure}
  \centering
  \includegraphics[width=8 cm]{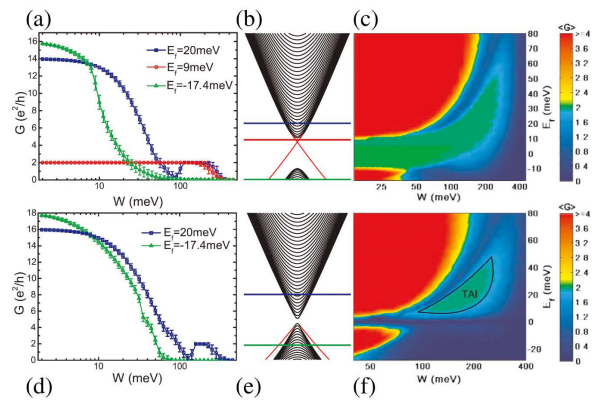}
  \caption{(color online). Conductance of disordered strips of HgTe/CdTe quantum wells for: (a) to (c) an inverted quantum well with $M = -10~$meV, and (d) to (f ) a normal quantum well with $M =1~$meV. (a) Conductance G as a function of disorder strength W at three Fermi energy values. The error bars show standard deviation of the conductance for 1000 samples. (b) Band structure calculated with the tight-binding model; its vertical scale (energy) is the same as in (c), and the horizontal lines correspond to the Fermi energy values of (a). (c) Phase diagram showing the conductance G as a function of both disorder strength W and Fermi energy $E_f$. Figs.~(d), (e) and (f ) are the same as (a), (b) and (c), but for $M>0$. The TAI phase regime is labeled. In all figures, the strip width $L_y$ is set to 500 nm; the length $L_x$ is 5000 nm in (a) and (d), and 2000 nm in (c) and (f ).
  Reprinted with permission from~[\onlinecite{TAI_QW_ShenSQ_09}], copyright 2009 by the American Physical Society.}
  \label{TAI_QW_ShenSQ_09_Fig2}
\end{figure}

It is known that, for normal 2D metals, the metallic phase is unstable and becomes an Anderson insulator under any weak disorder. Interestingly, an anomalous finding appears that, in HgTe quantum wells, the disorder can drive a topological phase transition from a trivial insulator to a topologically nontrivial insulator with a quantized spin-Hall conductance plateau~\cite{TAI_QW_ShenSQ_09}, as illustrated in Figs.~\ref{TAI_QW_ShenSQ_09_Fig2}(d)-(f). Numerical calculation has shown that the quantized conducting plateau arises from the dissipationless edge states in a disordered ribbon~\cite{TAI_QW_XieXC_09}. Although the transport properties of this phase share the same characteristics as those in the $\mathbb{Z}_2$ TI, their disconnected regions in the phase diagram of Fig.~\ref{TAI_QW_ShenSQ_09_Fig2}(f) imply that this may be a new topological phase different from the $\mathbb{Z}_2$ TI, and so has been given the name ``topological Anderson insulator"~\cite{TAI_QW_ShenSQ_09}. Later, this insulating phase was proved to be topologically equivalent to the $\mathbb{Z}_2$ TI phase of HgTe/CdTe in the inverted regime since it can be obtained by continuously varying the Dirac mass, Fermi energy, and disorder strength in the three dimensional parameter space~\cite{TAI_QW_Prodan_11}.

The topological phase transition from a trivial insulator to a TI is induced by the renormalization of the mass term due to the disorder via the quadratic  momentum term in the low energy continuum model Hamiltonian, which can convert the mass term from positive to negative~\cite{TAI_theory_Beenakker_09, TAI_QW_JiangX_12, TAI_QW_ShenSQ_12, TAI_QW_Kuramoto_12}. The quantized conductance plateau appears when the renormalized chemical potential/Fermi level lies in the band edge~\cite{TAI_theory_Beenakker_09}. Although the topological Anderson insulator shares the same topological properties as a TI, they are distinct in the bulk. Specifically, in a TI, the gapless edge modes live in the bulk band gap where no bulk states are present. However, in a topological Anderson insulator, localized bulk states may occur in the mobility gap where the localized bulk states appear~\cite{TAI_QW_JiangH_12,TAI_QW_ShenSQ_13}. Moreover, it is noteworthy that, when the Fermi-level is located at the valence band rather than the conduction band, there is no topological phase transition from the metallic phase to the 2D TI phase in both topological trivial and nontrivial phases, as shown in Fig.~\ref{TAI_QW_ShenSQ_09_Fig2}. Such an asymmetry behavior for a Fermi level lying in the conductance and valence bands arises from the particle-hole asymmetry. In a system with particle-hole symmetry, the Anderson disorder can induce the localization of the bulk states as well as the coexisting edge states, while a conductance plateau appears only when the Fermi level lies inside the topologically nontrivial band gap~\cite{TAI_G_ZhangYY_14}. Such an Anderson disorder induced topologically nontrivial phase can also be extended to a system with modified Dirac Hamiltonian by including the quadratic correction~\cite{TAI_G_WangJ_11}.

The effect of the hopping disorder has also been studied; it was found that the inter-cell hopping term can also lead to a topological Anderson insulator~\cite{TAI_QW_Li_13} while the intra-cell hopping cannot~\cite{TAI_G_XieXC_12}. In the presence of Rashba spin-orbit coupling, a mediated metallic phase between different topologically trivial and nontrivial insulating phases has been observed~\cite{TAI_Kuramoto_11,TAI_QW_Kuramoto_12,TAI_QW_Li_13}. Moreover, it was found that the bulk states can be effectively localized by the long-range disorder, but the edge states are much more robust~\cite{TAI_G_ZhangYY_14}. Furthermore, in contrast to the Anderson disorder without spatial correlation, the finite correlation length disorder was found to be detrimental to (or even totally suppress) the formation of the topological Anderson insulator~\cite{TAI_Rotter_13}. The presence of several types of disorder, Rashba spin-orbit coupling, the finite-size effect~\cite{TAI_SizeEffect_JiangY_11} as well as correlation make it challenging to observe topological Anderson insulator.

\subsection{Time-reversal symmetry breaking quantum spin-Hall effect} \label{TI_TRB-QSHE}
As long as the time-reversal symmetry is preserved, $\mathbb{Z}_2$ is a well-defined topological invariant no matter whether the spin is a good quantum number or not. For 2D systems with preserved mirror symmetry about the plane, $s_z$ is a good quantum number, leading to well-defined Chern numbers for each spin (i.e. $\mathcal{C}_{\uparrow}$ and $\mathcal{C}_{\downarrow}$), which are intimately related to each other by the time-reversal operation $\mathcal{C}_{\uparrow}=-\mathcal{C}_{\downarrow}$. Since the unit Chern number for a spin state corresponds to one gapless spin-polarized chiral edge state, the spin-Chern number defined as $\mathcal{C}_s=(\mathcal{C}_{\uparrow}-\mathcal{C}_{\downarrow})/2$ is fundamentally equivalent to $\mathbb{Z}_2$. The only difference between $\mathbb{Z}_2$ and $\mathcal{C}_s$ is that the former can only take the value of 0 or 1 while the latter can be any integer. When the mirror symmetry about the 2D plane is broken, $s_z$ is no longer a good quantum number and thus the above-mentioned spin-Chern number no longer has a valid definition and the resulting $\mathcal{C}_s$ is no longer exactly quantized. Nevertheless, the spin-Chern number can still be meaningful through an alternative definition, which is shown to be equivalent to the $\mathbb{Z}_2$ topological order whenever the time-reversal symmetry is preserved~\cite{QSHE_G_Haldane_06}.

However, when time-reversal symmetry breaks down, e.g. when the spontaneous anti-ferromagnetic order in a graphene nanoribbon is taken into account~\cite{QSHE_G_Xiong_13}, the QSHE may also exist with a well-defined spin-Chern number~\cite{QAHE_TRB_QSH_ShengDN_11,QSHE_TRB_Abanin_06,QSHE_TRB_Abanin_07}. Strictly speaking, this is not a topological phase since the Kramers degeneracy of the chiral edge states is lifted, signifying that the edge modes are no longer topologically protected from the elastic back-scattering by time-reversal symmetry~\cite{QAHE_TRB_QSH_ShengDN_11}. Although the backscattering is possible, the corresponding edge states can still give rise to a nearly quantized spin-Hall conductance and are quite robust against weak disorders due to the strong localization at their boundaries. Therefore, similar to $\mathbb{Z}_2$ TIs, they are still good candidates for practical applications~\cite{QAHE_TRB_QSH_ShengDN_11}. Additionally, time-reversal symmetry breaking QSHE have also been reported in monolayer graphene~\cite{QSHE_TRB_Abanin_06,QSHE_TRB_Abanin_07} and ferromagnetic metals in the presence of strong magnetic fields~\cite{QSHE_MagneticField_Ezawa_13, QSHE_MangeticField_Aldea_14, rev_QSHE_QAHE_Smith_12}, but these are beyond the scope of this review since we focus here on the topological phases without an applied external magnetic field.

\begin{table*}
  \caption{Possible materials for realizing TIs. 1st column, possible materials for realizing TIs based on theoretical proposals. 2nd column, the corresponding topological nontrivial band gaps. Here, the phrases ``dep. strain" or ``dep. E" indicates that the band gap is dependent on the external strain or electric field since some materials are not intrinsic TIs but a phase transition to TI is possible by applying strain or an electric field. 3rd column: some remarks. 4th column: corresponding references. Last column: the current state of experiments; ``Y" indicates experimental confirmation, ``Q" indicates some experimental discrepancy. The term ``fabricated" indicates that the hosting materials have been fabricated but the TI phases has yet been observed since the intrinsic Fermi energies do not lie inside the topologically nontrivial band gaps. Abbreviations used: G: graphene; BP: black phorsphorene;  BL: bilayer; TL: trilayer; Exp.: experiment realization; dep.: dependent on; BLG: bilayer graphene; MLG: multilayer graphene; Fun.: functionalized; QW: quantum wells. } \label{Tab_TI_Materials}
    \begin{tabular*}{\textwidth}{c@{\extracolsep{\fill}}|cccc}
      \hline \hline
      Material & Gap & Remark & Ref &  Exp. \\ \hline
      G & $\sim \mu$eV &  & \cite{QSHE_Mele_05,SOC_G_Fabian_09} &  \\ \hline
      In(Tl)/G & $\sim$7 (21) meV & $4\times 4$ supercell & \cite{QSHE_G_Wu_11} & \\ \hline
      5$d$ atom/G & $> 0.2$ eV &  & \cite{QSHE_G_Franz_12} & \\ \hline
      G/Re/SiC(0001) & $\sim$meV &  & \cite{QSHE_G_Duan_13} & \\ \hline
      Ru/G & $\sim$10 meV & only $2\times 2$ supercell & \cite{QSHE_G_Fazzio_14} & \\ \hline
      Bi$_2$Se$_3$/G/Bi$_2$Se$_3$ & 30 meV &  & \cite{QSHE_G_Frauenheim_13} & \\ 
      Bi$_2$Se$_3$/BLG/Bi$_2$Se$_3$ & 44 meV &  & \cite{QSHE_G_Chen_14} & \\ 
      G/BiTeX & 70-80 meV & pressure enhancement & \cite{QSHE_G-BiTeX_Yan_14} & \\ 
      Sb$_2$Te$_3$/G/Sb$_2$Te$_3$ & 1.5 meV &  & \cite{QSHE_G_Chen_15} & \\ 
      MoTe$_2$/G/MoTe$_2$ & 3.5 meV &  & \cite{QSHE_G_Chen_15} & \\  \hline
      2D triphenyl-Bi & 43 meV &  intrinsic TI & \cite{QSHE_Orgainc_Liu_13} & \\ 
      2D triphenyl-Pb & 8.6 meV & orgainic materials  & & \\ \hline
      Ni$_3$C$_{12}$S$_{12}$ & 22.7 and 9.5 meV & organic materials, Kagom\'e lattice,  & \cite{QSHE_Organic_kagome_WangZF_13} & fabricated \\
      Ni$_3$(C$_{18}$H$_{12}$N$_6$)$_2$ & 16.6 and 22.4 meV & two TI gaps both away from $E_f$ & \cite{QSHE_Organic_Yang_14} & fabricated \\ \hline
      s-triazines & 5.50 and 8.27 meV & honeycomb lattice with $p_{x,y}$ orbitals & \cite{QSHE_G_Zhao_14} & \\ \hline
      $\delta$-graphyne & 0.59 meV & sp-sp$^2$ hybridization enlarges SOC & \cite{QSHE_G_Wang_13} & \\ \hline
      1T$^\prime$ TMD & 10-100 meV & Dirac materials & \cite{QSHE_TMD_Li_14} & \\ 
      ZrTe$_5$/HfTe$_5$ & 0.1 eV &  in the absence of SOC & \cite{QSHE_Zr.HfTe5_Fang_14} & \\ 
      Bi(110) BL & 0.1 eV &  & \cite{QSHE_Bi_Wang_15} & Y \\ \hline
      silicene & $1.55~$meV & external tunability & \cite{QSHE_Si.Ge_Yao_11} & \\ 
      germanene & $23.9~$meV &  & \cite{QSHE_Si.Ge_Yao_11} & \\ 
      stanene & $73.5~$meV &  & \cite{QSHE_Si.Ge.Sn_Yao_11} & \\ \hline
      dumbbell stanene & 40 meV & strain engineering & \cite{QSHE_Sn_DB_Rubio_14} & \\ \hline
      MLG & dep. Rashba and electric field  &  valley polarized TI & \cite{QSHE_BLG_Qiao_11,QSHE_TLG_Qiao_12} & \\ \hline
      HgTe QW & $\sim$meV &  & \cite{QSHE_QW_HgTe_ZhangSC_06,QSHE_QW_HgTe_exp_ZhangSC_07,rev_QSHE_Zhang_08} & Y~\cite{QSHE_QW_HgTe_exp_ZhangSC_07} \\ 
      InAs/GaSb/AlSb QW &  &  & \cite{QSHE_InAs_ZhangSC_08} & Y ~\cite{QSHE_InAs_exp_DuRR_11} \\
      GaN/InN/GaN QW & $\sim 10$ meV &  & \cite{QSHE_QW_InN_Miao_12} &   \\
      GaAs/Ge/GaAs QW & $\sim 15$ meV &  & \cite{QSHE_QW_Ge_Zhang_13} &  \\ \hline
      4-layer BP & $\sim$5 meV, dep. E & electric field driven TI & \cite{QSHE_P_Zunger_15} & \\ 
      multilayer of BP or Sb (111) & dep. strain/E & extrinsically driven TI & \cite{QSHE_Sb_Wang_13,QSHE_P_Zunger_15} & \\ \hline
      Bi (111) BL & 0.2 eV &  & \cite{QSHE_Bi_Bihlmayer_11,QSHE_Bi_Murakami_06} & Q~\cite{QSHE_Bi_Takahashi_15,QSHE_Biexp_Yazdani_14} \\ \hline
      Bi on Si(111) & 0.8 eV & artificial  & \cite{QSHE_Bi-SiSurface_Liu_14} & \\ 
      Bi/Pb on H-Si(111) surface & $> 0.5$ eV &  honeycomb lattice & \cite{QSHE_Bi-SiSurface_Liu1_14} & \\ \hline
      Fun. germanene & dep. strain & -H, -F, -Cl, -Br, -CH$_3$ & \cite{QSHE_Ge_Duan_14,QSHE_GeCH3_Whangbo_14} & \\ 
                                & 0.3 eV & -I & \cite{QSHE_Ge_Duan_14} & \\ \hline
      Fun. stanene & $\sim$0.3 eV & -F, -Cl, -Br, -I, -OH & \cite{QSHE_Sn_Zhang_13} & \\ \hline
      Fun. BL or TL stanene & $\sim$ 0.244 eV, dep. strain & -H, strain driven TI & \cite{QSHE_Sn_Bansil_14} & \\ \hline
      Fun. Pb BL & $\sim$ 1 eV & -H, -F, -Cl, -Br, -I & \cite{QSHE_Ge_Duan_14} & \\ 
                     & 0.964 eV & CH$_3$ & \cite{QSHE_Bi_Heine_15} & \\ \hline
      Fun. Sb (111) BL & 0.41 eV & -H & \cite{QSHE_Bi.Sb_JiS_15} & \\
                              & 0.32-1.08 eV & -H, -F, -Cl, -Br & \cite{QSHE_Bi.Sb_QVHE_Yao_14,QSHE_Bi_functionalized_Yao_14,QSHE_Bi_QAHE_Jhi_15} & \\
                              & 0.386 eV & -CH$_3$ & \cite{QSHE_Bi_Heine_15} & \\ \hline
      Fun. Bi (111) BL & $\sim$ 1.03 eV & -H & \cite{QSHE_Bi.Sb_JiS_15,QSHE_Bi_QAHE_QVHE_Mokrousov_15} & \\
                              & 0.32-1.08 eV & -H, -F, -Cl, -Br & \cite{QSHE_Bi.Sb_QVHE_Yao_14,QSHE_Bi_functionalized_Yao_14,QSHE_Bi_QAHE_Jhi_15} & \\ 
                              & 0.934 eV & -CH$_3$ & \cite{QSHE_Bi_Heine_15} & \\ \hline
      Fun. GaBi & 650 meV & -Cl & \cite{QSHE_BiGaCl_Zhao_15} & \\ \hline
      GaAs, BBi, AlBi monolayer & dep. strain & extrinsically driven TI & \cite{QSHE_Bi-III_Bansil_14} & \\ \hline
      TlBi & 560 meV &  & \cite{QSHE_Bi-III_Bansil_14} & \\ 
      TlAs and TlSb & 131 and 268 meV &  & \cite{QSHE_III-V_HuangB_15} & \\ \hline
      Bi$_4$Br$_4$ & 0.18 eV & square lattice & \cite{QSHE_BiBr_Yao_14,QSHE_BiBr_Yao_15} & \\ 
      Bi$_4$F$_4$ & 0.69 eV &  & \cite{QSHE_BiF_Xiang_15} & \\ \hline
  \hline
\end{tabular*}
\end{table*}

\section{Quantum Anomalous Hall Effect (QAHE)} \label{QAHE}
In the presence of time-reversal symmetry, insulators can be classified into $\mathbb{Z}_2$ TIs and topologically trivial band insulators according to the $\mathbb{Z}_2$ topological invariant, as described above. In Sec.~\ref{introduction} we saw that for insulators with broken time-reversal symmetry, their topological properties are usually characterized by the first Chern number $\mathcal{C}$~\cite{Chern_Simon_83,Chern_Simon1_83},  which indicates a topologically trivial insulator when $\mathcal{C}=0$, and topologically nontrivial quantum Hall effects~\cite{QHE_Klitzing_04} and QAHE for a nonzero integer $\mathcal{C}$~\cite{QAHE_Haldane_88}. Here, the Chern number is closely related to the number of gapless chiral edge modes that emerge inside the bulk band gap of a finite-sized ribbon according to the ``\textit{bulk-edge correspondence}''~\cite{Bulk_edge_correspondence_Hatsugai_93}. Due to their one-way chiral propagation characteristic, these edge modes are robust against nonmagnetic, magnetic, short-range, long-range or any other kind of weak disorder~\cite{TAI_G_WangJ_11}. This is inherently superior to the spin-helical edge modes in $\mathbb{Z}_2$ TIs where backscattering is allowed by disorder that break time-reversal symmetry.

After the initial prediction of the QAHE, there was only little progress before the year 2004~\cite{QAHE_Nagaosa_03}. However, interest has been revived by the successful experiments on TIs and 2D atomic crystal layers, and in particular by two independent theoretical proposals in 2010 to realize the effect based on magnetic 3D TI thin films~\cite{QAHE_MagTI_FangZh_10} and graphene~\cite{QAHE_G_Qiao_10}. Even though these two schemes employ independent semiconductor and Dirac semi-metal materials, they are both based on the perpendicular ferromagnetic order and spin-orbit coupling that were the pioneer ideas involving semiconductor quantum wells and atomic crystal layers, as mentioned above in Sections~\ref{QAHE_MagTI} and \ref{QAHE_G}, respectively.
The above proposals based on a perpendicular Zeeman field and spin-orbit coupling with small Chern numbers can be regarded as the dominating ``conventional" QAHE structure. Recently, two new groups have been introduced into the ``conventional" family, i.e. the heterostructure quantum well which will be reviewed in Sec.~\ref{QAHE_QW} (heterostructures of ferromagnetic insulator films and other insulating thin films with strong spin-orbit coupling), and the transition metal oxide to be reviewed in Sec.~\ref{QAHE_TMO}.
In contrast, we shall also describe some ``unconventional" properties of the QAHE, e.g. a large Chern number corresponding to large anomalous Hall conductance (Sec.~\ref{QAHE_LargeC}), a system with in-plane ferromagnetism (Sec.~\ref{QAHE_InPlane}) or anti-ferromagnetic order (Sec.~\ref{QAHE_AFM}), and quantized Hall conductance produced by edge-engineering of a finite size sample (Sec.~\ref{QAHE_Edge}).

\begin{figure}
  \centering
  \includegraphics[width=8 cm]{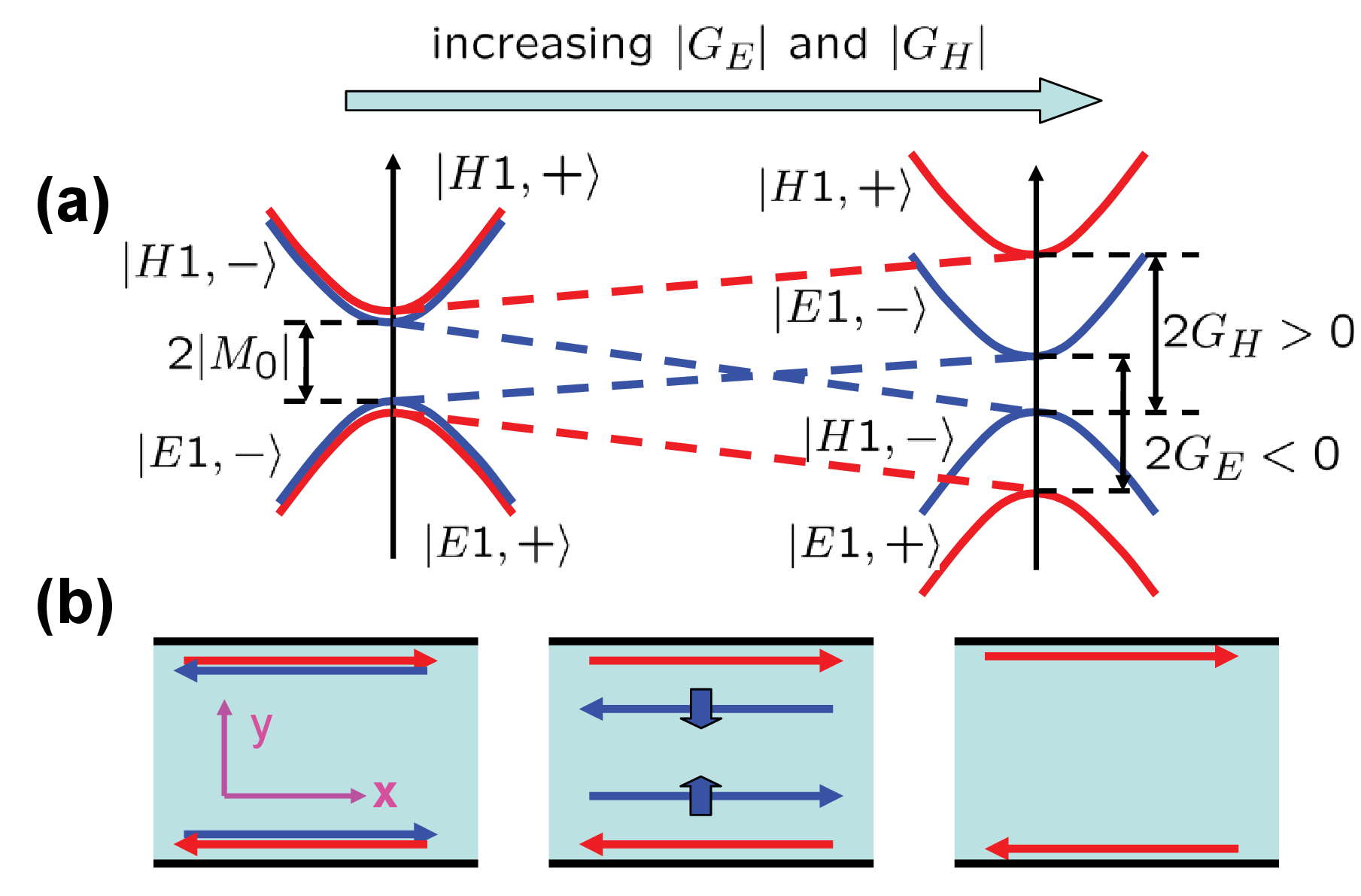}\\
  \caption{(color online). Evolution of band structure and edge states upon increasing the spin splitting. For (a) $G_E<0$ and $G_H>0$, the spin down states $|E1,-\rangle$ and $|H1,-\rangle$ in the same Hamiltonian block touch each other and then enter the normal regime. (b) Behaviour of the edge states during level crossing in the case of (a).
  Reprinted with permission from~[\onlinecite{QAHE_FMQW_ZhangSC_08}], copyright 2008 by the American Physical Society.}\label{QAHE_FMQW_ZhangSC_08_Fig1}
\end{figure}

\subsection{Magnetic doping in quantum well-based 2D-TIs and 3D-TI thin films} \label{QAHE_MagTI}
\subsubsection{Quantum well-based 2D-TIs} \label{QAHE_MagTI_2D}
The unique transport characteristic of 2D-TIs is the emergence of time-reversal symmetry protected spin-helical edge modes propagating along each boundary, which is actually a combination of two identical QAHE structures with exactly opposite spins. When one of the structures is eliminated, the QAHE is obtained. The first example was the Mn-doped HgTe quantum well~\cite{QSHE_QW_HgTe_exp_ZhangSC_07}. When ferromagnetism is introduced by doping magnetic atoms (e.g. Mn), the two-fold Kramers degeneracy of the conduction and valence bands is lifted. However, mixing of the spin angular momentum and orbital angular momentum makes the two QAHE structures respond differently to the Zeeman field from the magnetic order, i.e. the band gap of one model widens while that of the other closes and reopens to undergo a phase transition to become a trivial insulator~\cite{QAHE_FMQW_ZhangSC_08}. As a consequence, the QAHE is formed as half of the TI, as displayed in Fig.~\ref{QAHE_FMQW_ZhangSC_08_Fig1}(a). However, it has been shown that the ferromagnetic order is not favorable in such a system, making it unrealistic experimentally~\cite{QAHE_MagTI_LiuCX_14}. Within the same framework, it was later found that the ferromagnetism can be formed in magnetically doped InAs/GaSb quantum wells via enhanced Van Vleck paramagnetism from the strong interband coupling~\cite{QAHE_MagTI_LiuCX_14};  this has boosted hopes to realize the QAHE in quantum well-based 2D-TIs. Alternative methods have also been proposed for magnetically doped 2D-TIs within junction quantum wells~\cite{QSHE_QAHE_QW_Junction_ZhangSC_14}.

\subsubsection{3D-TI thin films} \label{QAHE_MagTI_3D}
The discovery of 3D $\mathbb{Z}_2$ TIs soon motivated the further exploration of QAHE in 3D-TI thin films through the establishment of stable ferromagnetism, e.g. by doping Cr/Fe-atoms in a Bi$_2$Se$_3$ thin film host material ~\cite{QAHE_MagTI_FangZh_10}. Inspired by this finding, Chang \textit{et al} finally observed the effect for the first time in Cr-doped (Sb,Bi)$_2$Te$_3$~\cite{QAHE_MagTI_exp_XueQK_13}. Subsequently, several other experimental groups independently reported observation in the same host material of (Sb,Bi)$_2$Te$_3$ but at an extremely low temperature (lower than $100~$mK). Below, we shall briefly describe how the QAHE is formed in a magnetic 3D TI thin film.

\begin{figure}
  \centering
  \includegraphics[height=8 cm,angle=-90]{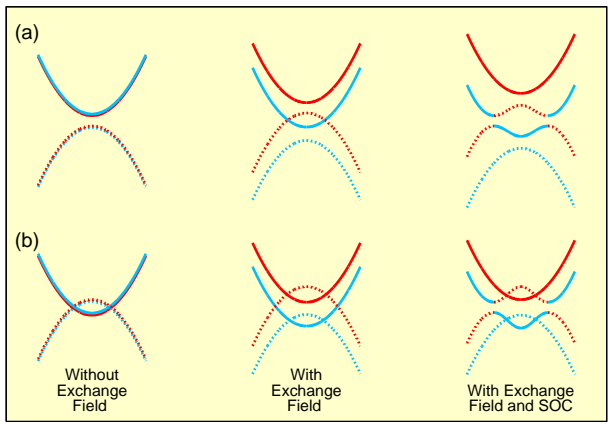}
  \caption{(color online). Evolution of the subband structure upon increasing the exchange field. Solid (dashed) lines denote the subbands that have even (odd) parity at the G point. The blue (red) color denotes the spin down (up) electrons. (A) The initial subbands are not inverted. When the exchange field is strong enough, a pair of inverted subbands appears (red dashed line and blue solid line). (B) The initial subbands are already inverted. The exchange field causes band inversion in one pair of subbands (red solid line and blue dashed line) and increases the inversion in the other pair (red dashed line and blue solid line). Reprinted with permission from ~[\onlinecite{QAHE_MagTI_FangZh_10}].}\label{QAHE_MagTI_FangZh_10_Fig3}
\end{figure}

Let us take Bi$_2$Se$_3$ as an example~\cite{rev_TI_Bernevig,rev_TI_Kane_10,QAHE_MagTI_FangZh_10,QAHE_MagTI_Freeman_11,3DTI_BiSe_Hasan_09,3DTI_BiTe_Hasan_09}. Similar to the 1D gapless edge modes of 2D-TIs, 2D gapless surface modes can be generated on the surface of 3D-TI thin films, where the spins are locked with the momenta preserving the time-reversal invariance~\cite{rev_TI_Bernevig,rev_TI_Kane_10}. Due to this spin-momentum locking, the elastic backscattering is completely suppressed when the 3D-TI thin film is thick enough to avoid direct coupling between the top and bottom surface states. When the film thickness is decreased, the coupling between top and bottom surface states results in a band gap at the Dirac cone of the surface states, forming a 2D insulator~\cite{QAHE_MagTI_FangZh_10}.

One intriguing property of this bulk 3D-TI is that the Bi$_2$Se$_3$ conduction and valence bands of arise from the bonding and anti-bonding $p$ orbitals, which is different from ordinary semiconductors where the contributions to the conduction and valence bands come mainly from the $s$ and $p$ orbitals, respectively. This feature greatly enhances the spin susceptibility via the Van Vleck paramagnetism. Moreover, the strong spin-orbit coupling induced band inversion further strengthens the spin susceptibility in an anisotropic manner. As a result, the magnetic doping prefers to form a spontaneous ferromagnetic order, aligning along the off-plane direction~\cite{QAHE_MagTI_FangZh_10}. Once the ferromagnetic order is established, the ferromagnetism splits the spin-degenerate and insulating bands from the coupled top and bottom surface states in the 3D-TI thin films. When band inversion occurs due to the spin splitting, the spin-orbit coupling reopens a band gap to induce a topological phase transition from a trivial insulator to quantum anomalous Hall insulator with a Chern number of $\mathcal{C}=1$~\cite{QAHE_MagTI_FangZh_10}. Such a delicate and precise theoretical prediction was later realized in Cr-doped (Bi,Sb)$_2$Te$_3$~\cite{QAHE_MagTI_exp_XueQK_13,rev_QAHE_MagTI_exp_XueQK_13,QAHE_MagTIexp_WangKL_14,QAHE_MagTI_exp_Checkelsky_14, QAHE_MagTI_exp.theo_ShenSQ_13,QAHE_MagTI_exp.theo_ChengSG_14} and V-doped (Bi,Sb)$_2$Te$_3$~\cite{QAHE_MagTI_exp_Moodera_15}. There are also similar proposals for other 3D-TI thin films, e.g. Cr-doped TlBiTe$_2$ and TlBiSe$_2$ films~\cite{QAHE_MagTI_Huang_11}. In addition, a similar band inversion can also be formed in magnetically-doped 2D topological crystalline insulator thin films~\cite{rev_TCI_FuL_15}, where the spin-orbit coupling induced band gap carries a Chern number of $\mathcal{C}=2$ rather than $\mathcal{C}=1$~\cite{QAHE_SnTe_Mele_13,QAHE_SnTe_Bernevig_14,QAHE_SnTe_Mokrousov_15}.

\subsection{Graphene and other honeycomb-lattice materials} \label{QAHE_G}

\begin{figure}
 \includegraphics[width=7cm,totalheight=6.cm,angle=0]{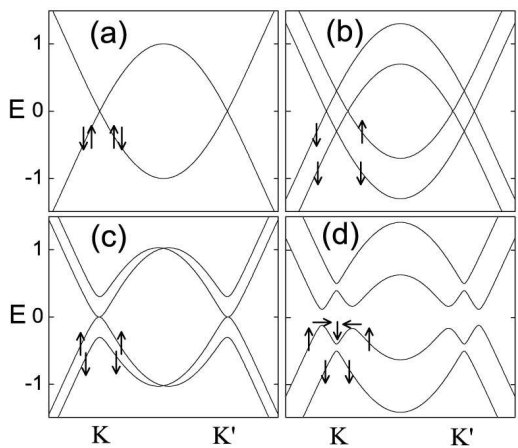}
 \caption{Evolution of band structures of bulk graphene along the profile of $k_y=0$; arrows represent the spin directions. (a) Pristine graphene: spin-up and spin-down states are degenerate; (b) When only a Zeeman field is applied, the spin-up/spin-down bands are upward/downward lifted with the four bands crossing near the $K$ and $K'$ points; (c) When only Rashba spin-orbit coupling is present, the spin-up and -down states are mixed around the band crossing points; (d) When both Zeeman field and Rashba spin-orbit coupling are present, a bulk gap is opened and all four bands become non-degenerate.
 Reprinted with permission from~[\onlinecite{QAHE_G_Qiao_10}], copyright 2010 by the American Physical Society.} \label{QAHE_G_Qiao_10_Fig1}
\end{figure}

\subsubsection{Monolayer graphene} \label{QAHE_G_M}
As mentioned in Sec.~\ref{TI_Haldane} above, the band crossover of graphene, i.e. the linearly dispersed  Dirac points K and K$^\prime$ shown in Fig.~\ref{QAHE_G_Qiao_10_Fig1}(a), originates from the sublattice/chiral symmetry. Therefore, in order to open a band gap at the Dirac points, this symmetry must be broken, for example by applying staggered AB sublattice potentials~\cite{QVHE_G_NiuQ_07} or by incorporating the intrinsic spin-orbit coupling from spin-dependent next-nearest neighbor hopping~\cite{QSHE_Mele_05}. However, the extrinsic Rashba spin-orbit coupling from spin-dependent nearest-neighbor hopping does not break the chiral symmetry but instead induces spin-mixing and lifts the four-fold degeneracy at the Dirac points. We see in Fig.~\ref{QAHE_G_Qiao_10_Fig1}(c) that the spin-degenerate linear dispersions around K and K$^\prime$ become quadratic band crossings where the up and down spins become mixed, which allows the band gap to open by applying a perpendicular Zeeman field, as shown in Fig.~\ref{QAHE_G_Qiao_10_Fig1}(d). Such a picture is closely related to the chirality of graphene's band structure. Various topologically nontrivial phases, such as $\mathbb{Z}_2$ TIs~\cite{QSHE_Mele_05}, QVHE~\cite{QVHE_G_NiuQ_07} and QAHE~\cite{QAHE_G_Qiao_10}, can be produced by breaking the sublattice/chiral symmetry.

On the other hand, graphene can also be regarded as a zero-gap semiconductor. When a perpendicular Zeeman field is applied, spin-splitting occurs to form crossing points between the spin-up and spin-down bands with the spin being a good quantum number~[see Fig.~\ref{QAHE_G_Qiao_10_Fig1}(b)]. The accidental degeneracy at the crossing points can be easily lifted by spin-mixing perturbation, e.g. Rashba spin-orbit coupling, as shown in Fig.~\ref{QAHE_G_Qiao_10_Fig1}(d)~\cite{QAHE_G_Qiao_10,QAHE_G_Guo_11}. Different from the understanding based on chiral symmetry, such a simple physical picture can be generally extended to other (quasi-)2D zero-gap or narrow-gap semiconductors, where a sufficiently large Zeeman field is required to induce crossing between the spin-up and -down bands by overcoming the bulk band gap. Moreover, the spin-mixing spin-orbit coupling is not merely limited to the Rashba type,for example, spin-orbit coupling in Bi$_2$Se$_3$ thin film may also result from bulk inversion-asymmetry~\cite{QAHE_MagTI_FangZh_10}.

\begin{figure}
 \includegraphics[width=7cm,totalheight=6.cm,angle=0]{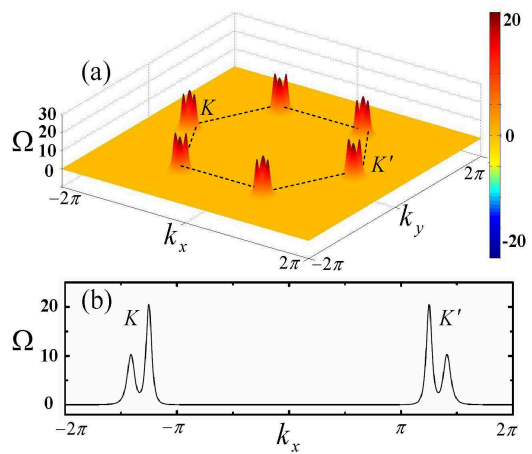}
 \caption{(color online). (a) Berry curvature distribution $\Omega$ (in units of $e^2/h$) of the valence bands in momentum space. The first Brillouin zone is outlined by the dashed lines, and two inequivalent valleys are labeled as $K$ and $K'$. (b) Profile of Berry curvature distribution along $k_y=0$.
 Reprinted with permission from~[\onlinecite{QAHE_G_Qiao_10}], copyright 2010 by the American Physical Society.} \label{QAHE_G_Qiao_10_Fig3}\end{figure}

In brief, the above explanation of graphene-based QA Hall effects from two distinct viewpoints can be regarded as two different limits since the effect is a joint consequence of the Rashba spin-orbit coupling and Zeeman field influence~\cite{QAHE_G_Qiao_12}. The first one corresponds to the large Rashba spin-orbit coupling limit, where the four-band low-energy continuum model Hamiltonian at the K or K$^\prime$ valley is effectively reduced to a two-band extended Haldane model (the detailed formula are presented in Ref.~[\onlinecite{QAHE_G_Qiao_12}]). The second corresponds to the large Zeeman field limit, where the QA Hall effect can be regarded as a consequence of the topological charges carried by skyrmions from the real-spin textures and merons from the AB sublattice pseudospin textures~\cite{QAHE_G_Qiao_12}. Since the merons from the lower two valence bands cancel each other, each valley carries a skyrmion and thus the total Chern number is $\mathcal{C}=2$, with equivalent contributions from both valleys, i.e. $\mathcal{C}_K=\mathcal{C}_{K^\prime}=1$. The equivalent contributions can be further understood from the Berry curvature, which is an analogy of the magnetic field in momentum space, as illustrated in Fig.~\ref{QAHE_G_Qiao_10_Fig3} where the Berry curvatures are peaked at the corners of the first Brilloin zone and have the same signs at the inequivalent valleys with $\Omega({\bm{k}})=\Omega ({-\bm{k}})$. In analogy to the formation of the QAHE in graphene with Dirac dispersion, similar proposals employing the joint effect of the Zeeman field and Rashba spin-orbit coupling have also been presented for Kagom\'e  ~\cite{QAHE_Kagome_Zhang_11}, checkerboard  ~\cite{QAHE_CheckBroad_Kumar_12}, star ~\cite{QAHE_StarLatt_WanSL_12}, and square lattices~\cite{QAHE_squareLett_Peng_13}.

\subsubsection{Experimental graphene-based QAHE prototypes}
It is noteworthy that both ferromagnetism and Rashba spin-orbit coupling do not exist in pristine graphene. Therefore, external means are required to induce these two effects. The most effective way is to adsorb magnetic 3$d$ transition-metal atoms on one side of a graphene sheet ~\cite{QAHE_G_adatom_Qiao_11,QAHE_G_Cluster_Wiesendanger_13}. For example, due to the magnetic proximity effect, graphene can be easily magnetized by the magnetic adatoms. Moreover, the charge transfer between graphene and the adatoms induces a charge redistribution in a very short distance (around 1.5 \AA) , which can generate a considerable electric field and result in a sizeable Rashba spin-orbit coupling by breaking the mirror symmetry about the graphene plane. In addition to 3$d$-adatoms, it is found that some 4$d$- and 5$d$-transition metal adatoms can also form ferromagnetic orders in graphene superlattices, e.g. Ru and W adatoms~\cite{QAHE_G_WuRQ_15,QAHE_G_5d_Mokrousov_12,QSHE_G_Fazzio_14}. In particular, for graphene with Ru adatoms, the QAHE with different Chern numbers can be obtained by adsorbing atoms in different supercells~\cite{QSHE_G_Fazzio_14,QAHE_G_WuRQ_15}.

\begin{figure}
 \includegraphics[width=7cm,angle=0]{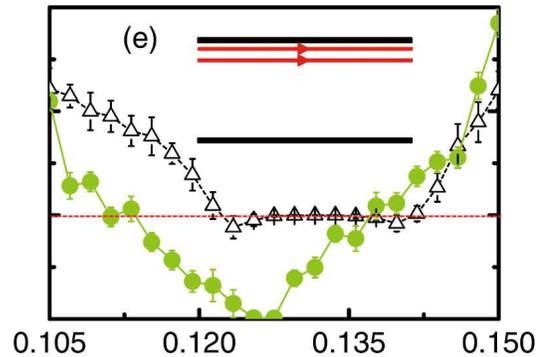}
 \caption{(colour online). Conductances $G$  of a two-terminal setup in the presence of periodically and randomly distributed adatoms as a function of Fermi level $\varepsilon_F$. Green solid line with solid circles: the case with periodic adsorption where an insulating regime occurs. Dashed black line with hollow triangles: the case of randomly distributed adatoms where the conductance shows a quantized plateau in some regime. Reprinted with permission from~[\onlinecite{QAHE_G_Qiao_randomAdat_12}], copyright 2012 by the American Physical Society.} \label{transport-valleyscattering}
\end{figure}

One of the important factors that may negatively influence the realization of QAHE is the possible existence of inter-valley scattering in $3\times3$ or $\sqrt{3}\times\sqrt{3}$ graphene supercells that couplesthe K and K$^\prime$ valleys by folding them into the $\Gamma$ point, which can open a sizeable trivial band gap in the hollow-adsorption case~\cite{QAHE_G_Qiao_12}, or form a quadratic band crossover in the top-adsorption case~\cite{QAHE_G_inter-valley_15}. In Ref.~[\onlinecite{QAHE_G_Qiao_randomAdat_12}], Jiang \textit{et al} employed a finite-size scaling method to show that the inter-valley coupling vanishes in a real sample with the adatoms being distributed in a completely random manner. Figure~\ref{transport-valleyscattering} displays the two-terminal averaged conductance as a function of the Fermi level in the presence of periodic-adsorption (i.e. one hollow-site adsorption in a $3\times3$ graphene supercell) and random adsorption with the same adsorption coverage, where the QAHE competes with the inter-valley scattering. We can see that in the periodic case there is an energy range with zero conductance $\langle G\rangle=0$, which corresponds to a trivial band gap [see Fig.~\ref{transport-valleyscattering} (a)], while when the adatoms become randomly distributed, a plateau of $\langle G \rangle=2e^2/h$ without fluctuation appears, indicating the formation of a QAHE band gap~[see Fig.~\ref{transport-valleyscattering} (b)]. This suggests that in a realistic sample with totally random distribution of the adatoms the inter-valley coupling will become vanishing but the real-spin related effects (e.g. magnetism and spin-orbit coupling) are not affected~\cite{QAHE_G_Qiao_randomAdat_12}.

It therefore seems that the QAHE should be easily realizable in graphene by randomly adsorbing some magnetic atoms. However, both later experiments and theories found that the adatoms in graphene cannot stabilize a dilute distribution but prefer to form clusters~\cite{QAHE_G_Cluster_MacDonald_13,QAHE_G_Cluster_Wiesendanger_13}, which is detrimental for QAHE generation. Within the same physical scheme, another promising method is to consider proximity-coupling with a ferromagnetic insulating substrate. For example, in Ref.~[\onlinecite{QAHE_G_AFM_Qiao_14}], graphene is coupled with the (111)-ferromagnetic plane of the anti-ferromagnetic insulator BiFeO$_3$, which is shown to be able to open a band gap larger than $1~$meV. Such a limited band gap originates from the extremely weak Rashba spin-orbit coupling since the van der Waals interaction between graphene and the substrate is weak due to their large separation of about 3 \AA. Thus the band gap can be further enlarged by applying an external stress to a certain degree. Another similar theoretical proposal is to place graphene on top of the (001) surface of RbMnCl$_3$, with a band gap in the order of 1-10~meV~\cite{QAHE_G_ZhangJ_15}. It is noticeable that although so far the QAHE has not yet been experimentally observed in graphene, considerable progress has been made, i.e. intrinsic ferromagnetism has been measured in graphene placed on top of LaMnO$_3$, and a large anomalous Hall conductance of $\sigma_{xy} \sim 0.2 e^2/h$ has been reported in a ferromagnetic insulating YIG thin film~\cite{QAHE_G_AFM_exp_Shi_15}, which is rather close to the final quantization observation because a finite Hall conductance only exists in a much narrower energy range~\cite{QAHE_G_AFM_Qiao_14}.

From the above analysis, it is reasonable to expect that, to realize the QAHE in graphene at a higher temperature (e.g. by engineering a large band gap), the most effective approach is still to dope magnetic atoms but not to use magnetic insulators. Fortunately, it has been shown that compensated $n$-$p$ codoping can be used to form long-range ferromagnetism in graphene by simultaneously codoping Ni and B atoms~\cite{QAHE_G_Qi_13}. This provides a valuable and practical route via magnetic doping. Moreover, ferromagnetic order can also be induced in nano-meshed graphene, where the $p_{xy}$ orbitals dominate the conducting electrons~\cite{QAHE_G_Zhao_15}.

\subsubsection{Low-buckled honeycomb-lattice systems of group-IV elements} \label{QAHE_G_Si}
The formation mechanism of the QAHE from the Zeeman field and Rashba spin-orbit coupling in graphene can also be applied to low-buckled honeycomb lattice systems, for example silicene, which is a single layer of silicon. The major difference between low-buckled systems and planar graphene is the generation of an intrinsic Rashba-type spin-orbit coupling which makes the extrinsic Rashba unnecessary for generating the QAHE. Since the intrinsic Rashba spin-orbit coupling is momentum-dependent and vanishes at the Dirac points K and K$'$, compared with the linearly dispersed edge states in graphene the nearly flat-band edge states are present in the absence of any extrinsic Rashba effect~\cite{QAHE_Si_Ezawa_12}.

Although both the intrinsic and the extrinsic Rashba spin-orbit couplings can give rise to the QAHE with a Chern number of $\mathcal{C}=2$ in the presence of a Zeeman field, their competition results in a new topological phase, i.e. a valley-polarized QAHE phase with a Chern number of $\mathcal{C}=1$ that originates from only one valley~\cite{QAHE_QVHE_Si_Yao_14}. This can be understood as an intermediate topological phase that occurs between the transition from the intrinsic to the extrinsic Rashba spin-orbit coupling induced QAHE phases ~\cite{QAHE_QVHE_Si_Yao_14,QAHE_QVHE_Si_YangSY_15}. Similar effects have also been proposed for bilayer silicene~\cite{QAHE_QVHE_Si_YangSY_15}. It is notable that the low-buckled structure or the top-bottom degree of freedom makes it possible to tune the band structure through applying an external electric field to induce a rich range of topological phases~\cite{QAHE_Si_Ezawa_12,QAHE_QVHE_Si_YangSY_15}. Although the Rashba spin-orbit coupling intrinsically exists in low-buckled systems, the Zeeman field has to be induced by some external means, such as 3$d$ and 4$d$ magnetic atom decoration~\cite{QAHE_Si_Schwingenschlogl_14,rev_QAHE.QSHE_LiuWM_15,QAHE_Si_LiuWM_14,QAHE_Si_yang_13}. Another most interesting topological phase is the QSHE-QAHE, wherein the chiral edge states are present in one valley while the spin-helical edge states appear in another valley through introduction of the sublattice-dependent Zeeman field~\cite{QAHE_QSHE_QVHE_Si_Ezawa_13}.

\subsubsection{Buckled honeycomb-lattice system of group-V elements} \label{QAHE_G_Bi}
Different from the Dirac semi-metal honeycomb lattices composed of group IV elements where the low-energy physics is determined by the half-filled $\pi$ band Dirac dispersions, in a honeycomb lattice Bi(111) bilayer the fully filled valence bands create an insulator where the $p_z$-dominated conduction band and the $p_{xy}$-dominated valence band edges possess odd and even parities, respectively, at the $\Gamma$ point~\cite{QSHE_Edge_Bi_FengJ_14}. The strong spin-orbit coupling can then induce inversion between the bands of opposite parities, and so generate a $\mathbb{Z}_2$ TI. When a small Zeeman field is introduced, a time-reversal symmetry breaking QS Hall effect results, characterized by a quantized spin-Chern number and spin-polarized edge states. Moreover, the presence of the Zeeman field also lifts the spin degeneracy of both the conduction and valence bands, shrinking the topological band gap.

When the Zeeman field is further increased, the band gap gradually closes and reopens, leading to a topological phase transition from a time-reversal symmetry breaking QSHE to one with a Chern number of $\mathcal{C}=-2$, where $M$ is the Zeeman field~\cite{QAHE_Bi(111)_Mokrousov_12,QAHE_Bi_Mokrousov_13}. This is different from the band inversion induced effect with a Chern number of $C=\sgn(M)$ from an ordinary insulator~\cite{QAHE_Bi_Mokrousov_13,QAHE_MagTI_FangZh_10}. The strong spin-orbit coupling plays a crucial role in driving the topological phase transition, i.e. the system evolves from a QAHE phase with a Chern number of $\mathcal{C}=1$ to one with $\mathcal{C}=-2$ mediated by a metallic phase with the lowest filled bands contributing a Chern number of $\mathcal{C}=2$~\cite{QAHE_Bi_Mokrousov_13}. So far, although the density-functional calculation has confirmed the existence of such a topological phase transition, the underlying reason why a single band inversion can induce various Chern numbers is still an open issue.

\subsubsection{Half-functionalized honeycomb-lattice systems of group-IV and -V elements} \label{QAHE_G_Fun}
In addition to doping magnetic atoms, spontaneous ferromagnetic order can also be induced by functionalization~\cite{QAHE_Ge_Mou_14,QAHE_Ge.Sn_YanBH_14}. Let us first take stanene as an example. Different from the full functionalization that can result in a large gap TI in stanene, the half-I-passivated stanene can establish a spontaneous ferromagnetic order due to the dangling bonds at one side~\cite{QAHE_Ge.Sn_YanBH_14}. Interestingly, because of the coupling between the functional group and the $p_z$-orbital of the tin atoms, the graphene-like Dirac bands at the K and K$^\prime$ points are pushed to high energy, whereas the \textit{s-p}-hybridized bands at the $\Gamma$ point become low-energy bands. In the absence of spin-orbit coupling, the ferromagnetism leads to asymmetry between the spin-up and -down bands, where the \textit{s-p$_{xy}$} band inversion for the spin-up bands disappears while that for the spin-down bands appears. Therefore, when the spin-orbit coupling is included, it only opens a gap between the spin-down bands at the $\Gamma$ point, which forms a QAHE phase with a Chern number of $\mathcal{C}=1$. It has been reported that the strong spin-orbit coupling can open a large band gap of about 340 meV, which is a good candidate for the realization of a room-temperature QAHE structure~\cite{QAHE_Ge.Sn_YanBH_14}. Similar results have alsobeen proposed for half-I-passivated germanene yet with a much smaller band gap of about 60 meV~\cite{QAHE_Ge.Sn_YanBH_14}. This kind of half functionalization at one side is geometrically equivalent to the case with stanene being placed on top of certain substrates, such as CdTe and InSb (111) surface~\cite{QAHE_Ge.Sn_YanBH_14}. In fractional functionalized silicene and germanene, QA Hall effects with a Chern number of $\mathcal{C}=-1$ or $2$ have also been predicted~\cite{QAHE_Ge_Mou_14}.

Similar to stanene, although the fully functionalized Bi(111) bilayer is shown to be a $\mathbb{Z}_2$ TI, the time-reversal symmetry is broken due to the formation of a spontaneous ferromagnetic order in the half-hydrogenated Bi(111) bilayer~\cite{QAHE_Bi-H_Yao_15,QSHE_Bi_QAHE_QVHE_Mokrousov_15} or Bi(HN) bilayer, with one side being hydrogenated and the other side decorated by nitrogen atoms~\cite{QSHE_Bi_QAHE_Jhi_15}. The induced ferromagnetism together with the strong spin-orbit coupling gives rise to a large gap QAHE with a Chern number of $\mathcal{C}=1$, which originates from the inversion symmetry breaking, giving different responses at the K and K$'$ valleys. As a consequence, band inversion occurs only at one valley to produce a valley-polarized QAHE~\cite{QAHE_Bi-H_Yao_15, QSHE_Bi_QAHE_QVHE_Mokrousov_15, QSHE_Bi_QAHE_Jhi_15}.

\subsubsection{Artificial honeycomb-lattice systems} \label{QAHE_G_Artificial}
The recent development of artificial lattices in cold atoms, photonic crystals, phononic crystals and so forth provides alternative platforms to study topological phases in media besides condensed matter~\cite{rev_ArtificialLattices_Pellegrini_13}. Apart from the realization of Haldane's model via externally controllable gauge fields~\cite{QAHE_ColdAtom_Haldane_WangZD_08}, cold atom systems can also be designed for studies related to $p_{x,y}$-orbital-dominated physics in contrast to the $p_z$-orbital-dominated $\pi$-bands in honeycomb lattice structures~\cite{QAHE_ColdAtom_WuCJ_08}. In a honeycomb optical lattice, the orbital momentum of the cold atoms makes it possible to split the $p_{x,y}$ degeneracy by rotating each lattice site around its center, which can result in a cold atom analogy of the QAHE in condensed matter, where the localized edge states are characterized by a Chern number of $\mathcal{C}=1$.

In addition to the $s$- and $p$-orbital-dominated bands, a $d$-orbital based QAHE has been predicted in an artificial structure composed of heavy transition metal atoms, e.g. W grown in a 1/3 monolayer of halogen-Si(111) that constitute a honeycomb lattice structure. The splitting of the $s$- and $d$-orbitals of W atoms due to the crystal field results in a spontaneous ferromagnetic order. This ferromagnetism and the strong spin-orbit coupling of the heavy metal atoms combine together to open up a large band gap of about 100 meV to create a QAHE structure with a Chern number of $\mathcal{C}=-1$. Such a complex structure is expected to be achievable based on current state-of-the-art technology~\cite{QAHE_TM_LiuF_14}.

In addition to the above inorganic materials, planar honeycomb lattices can also be artificially constructed by using organic molecules (e.g. triphenyl) and magnetic Mn atoms in a specific manner, which is shown to be able to realize the Kane-Mele QSHE when the time-reversal symmetry is preserved.
The magnetic Mn atoms can induce an extremely large intrinsic ferromagnetism, which can completely separate the spin-up and -down bands.
In contrast to the Rashba spin-orbit coupling induced gap at the crossing points of the spin-up and spin-down bands, the intrinsic spin-orbit coupling of the planar artificial honeycomb lattice gives rise to a fully spin-polarized QAHE system (or ``half" Kane-Mele type TI).

\subsection{Heterostructure quantum wells} \label{QAHE_QW}
Since the intrinsic ferromagnetism and spin-orbit coupling are two essential ingredients for realizing the QAHE, another possible route is to directly include these two factors in a heterostructure composed of a heavy atom insulator and a ferromagnetic insulator. For example, Garrity and Vanderbilt proposed doping heavy metal atoms in magnetically ordered MnTe, MnSe, or EuS surfaces ~\cite{QAHE_HM_Vanderbilt_13}. When the bands of the heavy atomic layer exhibit a gap that is located inside the large band gap of the magnetic insulator, the QAHE can in principle appear. Following this recipe, they predicted several such systems with large band gaps or large Chern numbers up to $\mathcal{C}=3$ by employing first-principles calculations. However, in these systems the requirement of periodic adsorption is currently beyond possibility of experimental realization.

Similarly, a heterostructure quantum well composed of CdO/EuO (both oxides have the rocksalt structure) provides another platform to realize QAHE. Here, EuO is a ferromagnetic semiconductor with a valence band mainly produced by the spin-polarized $f$-orbital of Eu while the conduction band is dominated by the $s$-orbital of Cd. In the presence of a suitable in-plane strain or an out-of-plane electric field, the strong spin-orbit coupling may lead to a band inversion between the conduction band of even parity and the valence band of odd parity, which can open up a bulk band gap to host the QAHE~\cite{QAHE_CdOEuO_QW_ZhangSC_14}. It is noteworthy that the time-reversal symmetry is broken by the spontaneous ferromagnetic order of EuO but not the magnetic doping that is used in InAs/GaSb quantum wells or 3D-TI thin films.

Comparable to the oxide heterostructure, a bilayer system composed of GdN and EuO (both topologically trivial ferromagnetic insulators) provides another scheme to realize the QAHE. The strong spin-orbit coupling of Gd and the spin-polarized conduction band minima from the $d$-orbital as well as the smaller lattice mismatch make this system a good candidate for the QAHE with a large band gap. Moreover, a bilayer of ferromagnetic insulators such as Cr-doped (Bi,Sb)$_2$Te$_3$ and GdI$_2$ has proved to be another possible candidate~\cite{QAHE_HeteroStru_ZhangSC_15}. In addition to the magnetic doping, the ferromagnetism can also be engineered through the proximity effect with the magnetic substrates~\cite{QAHE_MagTI_Lin_15}.

Additionally, another route is to start from the 3D Chern semi-metals, where the gapless dispersion at the Fermi point is determined by the band inversion from the spin-orbit coupling, as in CdO/EuO superlattices~\cite{QAHE_CdOEuO_QW_ZhangSC_14} and HgCr$_2$Se$_4$~\cite{QAHE_HgCr2Se4_FangZh_11}. In its corresponding thin film form, the characteristics of both the ferromagnetism and the spin-orbit coupling induced band inversion will still persist to harbour the QAHE. Interestingly, tuning the thickness may be an efficient way to produce a large Chern number [see Sec.~\ref{QAHE_LargeC}].

\subsection{Transition metal oxides} \label{QAHE_TMO}
So far, although the ferromagnetic order relies mainly on the presence of transition metal atoms, the host materials are based on the group-III, -IV, -V, and -IV elements or compounds with negligible electron-electron correlations. Recently, the search for topological materials, in part stimulated by experimental progress~\cite{QAHE_TMO_exp_Takagi_14}, has begun to target the transition metal oxides where the electron-electron correlation and strong spin-orbit coupling (especially the 5$d$ transition metal atoms with large atomic number) have led to many interesting discoveries~\cite{QAHE_IrO3_Kee_14,QAHE_IrO3_Vanderbilt_14}. Based on iridium oxide, two theoretical  predictions of the QAHE in (SrIr/TiO$_3$)$_n$ with $n=1$ or $2$~\cite{QAHE_IrO3_Kee_14} and the monolayer La$_2$MnIrO$_6$~\cite{QAHE_IrO3_Vanderbilt_14} have been reported. Different from the honeycomb lattices of graphene-like materials, these thin films possess square lattices, i.e. orthorhombic~\cite{QAHE_IrO3_Kee_14} or double perovskite~\cite{QAHE_IrO3_Vanderbilt_14} structures where the $t_{2g}$-orbitals dominate the low-energy physics around the Fermi level due to the crystal fields. In the orthorhombic case, the strong spin-orbit coupling splits the $t_{2g}$-orbitals into $J_{\rm{eff}}=1/2$ and $J_{\rm{eff}}=3/2$, where the spin is already combined with the orbital motions. In the case of a single IO$_2$ layer with $n=1$, the specific glide symmetry leads to the Dirac dispersion at the X and Y points in the Brillouin zone, which are the two time-reversal invariant momentum points. The TI phase can be formed when an external strain is applied to break the glide symmetry and open a band gap at the Dirac points. When the Zeeman field is further included to close and reopen the band gap at the X point without affecting the bands near the Y point, the QAHE with a Chern number of $\mathcal{C}=1$ appears.

Interestingly, in stacked bilayers, the inter-layer coupling drives the glide symmetry-protected Dirac cones away from the X and Y points. In this case, any kind of magnetism can open a band gap to host the QAHE phase with a Chern number of $\mathcal{C}=2$~\cite{QAHE_IrO3_Kee_14}. Different from SrIrO$_3$, the orbitals dominating the low-energy physics in the monolayer La$_2$MnIrO$_6$ come from the $t_{2g}$-orbitals of the $3d$-Mn atoms, whose inter-site hopping induced strong spin-orbit coupling opens a gap of about 26 meV when the Mn atoms establish a ferromagnetic order~\cite{QAHE_IrO3_Vanderbilt_14}. Apart from the $5d$-atom oxides, the QAHE has also been predicted in the oxidation of the $3d$-Cr element, i.e. CrO$_2$/TiO$_2$, where a smaller band gap of about 3 meV is induced due to the weak on-site spin-orbit coupling~\cite{QAHE_TMO_GongCD_13}.

\subsection{Large-Chern-number QAHE} \label{QAHE_LargeC}
The Chern numbers of any QAHE systems described in the previous sections are mainly limited to $\mathcal{C}=\pm 1$ and $\mathcal{C}=\pm 2$. Since the conductance in the Landau-level induced quantum Hall effect can be modified to have various integer values by changing the magnetic field or varying the Fermi levels, the possibility to obtain large-tunable large-Chern numbers that could provide strong currents and thus strong signals is of great interest from both the theoretical and practical aspects~\cite{QAHE_MagTI_Qiao_12,QAHE_MagTI_ZhangSC_13}.

In Bernal-stacked bilayer graphene, a Chern number of $\mathcal{C}=4$ is predicted in the presence of the Zeeman field and Rashba spin-orbit coupling~\cite{QAHE_G_Qiao_11, QAHE_G_Qiao_QSHE_QVHE_13}, where each valley contributes a Chern number of $\mathcal{C}=2$. Therefore, larger Chern numbers can be expected in Bernal-stacked multi-layer graphene. A large-Chern-number QAHE phase has also been reported in the graphene-like system $\beta$-graphyne where the carbon atom triple bond is inserted into graphene~\cite{QAHE_G_Juricic.Smith_14}. In such a system, the competition between the intrinsic and extrinsic Rashba spin-orbit couplings in this specific lattice structure allows us to tune the number and position of the Dirac cones. When the time-reversal invariance is broken, the resulting band structure can exhibit various QAHE phases with different Chern numbers ranging from $\mathcal{C}=-3$ to $\mathcal{C}=3$~\cite{QAHE_G_Juricic.Smith_14}.

\begin{figure}
  \centering
  \includegraphics[width=8 cm]{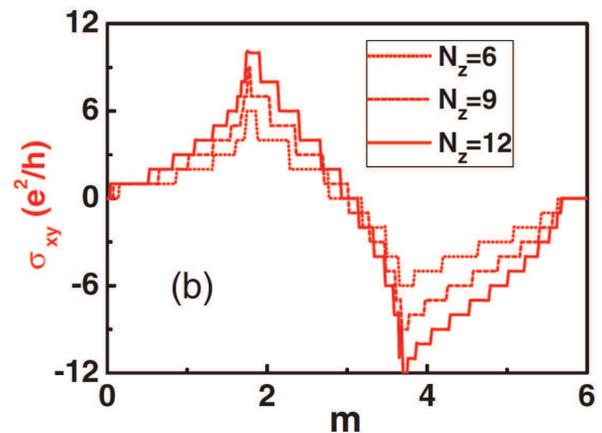}
  \caption{(color online). Hall conductance vs. Zeeman field $m$, for sample thicknesses of $N_z = $6, 9 and 12. Reprinted with permission from~[\onlinecite{QAHE_MagTI_Qiao_12}], copyright 2012 by the American Physical Society.} \label{QAHE_MagTI_Qiao_12_Fig5}
\end{figure}

Additionally, large Chern numbers have also been predicted in magnetic 3D-TI thin films beyond the 2D limit~\cite{QAHE_MagTI_Qiao_12,QAHE_MagTI_ZhangSC_13}. Within the 2D limit, the QAHE arises from the direct coupling between the top and bottom surface states which gives rise to the lowest Chern number of $\mathcal{C}=1$~\cite{QAHE_MagTI_FangZh_10}. When the films are thicker, the conduction and valence subbands due to the confinement along the $z$-direction become involved in the band inversion due to the Zeeman field and spin-orbit coupling. The Chern number is therefore strongly dependent on the relative magnitude of the Zeeman field to the sample thickness, which determines the separation between the subbands. As illustrated in Fig.~\ref{QAHE_MagTI_Qiao_12_Fig5}, for different sample thicknesses, increasing the Zeeman field can increase the Chern number to a rather large integer, which is different from the ordinary quantum Hall effect where increasing the magnetic field decreases the Hall conductance~\cite{QAHE_MagTI_Qiao_12,QAHE_MagTI_ZhangSC_13}. The recent experimental observation of the QAHE in (Cr$_{0.12}$Bi$_{0.26}$Sb$_{0.62}$)$_2$Te$_3$ samples demonstrates the possibility of realizing such large Chern numbers~\cite{QAHE_MagTIexp_WangKL_14}. In these systems, the nearly quantized Hall conductance was observed in a sample with a thickness of over 10 quintuple layerss, which is beyond the 2D hybridization thickness~\cite{QAHE_MagTIexp_WangKL_14}.

\subsection{In-plane magnetization induced QAHE} \label{QAHE_InPlane}
\begin{figure}
  \centering
  \includegraphics[width=6 cm]{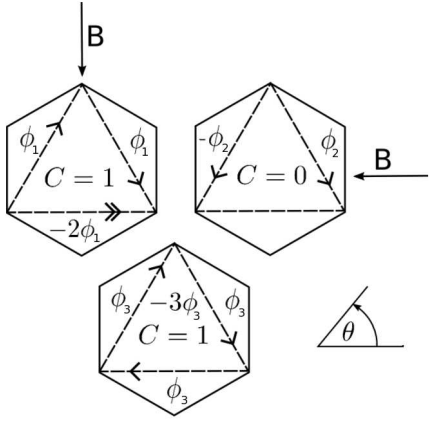}
    \caption{Hexagonal unit cells with zero net magnetic flux. The upper two unit cells are a buckled honeycomb lattice in an in-plane magnetic field. The lower figure is Haldane＊s unit cell. The direction of positive flux accumulation is indicated by the arrows along the bonds. For clarity, only the bonds along a single sublattice are shown. Reprinted with permission from~[\onlinecite{QAHE_Si_Wright_13}].} \label{QAHE_Si_Wright_13_Fig1}
\end{figure}

In addition to the out-of-plane Zeeman field, the in-plane field can also induce the QAHE ~\cite{QAHE_2DEG_ZhangCW_11, QAHE_HgMnTe_LiuCX13, QAHE_QW_InPlane_LiuCX_13}. In a 2D system, the presence of mirror reflection symmetry, i.e. symmetry under $M_x: (x,y) \rightarrow (-x,y)$, constrains the Hall conductance $\sigma_{xy}$ to be zero since this symmetry requires the same current along both the $x$ and $-x$ directions. Different from the out-of-plane Zeeman field which is a pseudoscalar, i.e. a scalar that changes sign under parity inversion and thus breaks any in-plane reflection symmetry, the in-plane Zeeman field is a pseudovector and does not break all the reflection symmetries; for example, the $M_x$ reflection symmetry is preserved when the field is applied along the $x$-direction. Therefore, the in-plane field can only induce a QAHE when the remaining mirror-reflection symmetry is broken~\cite{QAHE_QW_InPlane_LiuCX_13}, as in a patterned 2D electron gas with both Rashba and Dresselhaus spin-orbit couplings~\cite{QAHE_2DEG_ZhangCW_11}, or in magnetically-doped Bi$_2$Te$_3$ thin films with further inclusion of trigonal warping~\cite{QAHE_QW_InPlane_LiuCX_13}. Alternatively, in a honeycomb lattice, the mirror reflection transforms the states in the K to the K$'$ valley, hence the mirror reflection is effectively broken in a single valley. The QA Hall effect in honeycomb lattices is thus expected to be different from that in magnetically doped Bi$_2$Te$_3$ thin films.

On the other hand, in a buckled honeycomb lattice system, the in-plane magnetic field may be employed by considering only the magnetic flux induced orbital effect, in analogy to Haldane＊s model~\cite{QAHE_Si_Wright_13}. As displayed in the upper left diagram of Fig.~\ref{QAHE_Si_Wright_13_Fig1}, the in-plane magnetic field can generate a finite magnetic flux in the three outer triangles around the hexagon. Although the detailed flux configuration is different from that in the Haldane model, as displayed in the lower panel of Fig.~\ref{QAHE_Si_Wright_13_Fig1}, the total magnetic flux is zero in the whole system and a Chern number of $\mathcal{C}=1$ is induced for the spinless fermion, which is the same as that in the Haldane model. In realistic materials with a spinful fermion, the Chern number will double if the Zeeman splitting and the spin-orbit coupling do not close the bulk band gap. Note that, the magnetic flux configuration is strongly dependent on the angle of the magnetic field, and a topologically trivial phase occurs when this angle is rotated through 90 degrees as illustrated in upper right panel of Fig.~\ref{QAHE_Si_Wright_13_Fig1}. Therefore, a topological phase transition is easily manipulated by changing the direction of the magnetic field, which is the second major difference from that in Haldane＊s model.

\subsection{QAHE in a system with antiferromagnetic order} \label{QAHE_AFM}
It is known that time-reversal symmetry breaking is crucial in engineering QAHE, and the intrinsic ferromagnetism is usually adopted to play such a role. Recently, the effect was predicted in a [111] perovskite material, which has a buckled honeycomb lattice structure and antiferromagnetism in the presence of a perpendicular electric field~\cite{QSHE_TRB_Hu_13}. In the low-buckled honeycomb lattice system, the strong intra-atomic spin-orbit coupling is expected to generate a large Kane-Mele-type intrinsic spin-orbit coupling, giving rise to a QSHE. However, the presence of the anti-ferromagnetic order induces a spin-sublattice dependent site potential and breaks the time-reversal symmetry to drive the QSHE into a band insulator. The further application of a perpendicular electric field triggers a single band inversion at one valley while keeping the gap open at the other valley, leading to a valley-polarized QAHE with a  Chern number of 1.

\begin{table*}
  \caption{Possible materials for realizing the QAHE. 1st column: possible materials for realizing the QAHE. 2nd column: the corresponding Chern numbers. 3rd column: some remarks. 4th column: band gaps calculated from \textit{ab-initio}. Last column: relative references. In this table, $\lambda_{\rm{R}}^{\rm{int}}$ and $\lambda_{\rm{R}}^{\rm{ext}}$ indicate the strength of intrinsic and extrinsic Rashba spin-orbit couplings, respectively; Bi(BN) represents the functionalized Bi (111) bilayer where one side is saturated by B atoms while the other side is saturated by N atoms. Abbreviates used, G: graphene; BL: bilayer; TL: trilayer; BLG: bilayer graphene; Fun.: functionalized; QWs: quantum wells. } \label{QAHE_Table}
    \begin{tabular*}{\textwidth}{c@{\extracolsep{\fill}}|cccc}
      \hline \hline
      Material                        & $\mathcal{C}$          &              Remark                     &               Gap            & Ref. \\ \hline
            HgTe QWs             & 1                               &                                                  &                                 &     \cite{QAHE_FMQW_ZhangSC_08} \\ 
 InAs/GaSb QWs             & 1                                &   magnetic doping                       &                                 &     \cite{QAHE_MagTI_LiuCX_14} \\ 
 Junction QWs                & 1                                &                                                       &                                &     \cite{QSHE_QAHE_QW_Junction_ZhangSC_14} \\  \hline
  3D TI thin film             & 1                                 &   2D limit, magnetic doping             &                                &     \cite{QAHE_MagTI_FangZh_10} \\  
  3D TI thin film              & thickness dependent  &  beyond 2D limit, magnetic doping    &                                &     \cite{QAHE_MagTI_Qiao_12,QAHE_MagTI_ZhangSC_13} \\  
  3D TI thin film/FM       &  1                               &                                                          &    $\sim$ 10 meV      &     \cite{QAHE_MagTI_Lin_15} \\  \hline
      G                              & 2                                &     theory                                             &                                 & \cite{QAHE_G_Qiao_10} \\ 
  3$d$ atom/G                & 2                               &                                                            & $\sim$ 1-10 meV      & \cite{QAHE_G_Qiao_11} \\ 
  5$d$ atom/G                & 2                              &   electric field tunable                          & $\sim$ 10-100 meV & \cite{QAHE_G_5d_Mokrousov_12} \\ 
 Ru atom/G                      & -2                           &   4$d$ atom                                          & $\sim$ 10 meV       & \cite{QSHE_G_Fazzio_14} \\ 
 Co/Rh atom/G                 & 1                          &  impurity-band contribution                  & $\sim$ 50/100 meV & \cite{QAHE_G_WuRQ_15} \\ 
 G/BiFeO$_3$ (111)        & 2                           &                                                           & $\sim$ 1 meV          & \cite{QAHE_G_AFM_Qiao_14} \\ 
 G/RbMnCl$_3$ (001)     & 2                          &                                                          & $\sim$ 1-10 meV   & \cite{QAHE_G_ZhangJ_15} \\ \hline
 MLG                               &  4 or 2                       &       theory                                      &                              & \cite{QAHE_G_Qiao_11} \\ \hline
 silicene                             & 2                            &   theory                                            &                             & \cite{QAHE_Si_Ezawa_12} \\ 
 silicene                             & 1                          &   theory, competition between
                                                                                $\lambda_{\rm{R}}^{\rm{int}}$
                                                                                 and $\lambda_{\rm{R}}^{\rm{ext}}$ &                           & \cite{QAHE_QVHE_Si_Yao_14} \\ 
 3$d$ atom/silicene             & 2                          &                                                        & $\sim$ 1 meV        & \cite{QAHE_Si_Schwingenschlogl_14,rev_QAHE.QSHE_LiuWM_15,QAHE_Si_LiuWM_14} \\ 
 4$d$ atom/silicene              & -2  or 1                &   depend on atom type                  & $\sim$ 10 meV           & \cite{QAHE_Si_yang_13} \\ \hline
 Bi (111) BL                          & -2                         &   theory                                         &                                 & \cite{QAHE_Bi(111)_Mokrousov_12} \\ \hline
  I-stanene                              & 1                      &    half functionalization                    & $\sim$ 340 meV         & \cite{QAHE_Ge.Sn_YanBH_14} \\ 
  I-germanene                         & 1                         &    half functionalization                  & $\sim$ 60 meV         & \cite{QAHE_Ge.Sn_YanBH_14} \\ 
  Fun. silicene/germanene       & 2 or -1                & fractional one-side saturation        &                                    & \cite{QAHE_Ge_Mou_14} \\ \hline
  Fun. Bi (111)                        &       1                    & -H, half functionalization                &    $\sim$ 200 meV     & \cite{QAHE_Bi-H_Yao_15,QSHE_Bi_QAHE_QVHE_Mokrousov_15} \\ 
        Bi(BN)                          &       1                        &                                                       &    $\sim$ 100 meV     & \cite{QSHE_Bi_QAHE_Jhi_15} \\ \hline
  W atoms on halogen-Si(111) &       1                  & artificial honeycomb lattice            &    $\sim$ 100 meV     & \cite{QAHE_TM_LiuF_14} \\ \hline
  2D triphenyl-Mn                   &     1                         & organic material                         &    $\sim$ 9.5 meV     & \cite{QAHE_organic_LiuF_13} \\ \hline
  Heterostructure QWs         &  $-3$, $-2$, $-1$,  1     & material dependent                &    1-700 meV             & \cite{QAHE_HM_Vanderbilt_13} \\ \hline
  CdO/EuO or GdN/EuO       &            1                      &     intrinsic QAHE                &                                & \cite{QAHE_CdOEuO_QW_ZhangSC_14,QAHE_HeteroStru_ZhangSC_15} \\ \hline
  HgCr$_2$Se$_4$ thin film & thickness dependent     &   intrinsic QAHE                  &                                & \cite{QAHE_HgCr2Se4_FangZh_11}  \\ \hline
  (SrIr/TiO$_3$)$_n$ ($n=1,2$) &                      n       &                                             &                                 & \cite{QAHE_IrO3_Kee_14}  \\ \hline
  monolayer La$_2$MnIrO$_6$  &            1                  &                                              &  $\sim$ 26 meV     & \cite{QAHE_IrO3_Kee_14}  \\ \hline
  \hline
\end{tabular*}
\end{table*}

\subsection{Edge-state engineering} \label{QAHE_Edge}
Although the QAHE is a consequence of the bulk topology of band structure, its representative character that is required for application is the robust dissipationless chiral edge modes, which can also be engineered by simply manipulating a finite-sized ribbon~\cite{QAHE_in_QSHE_XingDY_13, QAHE_Edge_LiuXJ_14, QAHE_G_ZhangZhY_14}. The most effective approach to engineer the chiral edge modes at the boundaries is to destroy or remove half of the spin-helical edge modes of the QSHE, since their spin-up and -down states propagate in opposite directions so the Zeeman field is expected to lift the degeneracy between the Kramers pair. Li \textit{et al} found that when applied at the boundary the field can drive states with one spin away from the boundary, but the remaining gapless edge states with opposite spin are also localized at the boundaries. Therefore, only the gapless chiral edge states with a specific spin are present, which gives a quantized two-terminal conductance as well as the Hall conductance~\cite{QAHE_in_QSHE_XingDY_13}. Such a Zeeman field appearing at the edges can be induced from the proximity effect with a ferromagnetic insulator. In addition, it is found that the hopping amplitude between the topological insulator and a normal insulator can also lead to a spatial separation between the two copies of the counter-propagating edge states carrying opposite spins~\cite{QAHE_G_ZhangZhY_14}.

\section{Quantum Valley-Hall Effect and Topological Zero-Line Modes}\label{QVHE}
In this section we review the the QVHE and other valley-related topological phases in graphene and graphene-like honeycomb lattice systems. Different from the $\mathbb{Z}_2$ TIs and QAHE discussed above, the QVHE or topological 1D ZLMs do not rely on quantum manipulation of real-spin related effects such as spin-orbit coupling and ferromagnetism, but only require an external electric field, which is easily realizable in the lab. Therefore, superior to the QAHE that can only be experimentally observed at extremely low temperatures, multilayer graphene-based QVHE and topological 1D ZLMs should, in principle, be able to bring about a revolutionary development in room-temperature dissipationless electronics or valleytronics.

\subsection{Topological aspect of honeycomb lattice from inversion-symmetry breaking}
In graphene and graphene-like honeycomb-lattice systems, one of the most important properties is the structurally-induced linear Dirac dispersions at two inequivalent valleys K and K$^\prime$ points, which are closely related to each other by the time-reversal symmetry. In the absence of short-range scattering, the valleys are decoupled and possess a long valley lifetime due to their large-separation in momentum space. Therefore, the electron in either specific valley (K or K$^\prime$) effectively breaks the time-reversal symmetry~\cite{valleytronics_G_effectiveTRB_Guinea_06}, and can give rise to a finite magnetic momentum as a consequence of the local intrinsic Berry curvature when a bulk band gap is opened by breaking the inversion symmetry [see Figs.~\ref{QVHE_G_Edge_ZLM}(b$_1$) and (c$_1$)]~\cite{QVHE_G_NiuQ_07}. The presence of the nontrivial Berry curvature concentrated at valleys K and K$^\prime$ has led to many fascinating transport properties, e.g. the valley Hall effect where a valley current flows along the direction transverse to the applied longitudinal charge current~\cite{QVHE_G_NiuQ_07}. In the transition metal dichalcogenides materials, the inversion-symmetry breaking can also induce a large band gap at valleys K and K$^\prime$ where the strong spin-orbit coupling lifts the spin degeneracy of the bands~\cite{valleytronics_MoS2_YaoW_12}. However, the Kramers degeneracy is preserved due to time-reversal symmetry, which relates the spin-up states at valley K to a energy-degenerate-spin-down state at valley K$^\prime$. The locking of the spin and valley indices can not only result in a combined spin-valley Hall effect in the electron- or hole-doped region~\cite{valleytronics_MoS2_YaoW_12}, which has been recently observed~\cite{valleytronics_MoS2_McEuen_14}, but also gives rise to a valley-selective photo-excitation of carriers that provides the possibility of controlling spin and valley indices via an optical method. This kind of spin-valley coupling physics can also be applied to the low-buckled honeycomb-lattice structures (e.g. silicene and germanene) by externally breaking the inversion symmetry~\cite{valleytronics_Si_Ezawa_12,valleytronics_Si.Ge_Ezawa_14}. The analogy between valley and spin degrees of freedom as well as their interplay in a system with strong spin-orbit coupling provides superb properties and promising applications in the next generation of spintronics and valleytronics~\cite{rev_G_S.V.tronics_MacDonald_12,rev_MoS2_ElectricField_Heine_15, rev_MoS2_Kis_15, rev_MoS2_Yao_15,rev_2D_Goldberger_13,
valleytronics_G_effectiveTRB_Guinea_06, valleytronics_G_Beenakker_07, valleytronics_G_optics_NiuQ_08, valleytronics_G_VHE_exp_Geim_14, valleytronics_G_White_11, QVHE_G_NiuQ_07, valleytronics_MoS2_YaoW_12, valleytronics_MoS2_McEuen_14, valleytronics_BLG_Schomerus_09, valleytronics_Si_Ezawa_12, valleytronics_Bi_Behnia_11, valleytronics_Diamond_Twitchen_13, valleytronics_Si.Ge_Ezawa_14, valleytronics_TCI_Ezawa_14}. Several excellent reviews on the physics and possible applications based on the spin-valley locking have been published already, so they will not be mentioned further here~\cite{rev_G_S.V.tronics_MacDonald_12, rev_MoS2_ElectricField_Heine_15, rev_MoS2_Kis_15,rev_MoS2_Yao_15}. Apart from these fascinating properties pertaining to the Fermi levels lying inside the valance or conduction bands, when the Fermi energy is located inside the bulk gap, the bulk band topology also leads to other striking transport properties, i.e. the QVHE in analogy to the QSHE as will be discussed below.

\subsection{Quantum valley-Hall effect}
\begin{figure*}
  \centering
  \includegraphics[width=16 cm]{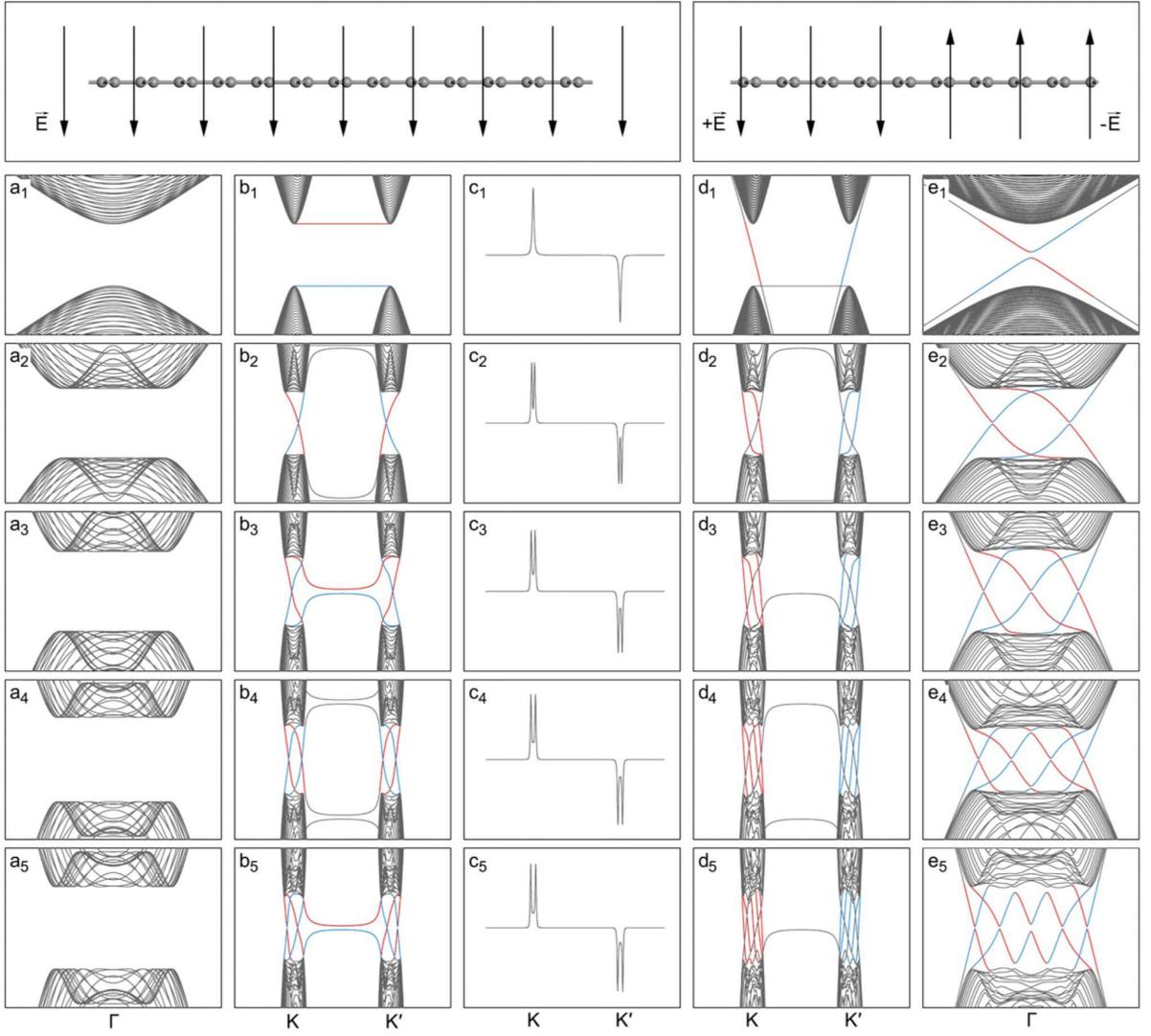}\\
  \caption{(color online). Left figure, upper panel (QVHE): the uniform staggered sublattice potential for a monolayer and perpendicular electric field for multilayer graphene; the corresponding Berry curvature profiles of the induced QVHE are plotted in the third column c$_i$, with $i=1$-$5$ indicating the layer number. The Chern number for a single valley is $0.5i$ ($i=1$-$5$) per spin as labeled by $\mathcal{C}_{K,K^\prime}$. The electronic structures of the armchair and zigzag ribbons are plotted in the first and second columns, labeled by a$_i$ and b$_i$ ($i=1$-$5$), respectively. The edge states are absent in the armchair terminated nanoribbon due to inter-valley scattering. For the zigzag nanoribbon, the K and K$^\prime$ valleys are well separated and edge states are present, coloured red and blue on different sides of the boundary. We see that the pair of edge states at two boundaries are the same and equal to $N/2$ for each spin for even $N$ layer graphene, while unbalanced edge state numbers occur for odd $N$-layer graphene.
  Right figure, upper panel (zero line modes): the varying site potentials for a monolayer or varying electric field for multilayer graphene. At the interface across which the staggered sublattice potential or electric field direction changes sign, 1D topological ZLMs are present for both armchair and zigzag terminated interfaces, as shown in the fourth and fifth columns labelled by d$_i$ and e$_i$ ($i=1$-$5$), respectively. The number of ZLMs propagating parallel or antiparallel to the interface is equal to $i$ per spin (labelled in blue or red, respectively). Apart from the similarities, there are two differences between the band structures in the fourth and fifth column: 1) Gapless ZLMs are present in the zigzag terminated interface while a relatively small band gap occurs for ZLMs along an armchair terminated interface due to strong inter-valley scattering; 2) gapless edge states occur for a zigzag terminated ribbon at the outer boundary rather than at the interface, as plotted in the fourth column in gray.}
  \label{QVHE_G_Edge_ZLM}
\end{figure*}
Both the monolayer and Bernal-stacked multilayer graphene are zero-gap semiconductors with distinct dispersions at the Dirac points K and K$^\prime$. A topologically nontrivial bulk gap can be opened to initiate the QVHE when the inversion-symmetry is broken, e.g. by introducing a staggered sublattice potential in monolayer graphene or by applying a perpendicular electric field in multilayer graphene. Distinct from the rigorous definition of the topological indices of the QAHE and the $\mathbb{Z}_2$ TIs, the QVHE is simply characterized by the valley Hall conductivity given by $\sigma_{xy}^v= (\sigma_{xy}^K - \sigma_{xy}^{K{^\prime}})/2 $ for the spinless case, where $\sigma_{xy}^{K,K{^\prime}}$ is obtained by integrating the Berry curvatures near valleys K and K$^\prime$ by using the low-energy continuum model. When the bulk band gap is small, the finite Berry curvatures are mainly concentrated at the Dirac points, with the two valleys being well-separated. In the absence of the inter-valley scattering, the resulting valley Hall conductivity assumes an integer (half-integer) for even (odd) N-layer graphene.

Such a bulk quantization only has edge correspondence at specific system boundaries without strong inter-valley interaction. For example, zigzag ribbon geometries with a large momentum separation between valleys can support gapless edge modes, manifesting the quantized valley-Hall conductivity of the bulk (an exception is the monolayer case where there are only flat bands connecting two valleys in the same conduction or valence band). For even N, there are N/2 pairs of valley-helical edge modes located at both zigzag boundaries, in consistence with the quantization of the valley Hall conductivity. However, for odd N, a qualitatively distinct feature is observed for the valley Hall edge modes that we shall discuss in detail later~\cite{QSHE_TLG_Qiao_12}. Since there must be no inter-valley-scattering, the quantum valley Hall effect can be considered as a ``weak" TI when compared with the topologically-protected quantum Hall effect. This scenario resembles the requirement of the time-reversal symmetry protection for the $\mathbb{Z}_2$ TI.

Nevertheless, the valley＊s binary degree of freedom is different from the electron spin and there is no rigid bulk-edge correspondence for the QVHE, which can be seen from the following facts: 1) there are no gapless valley Hall edge modes for ribbons where valleys K and K$^\prime$ are strongly coupled, especially in the armchair case [see the left column of Fig.~\ref{QVHE_G_Edge_ZLM}]; 2) there are no gapless edge states for monolayer graphene even in zigzag nanoribbons, as shown in Fig.~\ref{QVHE_G_Edge_ZLM}(b$_1$). Therefore, in the following, we centre on the topological properties of mono- and multilayer graphene and only discuss the edge states in zigzag nanoribbons.

\begin{figure*}
  \centering
  \includegraphics[width=12cm]{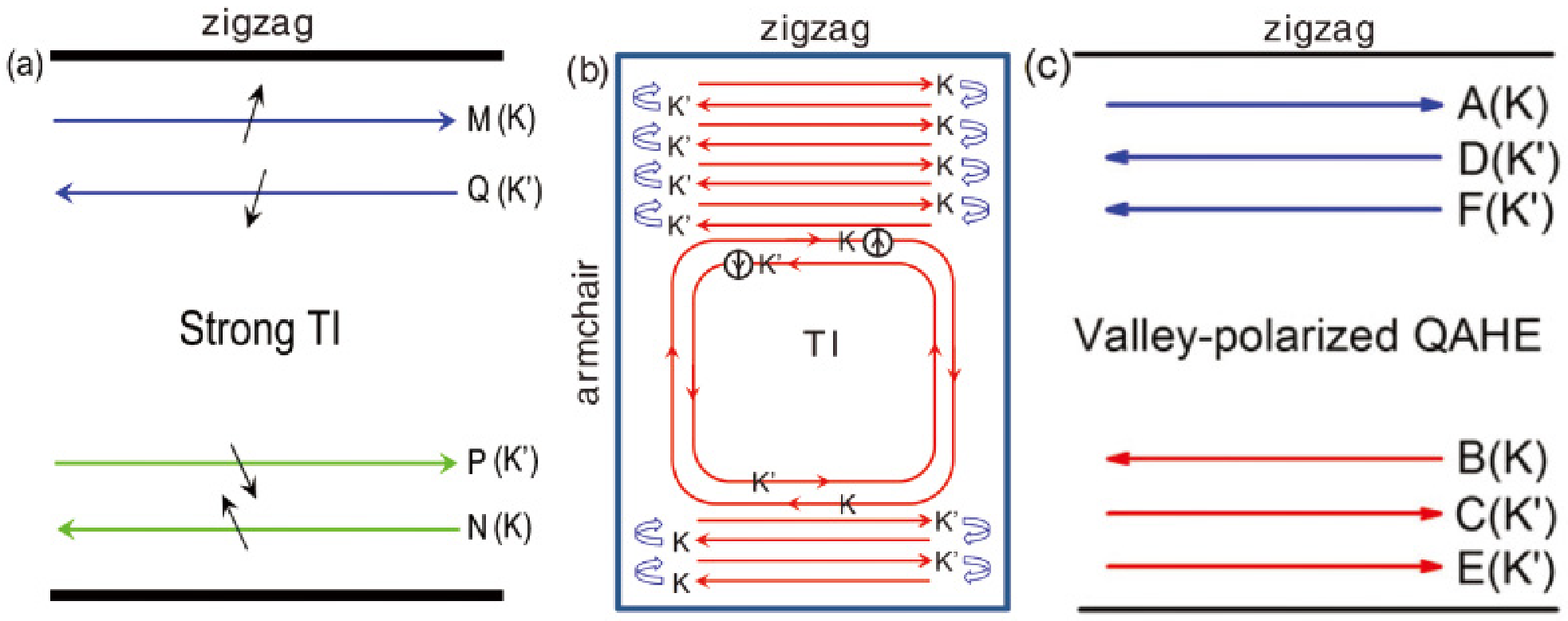}
  \caption{(colour online). Schematic of edge state propagation in the zigzag edge geometry for a TI in (a) bilayer (b) trilayer graphene. The arrows on the edge channels represent in-plane spin directions (out-of-plane spin component is zero). For trilayer graphene, the major difference from the QVHE in Fig.~\ref{QVHE_G_Edge_ZLM} is that one more pair of time-reversal invariance protected spin-helical edge states emerges at each boundary of the zigzag or armchair trilayer graphene ribbon. Note that in the zigzag geometry, all the edge modes are associated with both spin and valley degrees of freedom. (c) Schematic of edge states of the valley polarized QAHE.
  Fig.~(a) reprinted with permission from ~[\onlinecite{QSHE_BLG_Qiao_11}], copyright 2011 by the American Physical Society. Fig.~(b) reprinted with permission from~[\onlinecite{QSHE_TLG_Qiao_12}], copyright 2012 by the American Physical Society. Fig.~(c) reprinted with permission from~[\onlinecite{QAHE_QVHE_Si_Yao_14}], copyright 2014 by the American Physical Society.}
  \label{QSHE_QVHE_MLG}
\end{figure*}

For monolayer graphene, the inversion symmetry can be broken by using a substrate with inequivalent AB sublattice potentials, such as a hexagonal boron nitride (h-BN) monolayer~\cite{QVHE_G_hBN_Giovannetti_07,QVHE_G_hBN_Jung_15, ZLM_G_BN_Peeters_12}, or by adsorbing top-site adatoms in a certain type of sublattice~\cite{QAHE_G_adatom_Qiao_11}. When the inversion symmetry is broken, the Berry curvature has opposite signs at the K and K$^\prime$ valleys due to the time-reversal invariance $\Omega(-\bm{k})=-\Omega(\bm{k})$, as shown in Fig.~\ref{QVHE_G_Edge_ZLM}(c$_1$), which gives rise to opposite Chern numbers $\mathcal{C}_K=0.5$ and $\mathcal{C}_{K^\prime}=-0.5$ at different valleys for each spin~\cite{QVHE_G_NiuQ_07}. Therefore, when the spin degree of freedom is invoked, the Chern numbers for the K and K$^\prime$ valleys are, respectively, $\pm1$, which resemble the Chern numbers for the spin-up and -down copies in the QSHE. However, there are no gapless edge states but flat bands connecting valleys K and K$^\prime$ in the zigzag nanoribbons with well-separated valleys, as illustrated in Fig.~\ref{QVHE_G_Edge_ZLM}(b$_1$), in contrast to the gapless edge states in the QSHE. When the electron-electron interaction is included, the flat bands become dispersive and the spin-degeneracy is lifted, thus inducing spin-polarized edge modes with opposite spin-polarizations at opposite boundaries~\cite{QVHE_Edge_G_Son_06,QVHE_G_Edge_NiuQ_11}. This is also an approach to magnetize graphene for applications in spintronics.

In Bernal stacked multilayer graphene, the quasiparticles are chiral in the sublattice space. In the long wavelength limit, the \textit{effective} AB sublattice＊s degree of freedom in multilayer graphene is intimately related to the top/bottom layer＊s degree of freedom, which is different from that in monolayer graphene with AB sublattices in the same plane. This makes it possible to break the inversion symmetry by applying a perpendicular electric field (equivalent to introducing different site potentials in the top/bottom layers), which can also open a bulk band gap at valleys K and K$^\prime$, as displayed in the second column of Fig.~\ref{QVHE_G_Edge_ZLM}. In bilayer graphene, contrary to the half-Chern number contribution in monolayer graphene, each valley contributes to a unit Chern number. This leads to the formation of gapless valley-helical edge states in the zigzag nanoribbons. For an even number of layers, the edge states are balanced at each edge. However, they are unbalanced for an odd number of layers, as shown in Fig.~\ref{QVHE_G_Edge_ZLM} where the red lines indicate the edge states within one boundary while states in blue are localized on the other side~\cite{QSHE_BLG_Qiao_11,QSHE_TLG_Qiao_12}. Different from the QAHE chiral edge states, backscattering is possible due to the spatial overlapping of these counter-propagating edge states, as their large momentum separation protects them from the long-range scattering potential. If atomic short-range scattering does occur, e.g. through armchair termination of the graphene nanoribbon, then the valley-Hall conductance is no longer quantized and the edge states can even be destroyed, as in the case of time-reversal symmetry breaking scattering in $\mathbb{Z}_2$ TIs.

Similar results are expected in silicene~\cite{QAHE_Si_LiuWM_14, QSHE_Si_Yang_14, QSHE_Bi.Sb_QVHE_Yao_14}, germanene, and other artificial honeycomb lattice systems such as photonic crystals and optical lattices~\cite{rev_artificial_HoneyComb_Pellegrini_13}. This valley-related physics can also find application in square optical lattices, which have high tunability and so the QVHE may be realized by applying valley-dependent gauge fields~\cite{QVHE_ColdAtom_Zhu_14}.

\subsection{Coexistence of the QVHE and other topological phases}
Strictly speaking, an insulator without (with) time-reversal symmetry can be classified into several groups according to their different Chern numbers ($\mathbb{Z}_2$ index). In general, these topologically distinct phases cannot coexist simultaneously. For example, it is impossible to find a material that is both a QAHE structure and a $\mathbb{Z}_2$ TI, because the former requires breaking of time-reversal symmetry while the latter requires its preservation. However, the QVHE originates from the local Berry curvature around Dirac points that are well separated in momentum space. Therefore, in principle it does not conflict with the QAHE or a TI that originated from the global topology of the band structure, and should be compatible with either of them. This will now be discussed below.

\subsubsection{Time-reversal invariant systems}
In honeycomb lattice materials, there are two representative physical mechanisms that can give rise to a $\mathbb{Z}_2$ TI. One is to introduce intrinsic spin-orbit coupling into a monolayer graphene. It was shown that in Haldane＊s model or Kane-Mele＊s model, this intrinsic spin-orbit coupling competes with the inversion-symmetry breaking term, i.e. the staggered sublattice potential~\cite{QAHE_Haldane_88}. Therefore, this kind of TI based on intrinsic spin-orbit coupling cannot coexist with the QVHE. The other mechanism is to consider the Rashba spin-orbit coupling in gated Bernal stacked graphene multilayers ~\cite{QSHE_BLG_Qiao_11, QSHE_TLG_Qiao_12}. Contrary to the intrinsic spin-orbit coupling induced TI phase where the spin degenerate gapless edge states connect the conduction and valence band edges of the K and K$^\prime$ valleys, respectively, in the Rashba spin-orbit coupling induced TI the gapless edge modes link the conduction and valence band edges in the same K or K$^\prime$ valley [see Fig.~\ref{QSHE_BLG}(l)], which gives rise to a well-defined valley Chern number. Therefore, the Kramers-degenerate pairs at the edge possess opposite spins as well as opposite valleys, as displayed in Fig.~\ref{QSHE_QVHE_MLG}, thus the QVHE and $\mathbb{Z}_2$ TI can be realized simultaneously.

\subsubsection{Time-reversal symmetry breaking systems}
Although a well-defined valley index can also be given to the QAHE edge states in monolayer and Bernal-stacked multilayer graphene, the inversion symmetry relates the K and K$^\prime$ valleys and guarantees the equal contribution of these two valleys to the Chern number and hence the number of edge states. As a consequence, the edge modes from both valleys propagate along the same direction, exhibiting a chiral propagation nature and giving rise to a vanishing valley current ~\cite{QAHE_G_Qiao_10,QAHE_G_Qiao_11,QAHE_G_Qiao_12}, which is different from the case of a $\mathbb{Z}_2$ TI in multi-layer graphene where the edge states in different valleys are related via the time-reversal symmetry and hence propagate in opposite directions~\cite{QSHE_BLG_Qiao_11}. When the inversion symmetry is further broken, for example because of a low buckled structure~\cite{QAHE_QVHE_Si_Yao_14,QAHE_QVHE_Si_YangSY_15} or a staggered sublattice potential~\cite{QAHE_Bi-H_Yao_15,QSHE_Bi_QAHE_QVHE_Mokrousov_15}, a band inversion is induced in one valley while the band gap in the other valley is preserved, then the unbalanced contribution to the Chern number from these two valleys leads to a new type of topological phase, i.e. the valley-polarized QAHE phase~\cite{QAHE_QVHE_Si_Yao_14, QAHE_QVHE_Si_YangSY_15, QAHE_Bi-H_Yao_15, QSHE_Bi_QAHE_QVHE_Mokrousov_15}.

This new phase has been reported in monolayer silicene~\cite{QAHE_QVHE_Si_Yao_14, QAHE_QVHE_Si_YangSY_15} and half-hydrogenated Bi bilayers~\cite{QAHE_Bi-H_Yao_15, QSHE_Bi_QAHE_QVHE_Mokrousov_15}, which possess the charactersitics of both QVHE and QAHE. In monolayer silicene, the joint influence of the Zeeman field and Rashba spin-orbit coupling leads to a QAHE with a Chern number of $\mathcal{C}=2$, equally contributed from valleys K and K$^\prime$, similar to that in graphene. Unlike monolayer graphene, two types of Rashba spin orbit-coupling, intrinsic and extrinsic, can exist in silicene, either of which can induce the QAHE. However, their coexistence leads to competition, so that the contributions to the Chern number from valleys K and K$^\prime$ are, respectively, $\mathcal{C}_K=1$ and $\mathcal{C}_{K{^\prime}}=-2$, resulting in a Chern number of $\mathcal{C}=\mathcal{C}_K+\mathcal{C}_{K^\prime}=-1$ and a valley Chern number of $\mathcal{C}_V=\mathcal{C}_K-\mathcal{C}_{K^\prime}=3$. As shown in Fig.~\ref{QSHE_QVHE_MLG}(c), such a valley imbalance induces both a net valley current and a net charge current at each edge of the zigzag silicene nanoribbon~\cite{QAHE_QVHE_Si_Yao_14}. For the half-hydrogenated Bi bilayer, however, due to the strong sublattice imbalance, the Chern number contributions from the two different valleys are, respectively, $\mathcal{C}_K=1$ and $\mathcal{C}_{K^\prime}=0$, indicating a charge Chern number of $\mathcal{C}=1$ and a valley Chern number of $\mathcal{C}_V=1$~\cite{QAHE_QVHE_Si_Yao_14, QAHE_QVHE_Si_YangSY_15}. It is noteworthy that, although the $\mathbb{Z}_2$ TI and the QAHE cannot exist simultaneously, the time-reversal symmetry breaking QSHE can coexist with the QAHE, in the same manner as reported in Ref.~[\onlinecite{QAHE_QSHE_QVHE_Si_Ezawa_13}].

\subsection{Topological zero-line modes}
\subsubsection{Proposals and electronic structures}

\begin{figure}[h]
 \includegraphics[width=8cm]{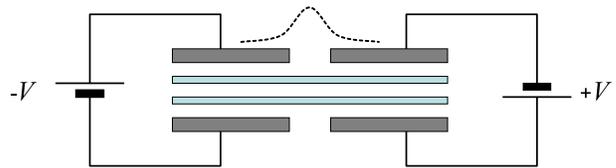}
  \caption[] {(color online). Side view of gated bilayer graphene configuration with a voltage kink.  The region where the interlayer voltage changes sign  (channel) supports bands of chiral zero-modes (dashed line). Conventional (non-topological) confinement would correspond to the same polarity of bias on both sides of the channel.
  Reprinted with permission from~[\onlinecite{ZLM_BLG_Morpurgo_08}], copyright 2008 by the American Physical Society.}
  \label{ZLM_BLG_Morpurgo_08_Fig1}
\end{figure}

When the valley Hall topologies are varied spatially, e.g. by applying electric fields in different directions, a topological ZLM arises along the interface between the regions with opposite valley-Chern numbers. These ZLMs based on topological confinement were first proposed by Martin \textit{et al} in a bilayer graphene continuum model, as illustrated in Fig.~\ref{ZLM_BLG_Morpurgo_08_Fig1}, where two electric fields opposite in sign are applied at two regions separated by a zero line~\cite{ZLM_BLG_Morpurgo_08}. Similar to the QVHE in multilayer graphene, chirally propagating gapless edge states are present in the zigzag terminated zero-line as displayed in Fig.~\ref{QVHE_G_Edge_ZLM}(d$_2$), where an equal number of ZLMs are present in the K and K$^\prime$ valleys propagating along opposite directions.
However, in the armchair-terminated zero-line, the chirally propagating edge states are also present despite a relatively small, avoided crossing band gap at the crossing points between counter-propagating edge states, as shown in Fig.~\ref{QVHE_G_Edge_ZLM}(e$_2$) , which is different from the QVHE electronic structure in the armchair nanoribbon plotted in Fig.~\ref{QVHE_G_Edge_ZLM}(a$_2$). Although in the QVHE there is no rigid bulk-edge correspondence at the boundary between the bulk and vacuum, the ZLMs are shown to be robust, with their number characterized by the difference in the valley Chern numbers across the interface~\cite{ZLM_BLG_BEC_Martin_10, ZLM_G_BEC_Morpurgo_12}. Similar results are present in Bernal-stacked multilayer graphene [see the 4th and 5th columns of Fig.~\ref{QVHE_G_Edge_ZLM}], while ZLMs can even form at the interface between multilayer graphene structures with different layer numbers, and the resulting pairs of ZLMs are determined by the difference in the Chern numbers of a single valley across the interface~\cite{ZLM_MLG_NiuQ_11}.

This scenario can also be extended to the monolayer graphene case~\cite{ZLM_G_ZhouF_08}, where the quantum valley Hall gap is opened by a staggered sublattice potential~\cite{ZLM_G_Niu_09, ZLM_G_BN_Peeters_12} rather than a perpendicular electric field since the AB sublattices reside in the same plane. The ZLMs in monolayer graphene with different edges have been investigated by employing a tight-binding model rather than a long wavelength low-energy continuum model. It is found that although gapless ZLMs are present at the zigzag-type zero-line interface [see Fig.~\ref{QVHE_G_Edge_ZLM}(d$_1$)], an avoided-crossing band gap is opened at the crossing points of ZLMs inside the bulk band gap in the armchair-type zero-line interface, as displayed in Fig.~\ref{QVHE_G_Edge_ZLM}(e$_1$)~\cite{ZLM_G_ZhouF_08}. These effects can also find application in silicene, where the low-buckled structure allows the realization of ZLMs via the application of a perpendicular electric field similar to that in bilayer graphene~\cite{ZLM_Si_Chan_14}.

\begin{figure}
 \includegraphics[width=6 cm]{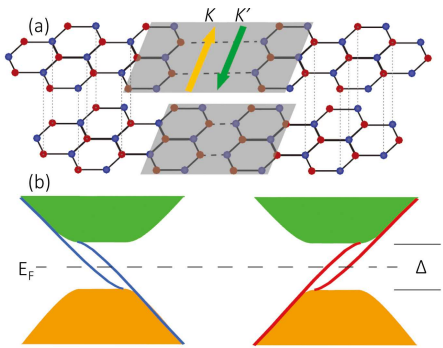}
  \caption[] {(color online). (a) Schematic of the zero line as a interface between AB and BA stacked bilayer graphene and (b) the corresponding band structure under electric field.
  Reprinted figure with permission from~[\onlinecite{ZLM_BLG_ABBA_exp_Wang_15}].}
 \label{ZLM_ABBA}
\end{figure}

Additionally, ZLMs can also be generated at the interface between bilayer graphene structures with different stacking orders (e.g. AB or BA stacking) under a uniform electric field, as illustrated in Fig.~\ref{ZLM_ABBA}(a)~\cite{ZLM_BLG_ABBA_Kim_13, ZLM_BLG_ABBA_Mele_13, ZLM_BLG_ABBA_exp_Wang_15}. Since the valley Chern numbers that characterize valley topologies change sign across the interface between AB and BA stacking layers, the same ZLMs arise along the interface [see Fig.~\ref{ZLM_ABBA}(b)]. Compared with the ZLMs induced by spatially manipulating the external electric fields, the different stacking order-induced ZLMs are rather easier to realize experimentally. Very recently, such a scheme was demonstrated in a suspended bilayer graphene system with a smoothly-varying interface that effectively suppressed the inter-valley scattering due to the absence of atomic scale disorders~\cite{ZLM_BLG_ABBA_exp_Wang_15}. The uniform electric field induced ZLMs at the interface separating different stacking orders can also find application in low-buckled honeycomb structures, e.g. silicene, where the zero lines form at the interface between two regions with opposite buckling configurations~\cite{ZLM_Si_Ihm_14}. Apart from the zero lines formed in some ribbons, the ZLMs can also be realized in superlattice structures, where periodic AB and BA stacking patterns are present ~\cite{ZLM_BLG_ABBA_exp_McEuen13}. The superlattices with adjacent AB and BA stacking orders, with boundaries that may support ZLMs, can also be realized in graphene on top of h-BN substrates as a result of lattice mismatch~\cite{ZLM_G-BN_ring_Miller_12}. In addition to the line defect exhibited when the stacking order of bilayer graphene changes, 1D ZLMs can also appear along the line defect of monolayer graphene if a suitable staggered sublattice potential is present~\cite{ZLM_G_Xie_12}.

In the above, we have mainly described ZLMs based on the time-reversal symmetric QVHE, where the time-reversal counterpart of the ZLMs at one valley K is located at the other valley K$^\prime$~\cite{ZLM_BLG_ABBA_Mele_13}. However, this scenario is not merely restricted to the interface between two such systems with different valley Chern numbers. In principle, it should exist at any interface separating two different topological orders, e.g. in QAHE systems with different Chern numbers (i.e. +$\mathcal{C}$/-$\mathcal{C}$, $+\mathcal{C}_1$/$+\mathcal{C}_2$, and $+\mathcal{C}_1$/$-\mathcal{C}_2$), in QSHE systems with different topologies, in hybrid systems composed of both effects ~\cite{ZLM_G_Yang_15} (or $\mathbb{Z}_2$ TI~\cite{ZLM_Si_Chan_14}), and in hybrid systems composed of the QAHE and a $\mathbb{Z}_2$ TI. To be specific, in monolayer graphene, the contributions of the Chern numbers from valleys K and K$^\prime$ are identical for the QAHE, but are opposite for the QVHE. Therefore, at the interface between the two, the difference between the Chern numbers is nonzero for the valley K while it vanishes for the other valley K$^\prime$. As a result, the ZLMs at the interface carry only the information of valley K and propagate chirally only along one direction, which is similar to the QAHE chiral edge modes but is valley-polarized.

In fact, the interface is not required to be a ``line", but may be slightly broadened due to mediation from the finite-size effect. For example, it is found that 1D ZLMs can also be hosted in a narrow graphene nanoroad embedded in h-BN sheets, where the boron (or nitrogen) atoms belong to different sublattices in two separate h-BN sheets~\cite{ZLM_G_Nanoroad_Qiao_12}. Due to the inversion-symmetry breaking from the unbalanced site-energies in h-BN, a bulk band gap is opened at the K and K$^\prime$ valleys, which each carry half a Chern number of opposite sign. The inverse of the sublattice topology across the graphene nanoroad changes the sign of the Chern number for each individual valley, leading to the formation of ZLMs throughout the graphene ribbon region.

\subsubsection{Transport properties of topological zero-line modes}
Based on either the low-energy continuum model in the long wavelength limit or the tight-binding model Hamiltonian, we have now established a fundamental understanding of the electronic structure of ZLMs. The gapless modes appear in the zigzag-type interfaces where the K and K$^\prime$ valleys are decoupled, while a relatively small but nonnegligible band gap opens in the armchair-type interfaces due to the atomic structure induced strong inter-valley scattering. For any type of zero lines, the resulting ZLMs are always spatially overlapped in real space, which suggests that these counter-propagating states should be easily backscattered, especially in the case where inter-valley scattering occurs. However, it is shown that the wide-spread wavefunctions mitigate the backscattering~\cite{ZLM_LuttingerLiquid_Paramekanti_10, ZLM_MLG_NiuQ_11} and these ZLMs also play an important role in the subgap conductance even in the presence of short-range disorder scattering~\cite{ZLM_BLG_Morpurgo_11} as reviewed in below.

To further verify the robustness of the ZLMs, electronic transport calculations have been performed using the Green's function technique implemented with the multi-terminal Landauer-B\"uttiker formalism~\cite{ZLG_BLG_MacDonald_11,ZLM_G_NiuQ_14,ZLM_G_Nanoroad_Qiao_12}. It has been shown that the ZLMs exhibit a chiral propagation characteristic for any kinds of zero lines, i.e. periodic, or curved with inter-valley scattering. Such a robustness against any specific zero-line geometry indicates their striking transport property of \textit{zero} bend-resistance, similar to the dissipationless transport property of the quantum Hall effect in some extent. In the presence of disorder, either short- or long-range, it is shown that the ZLMs are rather robust even under any variation of the path directions whenever the Fermi-level does not lie inside the avoided crossing band gap of the armchair-type zero lines. The corresponding mean free path under some weak disorder is estimated to be as large as several microns, which promises innumerable applications in low power electronics and valleytronics. The robustness against disorder or path directions can be attributed to the wide spread of the counter-propagating ZLMs carrying opposite valley degrees of freedom~\cite{ZLG_BLG_MacDonald_11}.

\begin{figure}
  \includegraphics[width=8 cm]{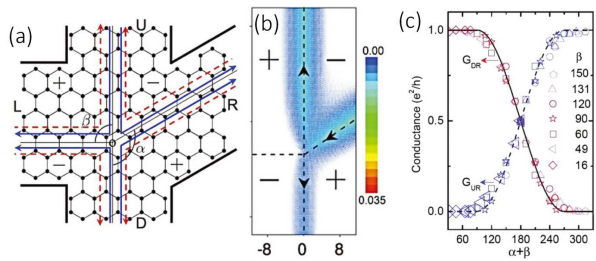}
  \caption[] {(color online). Current partition at the intersection of zero line modes. (a) Schematic of the zero line intersection.  L, D, R, and U represent the left, down, right, and up leads, respectively. The angle between left and up leads is denoted by $\alpha$, and that between right and down leads by $\beta$. The blue and red arrow lines indicate the chiral propagation of the ZLMs. (b) Partition of the current flowing from the right lead. (c) Conductance of the currents from the R zero line to the U and D zero lines as a function of $\alpha+\beta$ at the ZLM intersection in (a) for a series of $\beta$ values.
  Reprinted with permission from~[\onlinecite{ZLM_G_NiuQ_14}], copyright 2014 by the American Physical Society.}
 \label{ZLM_Current_Partition}
\end{figure}

According to the zero bend resistance characteristics of the ZLMs, when two zero-lines become crossed to form a topological intersection [see Fig.~\ref{ZLM_Current_Partition}(a)], the chiral propagating modes obey an interesting and counterintuitive current partition law at the intersection, which depends only on the angle of incidence~\cite{ZLM_G_NiuQ_14}. For example, when the incoming current from one terminal flows towards the intersection, the direct forward current cross the crossing point is forbidden due to the requirement of the chiral propagation, and the outgoing currents have only two possible directions, as illustrated in Fig.~\ref{ZLM_Current_Partition}(a). Moreover, a counterintuitive current partition is formed because the incoming current prefers to turn a larger angle, as shown in Fig.~\ref{ZLM_Current_Partition}(b)~\cite{ZLM_G_NiuQ_14}. For the Fermi level close to the charge neutrality point (i.e. $E_F=0.0$), the splitting of the current from any lead (e.g. the right lead) at the topological intersection simply depends on the combination of $\alpha+\beta$ [Fig.~\ref{ZLM_Current_Partition}(c)] with $\alpha$ ($\beta$) being the angle between the left (right) and up (down) leads, as shown in Fig.~\ref{ZLM_Current_Partition}(a). Such a counterintuitive current partition can be understood from the coupling between the wavefunctions at different paths forming the topological intersection~\cite{ZLM_G_NiuQ_14}.

However, although these ZLMs were theoretically predicted several years ago and were shown to have great application potential, experimental progress has been rather limited. The main difficulty is the design and fabrication of the devices, which for multilayer graphene require high precision alignment of the top and bottom gates, not to mention the control of eight gates in topological current splitter devices. With the development of state-of-the-art techniques for fine-tuning of the gates, it should be possible for such zero-lines to be be engineered in the future. For current splitters made of multilayer graphene, at the topological intersection the top and bottom electric gates are in principle fixed, so the angles can no longer be tuned to realize different current partitions. For fixed multilayer graphene systems we must therefore explore other approaches to achieve tunability, e.g. by tuning Fermi levels, applying some appropriate electric fields, or weak magnetic fields.

\section{Summary} \label{summary}
In summary, we have provided an overview of the most recent research on the topologically nontrivial phases in 2D systems, such as $\mathbb{Z}_2$ TI, QAHE, and QVHE. A typical 2D material is graphene, a Dirac semi-metal with a half-filled valence band. The Dirac dispersion around the K and K$'$ points are guaranteed by the sublattice symmetry and can be gapped by breaking this symmetry. Without spin-orbit coupling, a staggered sublattice potential can break the inversion symmetry to open up a band gap and form a QVHE with each valley spin carrying half a unit Chern number. In this effect, although there are no corresponding gapless edge state in monolayer graphene, midgap topological confinement states, i.e. ZLMs, can occur at the interface, across which the staggered potential changes sign. These ZLMs are protected from backscattering by the large momentum separation, but become gapped at the specific armchair zero lines. Moreover, it was shown that the ZLMs exhibit striking transport properties, e.g. zero bend resistance and counterintuitive current partition laws. Similar results can be applied to other honeycomb lattices and chirally stacked multilayer graphene. Recently, the ZLMs have already been experimentally observed in bilayer graphene through the application of a tunable electric field. These electric-field-tunable ZLMs provide the potential building blocks for constructing the next-generation electronics and valleytronics.

The Haldane model provides another scheme to break the chiral symmetry, i.e. by applying an alternating magnetic fluxes in a honeycomb lattice. The orbital effect of the magnetic field leads to the QAHE with a Chern number of $\mathcal{C}=1$. Although the time-reversal symmetry is broken, Landau levels are not formed due to the vanishing total magnetic flux. The orbital effect of the magnetic field induced QAHE may be seen in a buckled honeycomb lattice with an in-plane magnetic field. In addition, to form QAHE, there is another scheme that relies on spin-orbit coupling and out-of-plane ferromagnetism. In both semi-metals (e.g. graphene) and semiconductors (e.g. 3D-TI thin films), a ferromagnetic order Zeeman field can invert the conduction and valence bands and induce crossing points. The spin-orbit coupling can lift the accidental degeneracies at the crossing points, open up a bulk band gap, and thus give rise to the QAHE. Various semi-metals and semiconductors have been proposed as host materials, as listed in Table~\ref{QAHE_Table}, where the ferromagnetic order can be established by magnetic doping, considering magnetic insulating substrates, or functionalizing. Alternatively, the QAHE can also be realized in materials with spontaneous ferromagnetism, like transition metal oxides and heterostructures composed of magnetic thin films. However, since the basic requisite is just the breaking of time-reversal symmetry, ferromagnetic metal with Anderson disorders, and systems with in-plane magnetization or anti-ferromagnetism may also be possible. Although the QAHE has been extensively studied on elements with an outer $p$-shell, the explorations for transition metal compounds have been rather limited. The influence of the strong electron-electron correlations of these transition metals on the QAHE is also unclear.  So far, the QAHE has only been observed under extremely low temperatures in magnetically doped 3D-TI thin films with the Chern number of $\mathcal{C}=1$. The recent progress has shown the potential of realizing QAHE in graphene on insulating ferromagnetic substrate. For future applications, a large band gap (or high temperature) and simple experimental design are the motivations behind current research. Moreover, a large Chern number also benefits practical applications due to the strong quantized Hall-current density. In addition, artificial lattices with high tunability, like cold atoms in optical lattices, provide alternative platforms for simulation of the QAHE.

Another approach to break the chiral symmetry of graphene is to introduce the next-nearest-neighbour hopping related intrinsic spin-orbit coupling. The resulting QSHE can be regarded as a combination of two copies of the Haldane model with opposite spins and Chern numbers, which gives rise to spin-helical edge modes. These states are stable even when $s_z$ is not a good quantum number, reflecting the topological nature that is characterized by a binary-valued $\mathbb{Z}_2$ topological index. This is a brand-new classification of insulators with time-reversal invariance. Though the intrinsic spin-orbit coupling of graphene is extremely weak, it can be externally engineered via various methods, like adsorbing heavy atoms. The $\mathbb{Z}_2$ TI also exists in other Dirac materials, e.g. honeycomb lattices of group-IV elements, organic materials, \textit{etc}. as listed in Table~\ref{Tab_TI_Materials}. Except for these intrinsic Dirac materials with two inequivalent Dirac cones, strain and electric fields can also induce Dirac dispersions that can be gapped by the spin-orbit coupling to form a $\mathbb{Z}_2$ TI.

Apart from the Dirac materials, semiconductors with a small band gap can also generate a $\mathbb{Z}_2$ TI when spin-orbit coupling is strong enough to close the band gap, and hence heavy atoms such as Bi, Tl, Te, and Hg may play an important rule. The Bi (111) bilayer is a good example of a group-V element. Although the Bi bilayer has a honeycomb-lattice structure, the full-filled valence bands make it an insulator rather than semimetal with a bulk gap opened at the $\Gamma$ point. The strong spin-orbit coupling from Bi induces a band inversion to form a TI. Similar band inversion induced TIs can also be found in other atomic crystal layers, e.g. Bi or Tl based III-V compounds and Bi based V-VII compounds. Superior to the TI with light atoms, the heavy atom based TIs usually have large nontrivial band gaps, which offers the possibility for room-temperature applications. Additionally, functionalization is another useful approach to obtain TIs with a large band gap. So far, the $\mathbb{Z}_2$ TI has already been observed in CdTe/HgTe/CdTe and InAs/GaSb semiconductor quantum wells, based on band inversion. Such kind of band inversion is also expected to occur by disorders, leading to the ``topological Anderson insulator".

One of the most intriguing properties of these topological phases is the topologically protected edge modes that perform as perfectly conducting 1D wires. This holds great potential as building blocks in dissipationless quantum electronic devices, like interconnects between chips. The helical edge states of opposite spin (valley) in QSHE (QVHE) may also have potential applications in spin/valley-related electronics, for example, spin (valley) filter. Moreover, the proximity effect of $\mathbb{Z}_2$ TI on superconductors leads to the formation of an exotic quasi-particle of Majorana fermion~\cite{TI_MF_Kane_09,MF_QC_Alicea_12}, which may have promising applications in fault-tolerant quantum computation~\cite{MF_QC_Kitaev_03,MF_QC_Nayak_03}; the resulting Andreev reflection and crossed Andreev reflection also possess promising applications in quantum teleportation and quantum computation by making use of the spatially-separated electrons with entangled spin and momentum~\cite{Andreev_Beckmann_04, Andreev_Russo_05, MF_QC_Nielsen_00, MF_QC_Chtchelkatchev_02}. Similar proximity effects of the QAHE/QVHE proximity-coupled to superconductors are still open issues for future study.

\begin{acknowledgments}
Y.R. appreciates the valuable discussions with Fan Zhang and Yanyang Zhang as well as the help from Xinzhou Deng and Ke Wang on plotting figures. Z.Q. and Y.R. are financially supported by the 100-Talent Program of Chinese Academy of Sciences, China Government Youth 1000-Plan Talent Program, National Natural Science Foundation of China (NNSFC, Grant No. 11474265), and Anhui Provincial Natural Science Foundation. Q.N. is financially supported by the Welch Foundation (Grant No. F-1255), DOE (Grant No. DE-FG03-02ER45958, Division of Materials Science and Engineering), the MOST Project of China (Grant No. 2012CB921300, 2013CB921900), and NNSFC (Grant No. 91121004). The Supercomputing Center of the University of Science and Technology of China is gratefully acknowledged for high-performance computing assistance.
\end{acknowledgments}

\end{document}